\documentclass[a4paper,11pt]{article}
\pdfoutput=1 
\usepackage{jheppub} 
\usepackage[T1]{fontenc} 
\usepackage{amsmath,amssymb,epsfig,txfonts,relsize,slashed}
\usepackage[utf8]{inputenc}

\allowdisplaybreaks

\definecolor{nicered}{rgb}{0.7,0.1,0.1}
\definecolor{nicegreen}{rgb}{0.1,0.5,0.1}
\definecolor{niceblue}{rgb}{0.0,0.1,0.7}
\hypersetup{colorlinks,citecolor=nicegreen,linkcolor=nicered,urlcolor=niceblue}

\def \beq{\begin{equation}}
\def \eeq{\end{equation}}
\def \bea{\begin{eqnarray}}
\def \eea{\end{eqnarray}}

\usepackage{amsfonts}
\usepackage{graphicx}
\usepackage{mathtools}
\usepackage{braket}
\usepackage{enumerate}
\usepackage{bbold}
\usepackage{slashed}

\newcommand{\dbar}{{d\mkern-7mu\mathchar'26\mkern-2mu}}

\DeclareMathOperator{\Tr}{Tr}

\title{Sum rules for CP-violating operators of Weinberg type}

\author[1]{Ulrich Haisch}

\author[1]{and Amando Hala}

\affiliation[1]{Max Planck Institute for Physics, F{\"o}hringer Ring 6,  80805 M{\"u}nchen, Germany}   

\emailAdd{haisch@mpp.mpg.de}
\emailAdd{ahala@mpp.mpg.de}

\abstract{
\phantom{iii} We estimate the size of the hadronic matrix elements of CP-violating three-gluon and four-gluon Weinberg operators  using sum-rule techniques. In the three-gluon case, we are able to reproduce the expressions given in earlier works, while the four-gluon results obtained in this article are new. Our paper therefore represents the first systematic study of contributions to the electric dipole moment of the neutron due to CP-violating dimension-six and dimension-eight operators. We provide many details on both the derivation of the sum rules as well as the analysis of the uncertainties that plague our final predictions.  }

\preprint{}

\begin{document} 

\maketitle

\section{Introduction}
\label{sec:introduction}

Searches for electric dipole moments (EDMs) place stringent constraints on any beyond the standard model~(BSM) scenario with additional sources of CP violation~(see~\cite{Pospelov:2005pr,Li:2010ax,McKeen:2012av,Engel:2013lsa,Chang:2013cia,Jung:2013hka,Gripaios:2013lea,Brod:2013cka,Inoue:2014nva,Gorbahn:2014sha,Altmannshofer:2015qra,Dwivedi:2015nta,Chien:2015xha,Cirigliano:2016njn,Cirigliano:2016nyn,Yamanaka:2017mef,Yanase:2018qqq,Dekens:2018bci,Cesarotti:2018huy,Panico:2018hal,Brod:2018pli,Brod:2018lbf,Cirigliano:2019vfc} for reviews and recent discussions).  At present the strongest  limits are set  by measurements of the electron spin precession in thorium monoxid~\cite{Baron:2013eja,Andreev:2018ayy}, the EDM of the neutron~(nEDM)~\cite{Baker:2006ts,Afach:2015sja} and the mercury atom~\cite{Griffith:2009zz,Graner:2016ses}. While the thorium monoxid measurements can be interpreted as a probe of the electron EDM with small theoretical uncertainties~\cite{etde_22253800,Fleig:2014uaa}, nucleon, nuclear and diamagnetic EDMs receive contributions from several effective operators that are plagued by  theoretical uncertainties of different sizes. For instance, the EDM contributions from down and up quarks  to the nEDM have been calculated  with an accuracy of ${\cal O} (5\%)$ using lattice QCD~(LQCD)~\cite{Bhattacharya:2015esa,Bhattacharya:2015wna,Gupta:2018lv}, while sum-rule calculations~\cite{Pospelov:2000bw,Lebedev:2004va,Hisano:2012sc} allow to determine the nEDM contributions from the down-quark and up-quark chromomagnetic EDMs~(CEDMs) with  uncertainties of ${\cal O} (50\%)$.  To date only estimates of the hadronic  matrix element of the leading  operator of Weinberg type exist. These rely on either naive dimensional analysis~(NDA)~\cite{Weinberg:1989dx},  the vacuum insertion approximation~(VIA)~\cite{Bigi:1990kz} or  sum rules~\cite{Demir:2002gg}. The resulting uncertainties are hard to quantify, but commonly said to be of~${\cal O} (100\%)$.  LQCD computations of the contributions of the CEDMs and the leading Weinberg operator have gained significant momentum in recent years~\cite{Bhattacharya:2015rsa,Bhattacharya:2016rrc,Abramczyk:2017oxr,Dragos:2017wms,Rizik:2018lrz,Kim:2018rce,Bhattacharya:2018qat,Syritsyn:2019vvt}, and considering the ongoing efforts by several LQCD groups, calculations with uncertainties similar to those of the sum-rule estimates may be achievable within the next five years~\cite{Gupta:2019fex,Cirigliano:2019jig}. To fully exploit the expected increase in sensitivity of future EDM searches~(see for instance~\cite{Hewett:2012ns,Schmidt-Wellenburg:2016nfv,Ito:2017ywc} for discussions), improved calculations of the hadronic matrix elements of  CEDMs and Weinberg-type operators are direly needed.

The goal of this article is to determine the  hadronic matrix elements of the following effective operators of Weinberg type~\cite{Weinberg:1989dx,Morozov:1985ef,Braaten:1990gq,Braaten:1990zt,Chang:1991ry,Booth:1992iz} 
\beq \label{eq:o6o8}
O_{6} = f^{ABC} \hspace{0.5mm}  \tilde G_{\mu \nu}^A G^{B \, \nu \rho}  G^{C \, \mu}_{\rho} \,, \qquad 
O_{8} = c^{ABCD} \hspace{0.5mm} \tilde G_{\mu \nu}^A G^{B \, \mu \nu}  G_{\rho \lambda}^C G^{D \, \rho \lambda} \,.
\eeq
Here  $G^A_{\mu \nu}$ is the QCD field strength tensor, $\tilde G^{A \, {\mu \nu}} = 1/2 \hspace{0.5mm} \epsilon^{\mu \nu \rho \lambda} \, G^A_{\rho \lambda}$ with $\epsilon^{0123} = +1$ denotes its dual, $f^{ABC}$ are the fully anti-symmetric structure constants of $S\!U(3)$ and $c^{ABCD}$ denote the colour structures defined in~(\ref{eq:cABCD}). Lacking the expertise in LQCD as well as the needed computer resources, we will present estimates of the hadronic matrix elements of the operators in~(\ref{eq:o6o8}) using  QCD sum-rule techniques.  In the case of the dimension-six contribution $O_6$ such a calculation has already been performed in~\cite{Demir:2002gg}, but the latter publication does not provide details on the actual computation making an independent reevaluation worthwhile. Our determination of the hadronic matrix elements of the dimension-eight term $O_8$ is instead new. Both results will be used in a companion paper~\cite{Haisch:2019xyi}, where we derive model-independent bounds on CP-violating Higgs-gluon interactions in BSM scenarios with vanishing or highly suppressed light-quark Yukawa couplings. 

Our work is organised as follows. After briefly reviewing the basic idea behind the sum-rule determinations of the hadronic matrix elements of $O_6$ and $O_8$, we discuss in  Section~\ref{sec:pipheno}  the phenomenological side of the sum rules. The operator product expansion~(OPE) computation of the dimension-six and dimension-eight contributions is described in Section \ref{sec:piope6}  and Section \ref{sec:piope8}, respectively. The matching and the numerical analysis of the sum rules are performed in Section~\ref{sec:matchingandnumerics}. We conclude in Section~\ref{sec:conclusions}. Technical details are relegated to several appendices. 

\section{General idea behind the sum rules}
\label{sec:general}

The central object for the derivation of the sum-rule estimates for the hadronic matrix elements of operators of the type~(\ref{eq:o6o8}) is the following correlation function
\beq \label{eq:piq}
\Pi (q^2) = i \int \! d^4x \, e^{iqx} \, \left \langle \Omega \left |T\left[ \eta(x) \hspace{0.5mm} \bar{\eta}(0) \right] \right | \Omega \right \rangle_{{\rm EM}, O_k} \,,
\eeq
where $T$ denotes time ordering, $\eta(x)$ is an interpolating field that has the same quantum numbers as the neutron, $\ket{\Omega}$ represents the vacuum on a CP-conserving background and the subscripts ${\rm EM}$ and $O_k$ imply that the correlator is evaluated in the presence of a constant external electromagnetic~(EM) source and one of the operators introduced in~(\ref{eq:o6o8}). The basic idea is to calculate~(\ref{eq:piq}) using two different approaches and to match the results to obtain an analytic expression for the nEDM in terms of hadronic quantities. In the first approach, one defines a phenomenological form~$\Pi_{\rm phen}$ of the correlator, which incorporates the  wave function of the neutron, its EDM  and other parameters. The second approach relies instead on an OPE of the correlator leading to the object $\Pi_{\rm OPE}$ that depends on  the expectation values of effective operators, such as the three-gluon and four-gluon interactions introduced in~(\ref{eq:o6o8}). Matching the expressions for $\Pi_{\rm phen}$ and $\Pi_{\rm OPE}$ then yields the contribution of the effective operators of interest to the nEDM. To improve the accuracy of the sum rules, the correlators are, however, not matched themselves but their Borel transforms are considered. As we will explain in Section~\ref{sec:pipheno}, such a procedure removes higher-order polynomial terms and suppresses excited states. 

\section{Phenomenological side of the sum rules}
\label{sec:pipheno}

\subsection{Hadronic representation}
\label{sec:hadronicrep}

In this section, we derive the phenomenological form $\Pi_{\rm phen}$ of the correlator~(\ref{eq:piq}) following the argument presented in~\cite{Pospelov:2000bw,Hisano:2012sc,Pospelov:1999ha,Pospelov:1999mv}.  An often considered approach for the phenomenological side of two-point correlators is the use of dispersion relations~\cite{Shifman:1978bx,Ioffe:1983ju,Balitsky:1983xk,Novikov:1983gd,Reinders:1984sr,Colangelo:2000dp,Ioffe:2010zz}. Since we are interested in the correlator of two nucleon currents $\eta$ in an external EM field, we are, however, effectively dealing with a three-point correlation function. Dispersion relations for three-point correlators are less constraining than those of two-point correlators due to the lack of positivity constraints~\cite{Pospelov:1999mv}. Therefore, we  relate the correlator~(\ref{eq:piq})  to a perturbative expansion of the nucleon propagator in a non-zero and constant EM background.  We write 
\beq \label{eq:pisum}
\Pi_N (q^2) = \Pi_{N} ^{(0)}(q^2)  +  e \hspace{0.25mm} \Pi_{N} ^{(1)}(q^2)  + \ldots \,,
\eeq
where  $e$ is  the electron charge magnitude that serves as the expansion parameter. The first non-trivial term in~(\ref{eq:pisum}) describes the response of the nucleon states to the weak external perturbation and arises from a single insertion of the EM interactions 
\beq \label{eq:Lint}
\mathcal{L}_{\rm EM} (x) = J_\mu (x)  \hspace{0.5mm} A^\mu (x) \,, \qquad J_\mu (x)   = e \sum_{q=d,u} Q_q \, \bar{q} (x) \hspace{0.25mm} \gamma_\mu \hspace{0.25mm} q (x) \,.
\eeq
It takes the form 
\beq \label{eq:pi1}
 e \hspace{0.25mm}  \Pi^{(1)}_{N} (q^2) =  i\int \! d^4x \, d^4y \, e^{iqx} \left \langle \Omega \left | T\left[\eta(x)\hspace{0.5mm} \bar{\eta}(0)  \, i \mathcal{L}_{\rm EM}(y)\right] \right |  \Omega  \right \rangle   \,. 
\eeq 
Here $J_\mu$ denotes the EM current, $A^\mu$ is the photon field and $Q_q$~is the fractional electric charge of the relevant quark. Note that the EM field is a non-dynamical, classical field in this approach. 

In order to evaluate the first-order contribution to~(\ref{eq:pisum}),  we insert a complete set of hadronic states $N^\prime$ and $N^{\prime \prime}$  with the quantum numbers of the neutron into~(\ref{eq:pi1}),~i.e.~we make use of the identity $1 =\sum_{N^\prime} \ket{N^\prime}\bra{N^\prime}$ twice. Working in the so-called fixed-point gauge (see~Appendix~\ref{sec:fixedpointgauge}), which allows one to express the photon field through the QED field strength tensor employing $A_\mu(y) = -1/2 \, y^\nu F_{\mu \nu} (0)$ $\big($cf.~(\ref{eq:fexp1})$\big)$,  one obtains   the following expression for the first non-trivial term in the Taylor expansion~(\ref{eq:pisum}) of the nucleon propagator~$\Pi_N (q^2)\hspace{0.25mm}$:
\beq \label{eq:pi1sum2}
\begin{split}
 e \hspace{0.25mm} \Pi^{(1)}_{N} (q^2) =& \sum_{N^\prime,N^{\prime \prime}} \int \! d^4x \, d^4y \, e^{iqx} \, \theta(x_0 -y_0 ) \, \theta(y_0 )\,  \frac{1}{2} \, y^\nu F_{\mu \nu} (0) \\[2mm] 
& \phantom{xxxx} \times \bra{ \Omega } \eta(x) \ket{N^\prime} \bra{N^\prime} J_\mu(y) \ket{N^{\prime \prime}} \bra{N^{\prime \prime}} \bar{\eta}(0) \ket{ \Omega }  + \ldots \,.
\end{split}
\eeq
Here $\theta(z)$ denotes the Heavyside step function, the subscript $0$ indicates the time component and the ellipses represent the different combinations due to time ordering.  The double sum in~(\ref{eq:pi1sum2}) involves three types of matrix elements of the EM current. These correspond to nucleon transitions of $(i)$~ground state to ground state, $(ii)$~ground state to excited states and vice versa, and $(iii)$~excited states to excited states. 

Let us first focus on the ground-state contributions,~i.e.~the terms of the hadronic sums that involve only neutron states $\ket{n}$. Up to an arbitrary chiral phase $\chi$ the matrix elements involving  $\ket{n}$ can be parametrised by the coupling $\lambda$ between the physical neutron and the interpolating current~$\eta$ as follows 
\beq \label{eq:nmat}
\bra{ \Omega } \eta (x) \ket{n} = \lambda \hspace{0.5mm} U(\chi) \hspace{0.5mm}  u (p,s)  \,, \qquad  U(\chi) = e^{i \hspace{0.125mm} \frac{\chi}{2} \hspace{0.125mm} \gamma_5} \,.
\eeq
Here $u$ is the neutron  spinor which satisfies
\beq \label{eq:nspinor}
\big( \slashed{p} -m_n \big ) \, u (p,s) = 0 \,, \qquad \sum_s u(p,s) \hspace{0.5mm} \bar{u}(p,s) = \slashed{p}+m_n \,,
\eeq
with $\slashed{p} = p_\mu \gamma^\mu$, $m_n$ denoting the neutron mass and $\bar u (p,s)= u^\dagger (p,s) \hspace{0.25mm} \gamma_0$. Notice that for our correlator~(\ref{eq:pi1sum2}) a spin summation  is implicit in the sum over all hadronic states. 

The product of the matrix element involving the EM current and the photon field can be reduced to a set of four neutron form factors (see for instance~\cite{Chan:1997fw}) 
\beq \label{eq:matj}
\int \! d^4x \, e^{iqx}  \bra{n} J_\mu(x) \ket{n} \hspace{0.125mm} A^\mu = (2\pi)^{4} \, \delta^{(4)}\left (q-(p_2-p_1) \right ) \,  \bar{u}(p_2,s_2) \hspace{0.5mm} \Gamma_\mu   (p_1,p_2) \hspace{0.5mm} u(p_1,s_1) \hspace{0.5mm} A^\mu (q) \,,
\eeq
with
\beq \label{eq:gamman}
\Gamma_\mu (p_1,p_2) \hspace{0.5mm} A^{\mu} (q) = F_1 (q^2) \hspace{0.25mm} \gamma_\mu \hspace{0.25mm} A^{\mu}  -  \left ( F_2 (q^2) + F_3 (q^2) \hspace{0.5mm}  i  \gamma_5 \right ) \frac{\sigma_{\mu \nu} F^{\mu \nu}}{4 m_n} + F_4 (q^2)  \hspace{0.5mm} \gamma_\mu  \gamma_5 \hspace{0.5mm}  \partial_\nu F^{\mu \nu} \,.
\eeq
Here $q = p_2 - p_1$ is the outgoing momentum carried by the photon and $\sigma^{\mu \nu} = i/2 \,  [\gamma^\mu, \gamma^\nu]$ where the brackets represent the usual commutator. At~$q^2 = 0$, the form factors in~(\ref{eq:gamman}) can be identified with the fractional electric charge $Q_n$,  the magnetic moment $\mu_n$, the EDM  $d_n$ and the anapole moment $a_n$ of the neutron. Since the electric charge of the neutron is zero and  its anapole moment, as a result of the constant EM background, vanishes as well, one has explicitly 
\beq \label{eq:mndn}
\mu_n= \frac{F_2(0) }{2m_n}  \,, \qquad d_n = \frac{F_3(0)}{2m_n} \hspace{0.25mm}  \,.
\eeq 
It then follows that the tensor structures in~(\ref{eq:gamman}) associated with  $\mu_n$ and $d_n$ only differ by a factor~$i\gamma_5$, meaning that at zero-momentum transfer one can write
\beq \label{eq:formfactorstructure}
\left . \Gamma_\mu (p_1,p_2)  \hspace{0.5mm} A^{\mu} (q) \, \right |_{q^2 = 0} = - \frac{\mu_n}{ 2} \,  \left (1+\frac{d_n}{\mu_n} \, i\gamma_5 \right ) \sigma \cdot F = -\frac{1}{ 2} \,  \left (\mu_n  \hspace{0.25mm} \sigma \cdot F -  d_n  \hspace{0.25mm} \sigma \cdot \tilde F  \right ) \,,
\eeq
with $\sigma \cdot F = \sigma_{\mu \nu} F^{\mu \nu}$  etc. and we have used that $\gamma_5  \hspace{0.25mm} \sigma \cdot F = i \sigma \cdot \tilde F$.

Inserting~(\ref{eq:nmat}),~(\ref{eq:matj}) and~(\ref{eq:formfactorstructure}) into~(\ref{eq:pi1sum2}) and using~(\ref{eq:nspinor}), one obtains for the $\ket{n}$ contributions to the first-order correction~(\ref{eq:pi1sum2}) of the nucleon correlation function the following expression
\beq \label{eq:pi1ncont}
 e \hspace{0.25mm} \Pi^{(1)}_{N} (q^2) = -\frac{\lambda^2}{2 \left ( q^2 - m_n^2 \right )^2 }  \, U (\chi) \hspace{0.25mm} P \hspace{0.25mm}  U (\chi)  + \ldots \,,
\eeq
with 
\beq \label{eq:P} 
P = \left ( \slashed{q} + m_n  \right )  \left (\mu_n  \hspace{0.25mm} \sigma \cdot F -  d_n  \hspace{0.25mm} \sigma \cdot \tilde F  \right ) \left ( \slashed{q} + m_n  \right )   \,.
\eeq
Here the ellipses denote contributions due to excited states and other operators that turn out to be suppressed in the course of our analysis. Up to ${\cal O} (\chi)$ the Lorentz structure in~(\ref{eq:pi1ncont}) behaves under chiral transformations as 
\bea \label{eq:upu}
\begin{split}
U (\chi) \hspace{0.25mm} P \hspace{0.25mm}  U (\chi) & =  P + \left \{ P, i \, \frac{\chi}{2} \, \gamma_5 \right \} +{\cal O} (\chi^2) \\[2mm] 
& = m_n \left \{ \mu_n  \hspace{0.25mm} \sigma \cdot F -  d_n  \hspace{0.25mm} \sigma \cdot \tilde F , \slashed{q} \right \} + m_n^2  \left [ \mu_n  \hspace{0.25mm} \sigma \cdot F -  \left ( d_n  + \chi \hspace{0.25mm}  \mu_n \right )  \sigma \cdot \tilde F \right ] \\[1mm] 
& \phantom{xx}  + 
\slashed{q}  \left [ \mu_n  \hspace{0.25mm} \sigma \cdot F -  \left ( d_n  - \chi \hspace{0.25mm}  \mu_n \right )  \sigma \cdot \tilde F \right ]  \slashed{q} +{\cal O} (\chi^2) \,.
\end{split}
\eea
This result implies that the anti-commutators $\big \{ \sigma \cdot F , \slashed{q} \big \}$ and $\big  \{ \sigma \cdot \tilde F,  \slashed{q} \big \}$ are the only structures that are invariant under chiral rotations. 

\subsection{Phenomenological parametrisation}
\label{sec:phenoparam}

In calculating $d_n$ it should then be clear from the above discussion that one should study the operator~$\big  \{ \sigma \cdot \tilde F,  \slashed{q} \big \}$ as this structure is the unique choice with an unambiguous coefficient for what concerns the EDM. We thus make the following ansatz~\cite{Pospelov:2000bw,Hisano:2012sc,Pospelov:1999ha,Pospelov:1999mv} 
\beq \label{eq:piphen}
\Pi_{N}^{(1)} (q^2) = \frac{1}{2} \, f(q^2) \, \left \{ \sigma \cdot \tilde{F},\slashed{q} \right \} \,,
\eeq
with 
\beq \label{eq:fq2}
f(q^2) = \frac{\lambda^2 \hspace{0.25mm} m_n \hspace{0.25mm} d_n }{(q^2-m_n^2)^2}+ \sum_{N^\prime \neq n} \frac{f_{N}^\prime}{(q^2-m_n^2)(q^2-m_{N^\prime}^2)} + \sum_{N^{\prime} ,N^{\prime\prime} \neq n} \frac{f_{N^{\prime} N^{\prime \prime}}}{(q^2-m_{N^\prime}^2)(q^2-m_{N^{\prime \prime}}^2)} \,,
\eeq
for the first-order contribution to the nucleon propagator~(\ref{eq:pisum}). The first term in~(\ref{eq:fq2}) corresponds to the ground-state contribution. It matches the result that we have already derived in~(\ref{eq:pi1ncont}). The second and third term describe transitions of the ground state to excited states and vice versa and transitions of excited states to excited states, respectively. The corresponding form factors are called $f_{N^\prime}$ and $f_{N^{\prime} N^{\prime \prime}}$. They do not have  definite signs due to the lack of positivity constraints of the considered correlator~\cite{Pospelov:1999mv}.

Applying the Borel transformation defined in Appendix~\ref{sec:borel}, one finds that the numerically leading contributions of the Borel transforms of the three terms in~(\ref{eq:fq2}) are given by  
\beq \label{eq:boreltrans}
\begin{split}
c_n & = {\cal B} \left [\frac{\lambda^2 \hspace{0.25mm} m_n \hspace{0.25mm} d_n}{(q^2-m_{n}^2)^2}  \right ]  = \frac{\lambda^2 \hspace{0.25mm} m_n \hspace{0.25mm} d_n}{M^4} \, e^{-\frac{m_{n}^2}{M^2}} \,, \\[2mm]
c_{N^{\prime}} & = {\cal B} \left [\frac{f_{N^{\prime} }}{(q^2-m_{n}^2)(q^2-m_{N^{\prime}}^2)}  \right ]  \simeq \frac{f_{N^{\prime} }}{M^2 \left ( m_{N^\prime}^2 - m_n^2\right ) } \, e^{-\frac{m_{n}^2}{M^2}} \,, \\[2mm]
c_{N^{\prime} N^{\prime \prime}}  & = {\cal B} \left [\frac{f_{N^{\prime} N^{\prime \prime}}}{(q^2-m_{N^\prime}^2)(q^2-m_{N^{\prime \prime}}^2)}  \right ]  \simeq \frac{f_{N^{\prime} N^{\prime \prime}}}{M^4} \, e^{-\frac{m_{N^\prime}^2}{M^2}} \,.
\end{split}
\eeq
Here we have exploited that empirically $m_{N^\prime} \gg m_n $ and $m_{N^\prime} \simeq m_{N^{\prime \prime}}$. Compared to the ground-state contribution the mixed ground-state and excited-state contributions and the  excited-states only  contributions are  hence of approximate size 
\beq \label{eq:borelratios}
\frac{c_{N^{\prime}}}{c_n }  \simeq 0.3 \, \frac{f_{N^{\prime}} }{\lambda^2 \hspace{0.25mm} m_n \hspace{0.25mm} d_n} \left ( \frac{M}{0.6 \, {\rm GeV}} \right )^2 \,, \qquad 
\frac{c_{N^{\prime}N^{\prime \prime}}}{c_n } \simeq \frac{f_{N^{\prime} N^{\prime \prime}} }{\lambda^2 \hspace{0.25mm} m_n \hspace{0.25mm} d_n}  \, e^{-\frac{m_{N^\prime}^2 - m_n^2}{M^2}} \,,
\eeq
where we have used the  values $m_n \simeq 0.94 \, {\rm GeV}$ and $m_{N^\prime} \simeq 1.44 \, {\rm GeV}$~\cite{Tanabashi:2018oca} for the mass of the neutron and its lightest excitation  to obtain the quoted numerical prefactor.  Under the assumption that $|f_{N^{\prime}}| \simeq |f_{N^{\prime} N^{\prime \prime}}| \simeq \lambda^2 \hspace{0.25mm} m_n \hspace{0.25mm} d_n$ and setting $M = 2 \Lambda_{\rm QCD} \simeq 0.6 \, {\rm GeV}$ with $\Lambda_{\rm QCD}$ the QCD scale, the mixed ground-state and excited-state (excited-states only) contributions therefore naively amount to relative corrections of the order of $30\%$~$(5\%)$. In the following, we only include the ground-state contribution to~(\ref{eq:piphen}) in our sum-rule calculation, and estimate the uncertainties that are associated to this simplification by a  variation of the Borel mass $M$ (cf.~Section~\ref{sec:matchingandnumerics}).    

\begin{figure}
\centering
\includegraphics[width=0.95\textwidth]{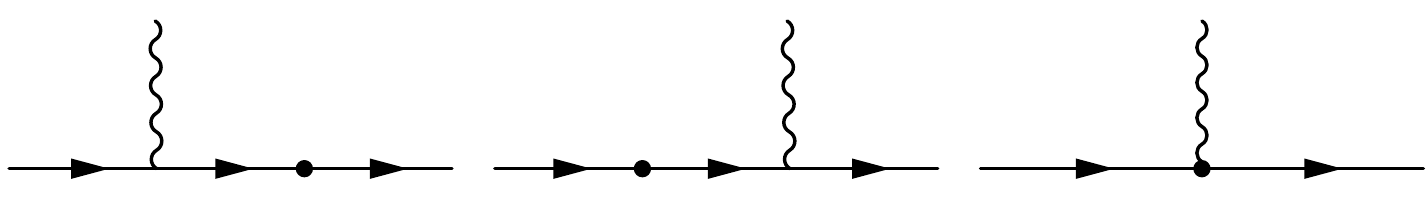}
\vspace{2mm} 
\caption{Contributions to the neutron EDM induced by CP-violating Weinberg-type operators. The  dotted vertices indicate  operator insertions and the solid lines represent propagators of the neutron $n$ or the excited states $N^\prime$ and $N^{\prime \prime}$. See text for further explanations.}
\label{fig:neutronpropexp}
\end{figure}

The appropriate form of the  phenomenological side of our sum rule can be established by realising that the contributions to the nEDM induced by CP-violating Weinberg-type operators~(\ref{eq:o6o8})  have a simple pictorial interpretation~\cite{Bigi:1990kz}. As illustrated in Figure~\ref{fig:neutronpropexp}, there are two types of graphs that one needs to consider in general. The first type of diagrams (left and middle) factorises into a propagator with a CP-violating mass insertion proportional to $i \gamma_5$ and into a part that couples to the external photon field. The effect of Weinberg-type operators in this context is to rotate the nucleon wave function by an amount proportional to $d_n/\mu_n$ as in~(\ref{eq:formfactorstructure}). The second type of diagrams (right) only exists if either an insertion of an operator  is considered that couples several gluons to a single photon or if at least one of the external legs corresponds to an excited state~\cite{Bigi:1990kz}. The former possibility is not viable at the dimension-six level, because there is no gauge-invariant operator that couples two gluons to a single photon. In the approximation that neglects the contributions of vertex diagrams and excitations, one can therefore use the following parameterisation  
\beq \label{eq:piphenfinal}
\begin{split}
\Pi_{\rm phen}(q^2) = -\frac{\lambda^2 \hspace{0.25mm} m_n^2 \hspace{0.75mm}  \mu_n}{2 \left (q^2 - m_n^2 \right )} \left ( 1 +  r (q^2) \hspace{0.5mm} i \gamma_5 \right ) \,,
\end{split}
\eeq
where the coefficient function $r(q^2)$ has to be determined by matching the phenomenological side of the sum rule to the corresponding OPE calculation.  Since from~(\ref{eq:formfactorstructure}) we know that the EDM and the magnetic moment of the neutron are simply related by a chiral rotation with  $ i \gamma_5$ though, the following relation holds
\beq \label{eq:dnmuntildeg}
d_n= \mu_n \hspace{0.25mm} r (q^2) \,.
\eeq
In physical terms this result means  that the Weinberg-type contributions to $d_n$ can be approximated  by calculating the $i \gamma_5$  rotation of the nucleon wave function and relating it to the corresponding chiral rotation of~$\mu_n$~\cite{Bigi:1990kz,Demir:2002gg}. In Section~\ref{sec:piope6} and Section~\ref{sec:piope8} we will use~(\ref{eq:piphenfinal}) and~(\ref{eq:dnmuntildeg}) to extract the hadronic matrix elements of $O_6$ and $O_8$, respectively. 

\section{OPE calculation for the dimension-six operator}
\label{sec:piope6}

\subsection{Interpolating current}
\label{sec:interpolationj}

We parameterise the  interpolating current introduced in~(\ref{eq:piq}) as  follows 
\beq \label{eq:eta}
\eta (x) = j_1 (x) + \beta \hspace{0.25mm}  j_2 (x)  \,,
\eeq
where the real parameter $\beta$ is kept arbitrary throughout our calculations. The two currents  
\beq \label{eq:j1j2}
j_1 (x) = 2 \epsilon_{abc} \left (d_a^T (x) \hspace{0.25mm} C \gamma_5 \hspace{0.25mm} u_b (x) \right ) d_c (x) \,, \qquad 
j_2 (x) = 2 \epsilon_{abc} \left (d_a^T (x) \hspace{0.25mm} C \hspace{0.25mm} u_b (x)\right  ) \gamma_5 \hspace{0.25mm} d_c (x) \,,
\eeq
form a basis for projection onto the neutron state in the case of a CP-conserving background. The current $j_1(x)$ is often used in LQCD simulations to describe the neutron wave function~(see for instance~\cite{Chung:1984gr,Leinweber:1994nm,Leinweber:1995fn}). While $j_2(x)$ vanishes in the non-relativistic limit, it should be included in the  interpolating field since we are dealing with light quarks. In~(\ref{eq:j1j2}) the symbols~$a$, $b$ and $c$ are colour indices,  $d$ and $u$ denote a down-quark and up-quark field, respectively, and $C$ is the charge conjugation matrix, which satisfies $C=C^\ast=-C^\dagger=-C^T=-C^{-1}$, $C  \hspace{0.25mm} \gamma_5^T  \hspace{0.25mm} C= -\gamma_5$ and $C^\dagger \gamma^0= \gamma^0 C$. 

Notice that in contrast to the publications~\cite{Pospelov:2000bw,Hisano:2012sc,Pospelov:1999ha,Pospelov:1999mv}, we do not need to consider the two additional currents $i_1 (x) = \gamma_5 j_2 (x)$ and $i_2 (x) = \gamma_5 j_1 (x)$, because in our case  the only source of CP-violation is provided by the Weinberg-type operators~(\ref{eq:o6o8}).  The vacuum $\ket{\Omega}$ appearing in correlators such as~(\ref{eq:piq})  is instead taken to be CP-conserving, which in particular means that we assume that the QCD theta term $\theta \, \tilde G^A_{\mu \nu} G^{A \, \mu \nu}$ vanishes either accidentally or  dynamically due to a Peccei-Quinn mechanism~\cite{Peccei:1977hh}.

\subsection{Weinberg contribution to the quark propagator}
\label{sec:o6quarkprop}

In the presence of a non-trivial EM background and the dimension-six operator $O_6$, the OPE of the correlator~(\ref{eq:piq}) can be formally  written as
\beq \label{eq:piope6}
\Pi_{\rm OPE} (q^2) = i \int \! d^4x \, e^{iqx} \, \left \langle \Omega \left |T\left[ \eta(x) \hspace{0.5mm} \bar{\eta}(0) \right] \right | \Omega \right \rangle_{{\rm EM}, O_6} =  \sum_k C_k (q^2) \braket{Q_k} \,,
\eeq
where $C_k$ are so-called Wilson coefficients and $ \braket{Q_k} = \langle \Omega | Q_k | \Omega \rangle$ are vacuum matrix elements or condensates of the operator $Q_k$. 

One important ingredient  to evaluate~(\ref{eq:piope6})  is  the quark propagator on the CP-conserving background including insertions of the EM field and $O_6$. In position space and suppressing colour and spinor indices the sought propagator  reads 
\beq \label{eq:s6start}
S(x) = S^{(0)}(x) +S^{q}(x) + S^{O_6} (x) \,,
\eeq
where the first term is the free propagator for a massless quark and the second term describes non-perturbative interactions with background quark fields. As shown in Appendix~\ref{sec:quarkOPE} at leading order~(LO) in the OPE these two quantities take the following form 
\beq \label{eq:firstsecond}
S^{(0)}(x) =  \frac{i \slashed{x}}{2 \hspace{0.25mm} \pi^2 x^4} \,, \qquad S^{q}(x) = - \frac{1}{12}  \braket{\bar{q}q} \,, 
\eeq
with $\braket{\bar{q}q} \simeq - (0.25 \, {\rm GeV})^3$~\cite{Colangelo:2000dp,Ioffe:2010zz,Gubler:2018ctz}  the quark condensate. 

\begin{figure}[t!]
\centering
\includegraphics[width=0.4\textwidth]{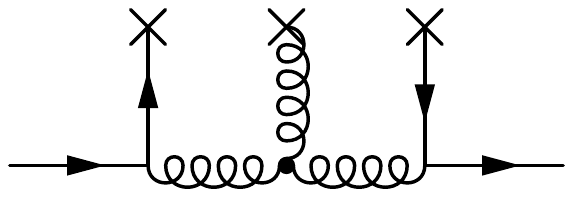}
\vspace{2mm}
\caption{Contribution of the dimension-six Weinberg operator  to the quark propagator. The dotted vertex represents the operator insertion,  the crosses indicate interactions with the background and the solid lines depict quark propagators. }
\label{fig:quarkweinbergdiag}
\end{figure}

The effective operator $O_6$ can be perturbatively inserted into the quark propagator~\cite{Demir:2002gg}. The corresponding Feynman diagram is shown in Figure~\ref{fig:quarkweinbergdiag}.  It follows that the Weinberg-induced contribution to~(\ref{eq:s6start}) can  be written as 
\beq \label{eq:SO6}
S^{O_6} (x) = \int \! d^4z_1 d^4z_2 \, S^{(0)}(x-z_1) \hspace{0.25mm} S^{O_6}_{\rm amp}(z_1-z_2)  \hspace{0.25mm}  S^{(0)}(z_2) \,,
\eeq
where the amputated two-point function $S^{O_6}_{\rm amp} (z)$ is given by
\beq \label{eq:quarkweinbergbasic}
\begin{split}
S^{O_6}_{\rm amp}(z) &= ig_s \hspace{0.25mm} t^A_{ac} \hspace{0.25mm} \gamma_{ik}^\mu \left \langle T\left[  q_c^k(x) \hspace{0.25mm} D_{\mu \nu}^{O_6 \, AB}(z)  \hspace{0.5mm} \bar{q}^l_d(0) \right] \right \rangle  ig_s \hspace{0.25mm} t^B_{db} \hspace{0.25mm} \gamma_{lj}^\nu \\[2mm]
& =  -g_s^2 \hspace{0.5mm}  t^A_{ac} \hspace{0.25mm} t^B_{db} \hspace{0.25mm} \gamma_{ik}^\mu \hspace{0.25mm} \gamma_{lj}^\nu \left \langle T\left[  q_c^k(0) \hspace{0.25mm} D_{\mu \nu}^{O_6 \, AB}(z)   \hspace{0.5mm} \bar{q}^l_d(0) \right] \right \rangle + \ldots \,.
\end{split}
\eeq
Here $g_s$ denotes the QCD coupling constant, $t^A$ are the colour generators of $S\!U(3)$ and we have expanded the quark wave function to zeroth order using~(\ref{eq:qexp}) to obtain the final result.  
The object $D_{\mu \nu}^{O_6 \, AB} (z)$ entering~(\ref{eq:quarkweinbergbasic}) represents the Weinberg-induced correction of the gluon propagator. Pictorially, one has
\beq \label{eq:gluonexp}
D_{\mu \nu}^{O_6 \, AB}(z) \; = \;  
\raisebox{-3mm}{\includegraphics[width=0.2\textwidth]{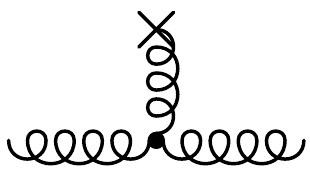}} \; +  \;  \ldots \;\;, 
\eeq 
where the dotted vertex represents the  insertion of the operator $O_6$ and  the cross indicates interactions with the classic background. 

In order to determine the form of~(\ref{eq:gluonexp}) we rely on standard background field techniques. We start by writing  the dimension-six Weinberg operator of~(\ref{eq:o6o8}) in a more convenient form, namely as~(see~for instance~\cite{Braaten:1990gq,Braaten:1990zt,Booth:1992iz})
\beq \label{eq:o6tr}
O_6 = -\frac{i}{16} \hspace{0.5mm}  f^{ABC} \hspace{0.5mm}  T^{\mu \nu \rho \lambda \sigma \tau} \hspace{0.5mm} G^A_{\mu \nu} G^B_{\rho \lambda} G^C_{\sigma \tau}   \, ,
\eeq
where $T^{\mu \nu \rho \lambda \sigma \tau}$ denotes the following trace
\beq \label{eq:tr6}
T^{\mu \nu \rho \lambda \sigma \tau} = \frac{i}{2} \Tr\left(\sigma^{\mu \nu} \sigma^{\rho \lambda} \sigma^{\sigma \tau} \hspace{0.25mm} \gamma_5\right) \,.
\eeq
Notice that this tensor is  anti-symmetric under  $\mu \leftrightarrow \nu$,  etc. as well as $\mu \nu \leftrightarrow \rho \lambda$ etc. By splitting the original gluon field $G_\mu^A = \bar G_\mu^A + \hat G_\mu^A$ into a classical  field $\bar G_\mu^A$ and quantum  field $\hat G_\mu^A$, one can then expand the QCD field strength tensor around its classical configuration to obtain 
\beq \label{eq:classfieldstrengthexp}
G_{\mu \nu}^A = \bar{G}_{\mu \nu}^A + \bar{D}_\mu \hat G_\nu^A - \bar{D}_\nu \hat G_\mu^A + g_s f^{ABC} \hat G_\mu^B \hat G_\nu^C \,, \qquad \bar{D}_\mu \hat G_\nu^A = \partial_\mu \hat G_\nu^A + g_s f^{ABC} \bar{G}_\mu^B \hat G_\nu^C \,. 
\eeq
When one now expands~(\ref{eq:o6tr}) using~(\ref{eq:classfieldstrengthexp}), one is only interested in terms that are linear in $\bar G_\mu^A$ and bilinear in $\hat G_\mu^A$. Using the anti-symmetric properties of $f^{ABC}$ and that of~(\ref{eq:tr6}), we find that the relevant terms are 
\beq \label{eq:O6FR}
O_6   = -\frac{3 i}{2} \hspace{0.5mm}  f^{ABC} \, T^{\mu \nu \rho \lambda \sigma \tau}\, \partial_\mu \bar{G}_\nu^A \partial_\rho \hat G_\lambda^B \partial_\sigma \hat G_\tau^C  + \ldots \,.
\eeq

Employing the result~(\ref{eq:O6FR}) one can now calculate the Weinberg-induced corrections~(\ref{eq:gluonexp}) to the gluon propagator. By performing all possible contractions of the time-ordered product, we obtain the expression 
\bea \label{eq:DO6}
\begin{split}
D_{\mu \nu}^{O_6 \, AB}(z)  & = \int \! d^4y \, \left \langle T\left[ \hat G_\mu^A(z) \hspace{0.25mm} \hat G_\nu^B(0)  \, i O_6 (y)  \right] \right \rangle \\[2mm] 
& =  \frac{3}{2}\hspace{0.5mm}  f^{CDE} T^{\alpha \beta \gamma \delta \varphi \pi} \int \! d^4y \; \bigg \{ \, \partial_\alpha^y \bar{G}_\beta^C (y) \left[ \hspace{0.25mm} \partial_\gamma^y D_{\mu \delta}^{(0)AD}(z-y) \hspace{0.5mm} \partial_\varphi^y D_{\nu \pi}^{(0)BE}(y) \right . \\[1mm] 
& \hspace{5.5cm} \left . + \, \partial_\varphi^y D_{\mu \pi}^{(0)AE}(z-y) \hspace{0.5mm}   \partial_\gamma^y D_{\nu \delta}^{(0)BD}(y) \right] \bigg \}  \,, \hspace{6mm}
\end{split}
\eea
where $D_{\mu \nu}^{(0)AB}(x)$ denotes the free gluon propagator in position space. In Feynman gauge it takes the following simple form 
\beq \label{eq:Dgluon0}
D_{\mu \nu}^{(0)AB}(x) = \int \! \dbar^{\;  4} p \, e^{-ipx}\left[ -i \hspace{0.25mm} \delta^{AB}\,  \frac{g_{\mu \nu}}{p^2} \right] = \frac{1}{4\hspace{0.25mm} \pi^2 x^2} \hspace{0.75mm} \delta^{AB}g_{\mu \nu} \,,
\eeq
a result that can be gleaned by inspection of~(\ref{eq:FT1}).  Here $g_{\mu \nu} = {\rm diag} \left (1, -1, -1, -1 \right )$ denotes the Minkowski metric and we have introduced the abbreviation~${\dbar}^{\; 4} p = dp^4/(2\pi)^4$. 

In order to simplify~(\ref{eq:DO6}) we use the following two relations 
\beq \label{eq:D06simp}
\partial_\mu \bar{G}_\nu^A (x) = \frac{1}{2} \hspace{0.25mm} G^A_{\mu \nu} (0) + \ldots \,, \qquad 
\partial_\mu D_{\nu \rho}^{(0)AB}(x) =  -\delta^{AB }g_{\nu \rho} \int \! \dbar^{\; 4}p \,  e^{-i p x} \, \frac{p_\mu}{p^2}\,, 
\eeq
which follow from the expansion~(\ref{eq:fexp2}) and the  explicit form~(\ref{eq:Dgluon0}) of the free gluon propagator  in momentum space, respectively.  Using~(\ref{eq:D06simp}) yields
\bea \label{eq:D06first}
\begin{split}
D_{\mu \nu}^{O_6 \, AB}(z) & = -\frac{3}{4} \hspace{0.5mm}   f^{ABC}  G^C_{\alpha \beta} (0) \int \! d^4y \hspace{0.25mm} \dbar^{\; 4}p \hspace{0.25mm} \dbar^{\; 4}q  \hspace{0.25mm} e^{-i p (z-y)} e^{-i q y}  \left[ T^{\alpha \beta \gamma \mu \varphi \nu} \, \frac{p_\gamma q_\varphi}{p^2 q^2} -  T^{\alpha \beta \gamma \nu \varphi \mu} \, \frac{p_\varphi q_\gamma}{p^2 q^2} \right] \hspace{8mm} \\[2mm]
& =  -\frac{3}{2} \hspace{0.5mm}   f^{ABC}  G^C_{\alpha \beta} (0) \, T^{\alpha \beta \gamma \mu \varphi \nu}  \int \! \dbar^{\; 4}p  \hspace{0.25mm} e^{-i p z} \, \frac{p_\gamma p_\varphi}{p^4} \\[2mm]
& = \frac{3  \hspace{0.25mm} i}{8 \hspace{0.25mm}  \pi^2} \hspace{0.5mm}   f^{ABC}  G^C_{\alpha \beta} (0)  \, T^{\alpha \beta \gamma \mu \varphi \nu}   \hspace{0.25mm}  \frac{1}{z^4} \left [ z_\gamma z_\varphi  - \frac{g_{\gamma \varphi }}{2} \hspace{0.25mm} z^2 \right ] \,,
\end{split}
\eea
where in the last step we have employed the Fourier integral given in~(\ref{eq:FT4}). 

Plugging this into~(\ref{eq:quarkweinbergbasic}) we find
\beq \label{eq:S06finpos}
\begin{split}
S^{O_6}_{\rm amp}(z) &=  -\frac{3  \hspace{0.25mm}  i g_s}{8 \hspace{0.25mm}  \pi^2} \hspace{0.5mm}   f^{ABC}  t^A_{ac} \hspace{0.25mm} t^B_{db} \hspace{0.25mm} \gamma_{ik}^\mu \hspace{0.25mm} \gamma_{lj}^\nu \braket{ q_c^k(0)  g_s \bar{G}_{\alpha \beta}^C(0) \bar{q}^l_d(0)}  \, T^{\alpha \beta \gamma \mu \varphi \nu}  \hspace{0.25mm} \frac{1}{z^4}\left [ z_\gamma z_\varphi  - \frac{g_{\gamma \varphi }}{2} \hspace{0.25mm} z^2 \right ]  \\[2mm]
&=  \frac{i g_s}{512 \hspace{0.25mm}  \pi^2} \hspace{0.5mm} f^{ABC} \hspace{0.25mm}  ( t^A \hspace{0.25mm} t^C \hspace{0.25mm} t^B)_{ab}  \hspace{0.25mm}  (\gamma^\mu \sigma_{\alpha \beta} \gamma^\nu)_{ij}  \, T^{\alpha \beta \gamma \mu \varphi \nu} \hspace{0.25mm} \frac{1}{z^4} \left [ z_\gamma z_\varphi  - \frac{g_{\gamma \varphi }}{2} \hspace{0.25mm} z^2 \right ]   \hspace{0.25mm}  \braket{ \bar q g_s \sigma \cdot G  q} \\[2mm]
&=  \frac{3  \hspace{0.25mm} i g_s \hspace{0.25mm} \delta_{ab} }{32\hspace{0.25mm}  \pi^2 \hspace{0.25mm} z^2}  \hspace{0.75mm}  \gamma^5_{ij}  \braket{ \bar q g_s \sigma \cdot G  q}  \,.
\end{split}
\eeq 
Here we have used that the non-perturbative quark-gluon condensate appearing in the first line simplifies as follows~(see for instance~\cite{Leinweber:1995fn,gubler})
\beq \label{eq:vacuum2}
 \braket{ q_c^k(0)  g_s \bar{G}_{\alpha \beta}^C(0) \bar{q}^l_d(0)} =  -\frac{1}{192} \, \sigma^{kl}_{\alpha \beta} \, t^C_{cd} \hspace{0.25mm} \braket{ \bar q g_s \sigma \cdot G  q} \,,
\eeq
with $\sigma \cdot G = \sigma_{\mu \nu} \hspace{0.25mm} G^{A \, \mu \nu}  t^A$. Furthermore, the colour factor in the second line evaluates to 
\beq \label{eq:colour6}
 f^{ABC} \hspace{0.25mm}  ( t^A \hspace{0.25mm} t^C \hspace{0.25mm} t^B)_{ab} = - \frac{i}{2} \hspace{0.25mm} C_A C_F \hspace{0.25mm} \delta_{ab} \,, 
\eeq
with the Casimir  operators given by $C_A =3$ and $C_F = 4/3$ for $S\!U(3)$.  

From~(\ref{eq:FT1add}) one sees  that the Fourier transform of~(\ref{eq:S06finpos}) reads 
\beq \label{eq:SO6ampmom}
S^{O_6}_{\rm amp}(p) =  \frac{3 g_s}{8 p^2}   \hspace{0.5mm}  \gamma_5  \braket{ \bar q g_s \sigma \cdot G  q} \,,
\eeq 
where we have dropped  colour and spinor indices. Inserting this result into~(\ref{eq:SO6}) leads to 
\beq \label{eq:SO6pos} 
\begin{split}
S^{O_6}(x) & = \frac{3 g_s}{8} \braket{ \bar q g_s \sigma \cdot G  q}  \int \! d^4 z_1  d^4 z_2 \dbar^{\;4} p \hspace{0.25mm} \dbar^{\;4} q \hspace{0.25mm} \dbar^{\;4} r \hspace{0.5mm} e^{i (q-p) z_1}   e^{i (p-r) z_2}  e^{-i q x}  \, \frac{\slashed{q} \gamma_5 \slashed{r}}{q^2 p^2 r^2} \\[2mm]
& =  -\frac{3 g_s}{8} \, \gamma_5 \braket{ \bar q g_s \sigma \cdot G  q}  \int \!  \dbar^{\;4} p \hspace{0.5mm}   e^{-i p x}  \hspace{0.5mm} \frac{1}{p^4}  =  \frac{3 g_s}{128 \hspace{0.25mm} \pi^2} \, i \gamma_5 \braket{ \bar q g_s \sigma \cdot G  q}    \hspace{0.25mm} \ln \left (- \frac{\mu_{\rm IR}^2 x^2}{4} \right )   \,. 
\end{split}
\eeq
To arrive at the final result we have made used of the Fourier integral~(\ref{eq:FT2}) dropping infrared~(IR) poles and constant pieces, because such terms  vanish after Borel transformation. The appearance of the scale $\mu_{\rm IR}$  signals that the $O_6$ contribution to~(\ref{eq:piope6}) will  depend logarithmically on an IR cut-off. Notice finally that  the second result in~(\ref{eq:SO6pos}) implies that  the Weinberg-induced correction to the quark propagator takes the following form in momentum space 
\beq \label{eq:SO6mom}
S^{O_6}(p) =  -\frac{3 g_s}{8 p^4} \, \gamma_5 \braket{ \bar q g_s \sigma \cdot G  q} \,.
\eeq 
This result  agrees  up to a factor of $i$ with the corresponding expression reported in~\cite{Demir:2002gg} after taking into account that the definition of $O_6$ used in this paper differs from the one employed in~(\ref{eq:o6o8}) by a factor of $1/3$. 

\subsection{OPE including the Weinberg operator}
\label{sec:OPEwein6}

In terms of~(\ref{eq:s6start}) and 
\beq \label{eq:Sc}
S^c(x) = C S^T (x) \hspace{0.25mm} C = S^{(0)}(x) - S^q(x) - S^{O_6}(x) \,,
\eeq
the general form of the correlator~(\ref{eq:piope6}) can be written as 
\beq \label{eq:opegen}
\begin{split}
\Pi_{\rm OPE} (q^2) & = 24 \hspace{0.25mm} i \int \! d^4x \, e^{iqx} \, \bigg \{ \Tr\left(S^c \gamma_5 S \gamma_5 \right) S + S\gamma_5 S^c \gamma_5 S \\[1mm]
& \hspace{1.5cm} + \beta \, \left[\Tr\left(S^c S \gamma_5 \right) \gamma_5 S +\gamma_5 S\gamma_5 S^c S +\Tr\left(S^c \gamma_5 S \right)  S \gamma_5 + S S^c \gamma_5 S \gamma_5 \right]\\[1mm]
& \hspace{1.5cm}  + \beta^2 \, \big[ \Tr\left(S^c S \right) \gamma_5 S \gamma_5 + \gamma_5 S S^c S \gamma_5 \big] \bigg\}  \,,
\end{split}
\eeq
where we have  performed all possible Wick contractions. Here $\beta$ is the real parameter that appears in our definiton~(\ref{eq:eta}) of the interpolating current for the neutron. 

\begin{figure}[t!]
\centering
\includegraphics[width=0.65\textwidth]{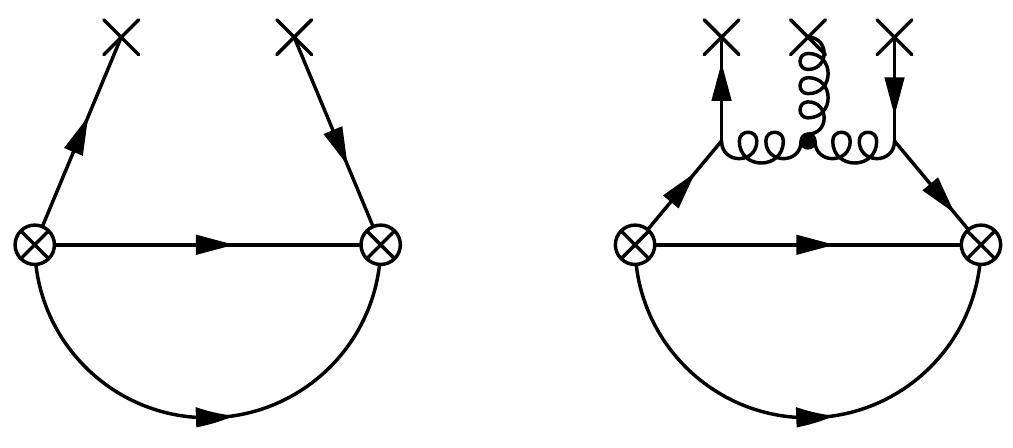}
\vspace{2mm}
\caption{One-loop and two-loop contributions to the OPE correlation function~(\ref{eq:opegen211}). Insertions of the interpolating currents $\eta(x)$ or $\bar \eta(0)$  are denoted by circled crosses. The lower quark lines correspond to the free propagator $S^{(0)} (x)$ of a massless quark, while the crosses indicate interactions with either quark or gluon background fields. In the right graph the  dotted vertex represents the  insertion of the Weinberg-type operator~$O_6$. }
\label{fig:nedmweinbergdiag6}
\end{figure}

We are only interested in the LO result of the OPE, in other words in terms that are linear either in the quark condensate~(\ref{eq:firstsecond}) or  the Weinberg-induced correction~(\ref{eq:SO6pos}) to the quark propagator. The relevant contributions are given by the following expression 
\beq \label{eq:opegen211}
\begin{split}
\Pi_{\rm OPE} (q^2) & = -24 \hspace{0.25mm} i \int \! d^4x \, e^{iqx} \, \left [ S^{(0)}(x) \right ] ^2  \Big [ f_q (\beta) \, S^q(x) +  f_{O} (\beta) \, S^{O_6}(x) \Big ] \,,
\end{split}
\eeq
with 
\beq \label{eq:fqf6}
f_q (\beta ) = 7-2\beta-5\beta^2  \,, \qquad 
f_{O} (\beta ) = 5+2\beta-7\beta^2 \,.
\eeq
This result can be interpreted in terms of the two Feynman diagrams depicted in Figure~\ref{fig:nedmweinbergdiag6}. The left graph shows the $S^q (x)$ part of~(\ref{eq:opegen211}) which corresponds to a one-loop diagram because the background quark fields are non-dynamical. The right diagram has instead  a  dynamically and perturbative gluon that closes the second loop. In the case of the $S^{O_6}(x)$ correction one therefore has to deal with a two-loop contribution. 

To evaluate~(\ref{eq:opegen211}) we now parameterise the mixed quark-gluon condensate as~\cite{Belyaev:1982sa}
\beq \label{eq:cond2}
\braket{ \bar q g_s \sigma \cdot G  q} = m_0^2  \braket{\bar q q}  \,,
\eeq
where $m_0^2 \simeq 0.8 \, {\rm GeV}^2$ is a QCD parameter. Inserting~(\ref{eq:firstsecond}) and~(\ref{eq:SO6pos}) into~(\ref{eq:opegen211}), we then find 
\bea \label{eq:o6final}
\begin{split}
\Pi_{\rm OPE} (q^2) & = \frac{6 \hspace{0.25mm} i  \braket{\bar q q}}{\pi^4} \int \! d^4x \, e^{iqx} \, \frac{1}{x^6} \left [-\frac{f_q (\beta)}{12} +  \frac{3 g_s \hspace{0.25mm} m_0^2  \hspace{0.25mm} f_{O} (\beta)}{128 \hspace{0.25mm} \pi^2}   \hspace{0.25mm} \ln \left (- \frac{\mu_{\rm IR}^2 x^2}{4} \right ) \,   i \gamma_5  \right ]  \\[2mm]
& = \frac{q^2}{16 \hspace{0.25mm} \pi^2} \, f_q (\beta)  \ln \left ( -\frac{\mu_{\rm UV}^2}{q^2} \right )  \braket{\bar q q} \left [ 1  - \frac{9 g_s  \hspace{0.25mm} m_0^2}{32 \hspace{0.25mm} \pi^2}   \hspace{0.25mm} \frac{f_{O} (\beta)}{f_q (\beta)} \hspace{0.25mm}  \ln \left ( -\frac{\mu_{\rm IR}^2}{q^2} \right )   \hspace{0.5mm} i \gamma_5 \right ] \,.
\end{split}
\eea
Here we have used the Fourier integrals given in~(\ref{eq:FT3add}) and~(\ref{eq:doubleFTneeded}), respectively, to obtain the second line. 

\subsection{Matching and discussion}
\label{eq:matchingdiscussion6}

In order to derive the  sum rules for the $O_6$ contribution to the nEDM, we match the phenomenological and the OPE correlators, i.e.~we set~(\ref{eq:piphenfinal}) and~(\ref{eq:o6final}) equal and determine the coefficient $r (q^2)$ that appears in front of the $i \gamma_5$ term in $\Pi_{\rm phen} (q^2)$. After Borel transformation and identifying the IR cut-off  with the QCD scale, we obtain 
\beq \label{eq:r6}
r_6 (\beta) = \frac{9 g_s  \hspace{0.25mm} m_0^2}{32 \hspace{0.25mm}  \pi^2}   \hspace{0.25mm} \frac{f_{O} (\beta)}{f_q (\beta)}  \hspace{0.25mm} \ln \left ( \frac{M^2}{\Lambda_{\rm QCD}^2}  \right ) \,.
\eeq

With this result at hand, we now discuss the appropriate choice for the mixing parameter~$\beta$ introduced in~(\ref{eq:eta}) for our sum rule. There are two commonly used ways for fixing this parameter: ($i$) at a value where the leading terms of the OPE are stationary under variations of $\beta$ or ($ii$) at a point  that maintains a balance between OPE convergence and contributions of excited states. Both methods are not applicable in our case, because~$(i)$ the result~(\ref{eq:r6}) does not possess an extremum and $(ii)$ contributions of excited states have been ignored in our sum rule~(cf.~Section~\ref{sec:hadronicrep}). The procedure advocated in~\cite{Pospelov:2000bw,Hisano:2012sc,Pospelov:1999ha,Pospelov:1999mv} where $\beta$ is chosen such that subleading IR logarithms are cancelled in the QCD $\theta$-term and CEDM contributions to the nEDM is also not useful, since in the OPE side~(\ref{eq:o6final}) of our sum rule IR logarithms appear already at LO. 

Our choice of $\beta$ is instead based on the observation that the function $f_q (\beta)$ introduced in~(\ref{eq:fqf6}) appears in the numerator of Ioffe's formula~\cite{Ioffe:1981kw}, which for arbitrary $\beta$ takes the following form~\cite{Leinweber:1995fn}
\beq \label{eq:ioffeformula}
m_n = - \frac{7-2\beta-5\beta^2  }{5+2\beta+5\beta^2 } \hspace{0.5mm} \frac{4 \left (2 \pi \right )^2}{M^2}   \braket{\bar q q} \,.
\eeq
This relation   connects the neutron mass~$m_n$ to the quark condensate~$\braket{\bar q q}$.  While it is not an exact relationship, one observes that for $\beta = 1$, Ioffe's formula~(\ref{eq:ioffeformula}) predicts $m_n = 0$ in gross disagreement with observation. For the second standard choice of  the mixing parameter,~i.e.~the so-called Ioffe interpolating current with $\beta = -1$, one instead has $m_n \simeq 1.2 \,{\rm GeV}^3/M^2$ if  the numerical value $\braket{\bar q q} \simeq -(0.25 \, {\rm GeV})^3$ for the quark condensate is used. For $\beta = -1$  and a Borel mass of $M \simeq 1 \, {\rm GeV}$, the formula~(\ref{eq:ioffeformula}) thus predicts a neutron mass that is in the ballpark of  the experimental measured value $m_n \simeq 0.94 \, {\rm GeV}$.  We~conclude from this that the appropriate choice for the mixing parameter in the case of our sum rule~(\ref{eq:o6final}) is $\beta = -1$. In fact, this choice is the one that has been employed in essentially all CP-even sum rules~\cite{Ioffe:1981kw,Belyaev:1982sa,Ioffe:1982ce}, including the sum-rule calculations of the anomalous magnetic moment $\mu_n$ of the neutron~\cite{Ioffe:1983cj,Ioffe:1983ju,Balitsky:1983xk,Aw:1999gz}. We believe that the Ioffe interpolating current has also been used in~\cite{Demir:2002gg}. 

\section{OPE calculation for the dimension-eight operators}
\label{sec:piope8}

\subsection{Weinberg contribution to the quark propagator}
\label{sec:o8quarkprop}

\begin{figure}[t!]
\centering
\includegraphics[width=0.4\textwidth]{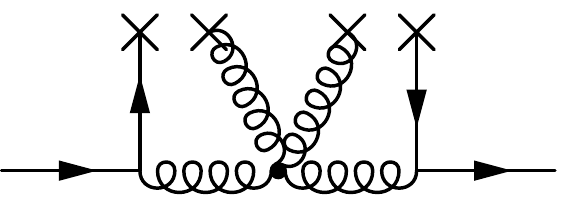}
\vspace{2mm}
\caption{Contribution of the dimension-eight Weinberg operator  to the quark propagator. The  operator insertion corresponds to the dotted vertex, interactions with the background are represented by  crosses and quark propagators are depicted by solid lines. }
\label{fig:quarkweinbergdiag8}
\end{figure}

In this section, we derive the contribution to the nEDM from the dimension-eight Weinberg operator~$O_8$ introduced in~(\ref{eq:o6o8}). Like in Section~\ref{sec:o6quarkprop} we will treat the CP-violating four-gluon operator as a perturbative insertion into the quark propagator. The corresponding Feynman graph is shown in Figure~\ref{fig:quarkweinbergdiag8}. In analogy to~(\ref{eq:s6start}), (\ref{eq:SO6}) and (\ref{eq:quarkweinbergbasic}) we write 
\beq \label{eq:s8start}
S(x) = S^{(0)}(x) +S^{q}(x) + S^{O_8} (x) \,,
\eeq
and 
\beq \label{eq:SO8}
S^{O_8} (x) = \int \! d^4z_1 d^4z_2 \, S^{(0)}(x-z_1) \hspace{0.25mm} S^{O_8}_{\rm amp}(z_1-z_2)  \hspace{0.25mm}  S^{(0)}(z_2) \,, 
\eeq
with
\beq \label{eq:quarkweinbergbasic8}
S^{O_8}_{\rm amp}(z)  = -g_s^2 \hspace{0.5mm}  t^A_{ac} \hspace{0.25mm} t^B_{db} \hspace{0.25mm} \gamma_{ik}^\mu \hspace{0.25mm} \gamma_{lj}^\nu \left \langle T\left[  q_c^k(0) \hspace{0.25mm} D_{\mu \nu}^{O_8 \, AB}(z)   \hspace{0.5mm} \bar{q}^l_d(0) \right] \right \rangle + \ldots \,.
\eeq
The explicit LO expressions for $S^{(0)}(x)$ and $S^{q}(x)$ can be found in~(\ref{eq:firstsecond}).  The function $D_{\mu \nu}^{O_8 \, AB} (z)$ in~(\ref{eq:quarkweinbergbasic8}) corresponds to the correction of the gluon propagator due to an insertion of $O_8$, namely 
\beq \label{eq:gluonexp8}
D_{\mu \nu}^{O_8 \, AB}(z) \; = \;  
\raisebox{-3mm}{\includegraphics[width=0.2\textwidth]{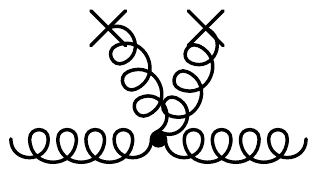}}  \; + \; \dots \;\;.
\eeq 

To determine the analytic expression corresponding to~(\ref{eq:gluonexp8}), we proceed as in Section~\ref{sec:o6quarkprop} and write the operator $O_8$ in the more convenient form
\beq \label{eq:o8tr}
O_8 = \frac{i}{16} \hspace{0.5mm}  c^{ABCD} \hspace{0.5mm} T_1^{\mu\nu\lambda\rho}T_2^{\alpha\beta\gamma\delta}  \hspace{0.5mm} G^A_{\mu\nu}G^B_{\lambda\rho} G^C_{\alpha\beta}G^D_{\gamma\delta}  \,,
\eeq
with
\beq \label{eq:T1T2}
T_1^{\mu\nu\lambda\rho} =  \frac{i}{2} \Tr\left(\sigma^{\mu\nu}\sigma^{\lambda\rho}\right) \,, \qquad  
T_2^{\alpha\beta\gamma\delta}=  \frac{i}{2} \Tr\left(\sigma^{\alpha\beta}\sigma^{\gamma\delta}\gamma_5\right) \,.
\eeq
These traces are anti-symmetric under the exchanges $\mu \leftrightarrow \nu$ etc. but symmetric under the exchanges $\mu  \nu \leftrightarrow \lambda\rho$ etc.  The effective operator $O_8$ is then   expanded  in terms of partial derivatives and gluon fields using~(\ref{eq:classfieldstrengthexp}). Picking out the terms that are bilinear in both the background field $\bar G_\mu^A$ and the quantum field $\hat G_\mu^A$, we obtain 
\beq \label{eq:O8FR}
\begin{split}
O_8  & =  i \hspace{0.5mm} c^{ABCD} \hspace{0.5mm} T_1^{\mu \nu \lambda \rho} T_2^{\alpha\beta\gamma\delta} \, \left[ \hspace{0.25mm} \partial_\mu \bar G^A_\nu  \partial_\lambda \bar G^B_\rho \partial_\alpha \hat G^C_\beta  \partial_\gamma \hat G^D_\delta +  \partial_\mu \bar G^A_\nu  \partial_\lambda \hat G^B_\rho \partial_\alpha \bar G^C_\beta  \partial_\gamma \hat G^D_\delta  \right .  \\[1mm] 
& \hspace{3.75cm} \left. + \, \ldots \, + \, \partial_\mu \hat G^A_\nu  \partial_\lambda \hat G^B_\rho \partial_\alpha \bar G^C_\beta  \partial_\gamma \bar G^D_\delta  \hspace{0.25mm} \right]   + \ldots \,,
\end{split}
\eeq
where the ellipses in the bracket represent the other three terms that are quadratic in $\bar G_\mu^A$ and $\hat G_\mu^A$. 

With the expression~(\ref{eq:O8FR}) at hand  it is a matter of simple algebra to calculate~(\ref{eq:gluonexp8}). Using the relations in~(\ref{eq:D06simp}) we find
\beq \label{eq:DO8}
\begin{split}
D_{\mu \nu}^{O_8 \, AB}(z)  & = \int \! d^4y \, \left \langle T\left[ \hat G_\mu^A(z) \hspace{0.25mm} \hat G_\nu^B(0)  \, i O_8 (y)  \right] \right \rangle \\[2mm]
& = - 4 \hspace{0.25mm} i \hspace{0.5mm} c^{CDEF}  \int \! \dbar^{\; 4}p  \hspace{0.25mm} e^{-i p z} \, \frac{1}{p^4} \, \bigg [  \left (  \delta^{AD}\hspace{0.25mm} \delta^{BF}  \hspace{0.25mm} \tilde G_{\nu \lambda}^E  \hspace{0.25mm}  G^C_{\mu \rho}   +  \delta^{AF}\hspace{0.25mm} \delta^{BD}   \hspace{0.25mm}\tilde G_{\mu \lambda}^E  \hspace{0.25mm}  G^C_{\nu \rho}   \right )   p^\lambda p^\rho  + \, \ldots  \\[1mm]
& \hspace{4.5cm} -\frac{1}{2} \left ( \delta^{AC} \hspace{0.25mm} \delta^{BD} +  \delta^{AD} \hspace{0.25mm} \delta^{BC} \right )  \tilde G^E  \cdot G^F  \left ( p_\mu  p_\nu - g_{\mu \nu} p^2 \right ) \bigg ] \,.
\end{split}
\eeq
Here we have  used the shorthand notation $\tilde G^A \cdot G^{B} =  \tilde G_{\mu \nu}^A G^{B \, \mu \nu}$ and  all field strength tensors and duals are evaluated at $0$. The ellipses represent  three additional  terms that have a structure that is similar to that of the  contribution proportional to $p^\lambda p^\rho$. 

Inserting~(\ref{eq:DO8}) into~(\ref{eq:quarkweinbergbasic8}) it then turns out that  to obtain the amputated two-point function $S^{O_8}_{\rm amp}(z)$, one has to calculate objects of the form 
\beq \label{eq:dim8condensates}
\begin{split}
M_{\lambda \rho}^{AB} & =  t^A_{ac} \hspace{0.25mm} t^B_{db} \hspace{0.25mm} \gamma_{ik}^\mu \hspace{0.25mm} \gamma_{lj}^\nu \; X^{CD} \left \langle q_c^k (0) \hspace{0.5mm} g_s^2  \hspace{0.25mm}  \tilde G_{\mu \lambda}^C (0 ) \hspace{0.5mm}  G_{\nu \rho}^D (0)  \hspace{0.5mm}  \bar q_d^l (0) \right \rangle \,, \\[2mm]
N_{\mu \nu }^{AB} & =  t^A_{ac} \hspace{0.25mm} t^B_{db} \hspace{0.25mm} \gamma_{ik}^\mu \hspace{0.25mm} \gamma_{lj}^\nu \; X^{CD} \left \langle q_c^k (0) \hspace{0.5mm} g_s^2  \hspace{0.25mm}  \tilde G^C (0 ) \cdot G^D (0)  \hspace{0.5mm}  \bar q_d^l (0) \right \rangle \,,
\end{split}
\eeq
where $X^{CD}$ is a symmetric tensor in colour space. To achieve this we expand the quark current in terms of the set of basis matrices
\beq \label{eq:lorentzbasis}
\Gamma_n  = \left \{\hspace{0.25mm} \mathbb{1}, \gamma_5,\gamma_\alpha, i \gamma_5 \gamma_\alpha, \sigma_{\alpha \beta} \hspace{0.25mm} \right \} \,,
\eeq
using the Fierz identity (see for instance~\cite{Nieves:2003in,Nishi:2004st})
\beq \label{eq:qqexpansion}
q_a^i \hspace{0.25mm} \bar{q}^l_d = -\frac{\delta_{ad}}{12} \, \Gamma_n^{il} \, \big(\bar{q} \hspace{0.25mm} \Gamma_n  \hspace{0.25mm} q\big) -\frac{t^C_{ad}}{2} \, \Gamma_n^{il} \, \big (\bar{q}  \hspace{0.25mm}t^C\Gamma_n q  \hspace{0.25mm} \big ) \,.
\eeq
In the case of the structure $M_{\lambda \rho}^{AB}$, we obtain 
\beq \label{eq:MAB1}
M_{\lambda \rho}^{AB} = \gamma^\mu \hspace{0.25mm} \Gamma_n \hspace{0.25mm} \gamma^\nu \left [ - \frac{t^A t^B}{12} \, X^{CD}  \left \langle \bar q  \hspace{0.25mm}  \Gamma_n  \hspace{0.25mm}  q  \hspace{0.25mm}  g_s^2  \hspace{0.25mm}   \tilde G_{\mu \lambda}^C \hspace{0.25mm}  G_{\nu \rho}^D  \right \rangle  - \frac{t^A t^E t^B}{2} \, X^{CD}  \left \langle \bar q  \hspace{0.25mm}  t^E \hspace{0.25mm} \Gamma_n  \hspace{0.25mm}  q  \hspace{0.25mm}  g_s^2  \hspace{0.25mm}   \tilde G_{\mu \lambda}^C \hspace{0.25mm}  G_{\nu \rho}^D  \right \rangle   \right ]  \,,
\eeq
where a sum over the five different Lorentz structures in~(\ref{eq:lorentzbasis}) is implicit. 

Recalling from~(\ref{eq:S06finpos}) that we are only interested in the pieces of  $S^{O_8}_{\rm amp}(z)$ that are proportional to $i \gamma_5$ it is readily seen that these contributions all arise from the term 
\beq \label{eq:g5g}
\gamma^\mu \hspace{0.25mm} \gamma_5 \hspace{0.25mm} \gamma^\nu = -g^{\mu \nu} \gamma_5 + \frac{1}{2} \hspace{0.5mm} \epsilon^{\mu \nu \lambda \rho}  \hspace{0.25mm} \sigma_{\lambda \rho }\,,
\eeq
 in~the above expression for $M_{\lambda \rho}^{AB}$. The colour factors appearing in~(\ref{eq:MAB1}) can be decomposed in the following way 
\beq \label{eq:t2t3}
\begin{split}
t^At^B & = \frac{1}{6} \hspace{0.5mm} \delta^{AB} + \frac{1}{2} \hspace{0.5mm}  d^{ABC} t^C + \frac{i}{2} \hspace{0.5mm}  f^{ABC} t^C \,, \\[2mm]
t^A t^C t^B & = \frac{1}{12} \hspace{0.5mm}  d^{ABC} - \frac{i}{12} \hspace{0.5mm}  f^{ABC} + \frac{1}{6} \hspace{0.5mm}  \delta^{BC} t^A + \frac{1}{4} \hspace{0.5mm} \left ( d^{BCD}- if^{BCD} \right ) \left (d^{ADE} + if^{ADE} \right ) t^E \,,
\end{split} 
\eeq
with 
\beq \label{eq:dABC}
d^{ABC}= 2 \Tr\left[ \hspace{0.125mm} t^A\left\{t^B,t^C\right\}\right] \,,
\eeq
the totally symmetric structure constants of $S\!U(3)$ colour.   Realising that only the structures that~are proportional to $ \delta^{AB}$ and  $d^{ABC}$  in~(\ref{eq:t2t3}) can lead to $S\!U(3)$ invariant condensates, we find that the terms relevant for our sum-rule calculation of the $O_8$ contributions are 
\beq \label{eq:MAB2}
\begin{split}
M_{\lambda \rho}^{AB} & = -g^{\mu \nu} \, \gamma_5 \, \left [ - \frac{\delta^{AB}}{72} \, X^{CD}  \left \langle \bar q  \hspace{0.25mm}  \gamma_5 \hspace{0.25mm}  q  \hspace{0.25mm}  g_s^2  \hspace{0.25mm}   \tilde G_{\mu \lambda}^C \hspace{0.25mm}  G_{\nu \rho}^D  \right \rangle  - \frac{d^{ABE}}{24} \, X^{CD}  \left \langle \bar q  \hspace{0.25mm}  t^E \hspace{0.25mm} \gamma_5  \hspace{0.25mm}  q  \hspace{0.25mm}  g_s^2  \hspace{0.25mm}   \tilde G_{\mu \lambda}^C \hspace{0.25mm}  G_{\nu \rho}^D  \right \rangle   \right ]  + \, \ldots \\[2mm]
& = g_{\lambda \rho} \gamma_5 \, \left [  \frac{\delta^{AB}}{288} \, X^{CD}   \left \langle Q^{CD} \right \rangle  + \frac{d^{ABE}}{96} \, X^{CD}   \left \langle Q^{ECD} \right \rangle   \right ] +  \, \ldots \;  \,.
\end{split}
\eeq
Here we have used the identity 
\beq \label{eq:amandostrick}
X^{AB} \hspace{0.25mm} \tilde G_{\mu \lambda}^A {G^{B \, \mu}}_{\rho} = \frac{1}{4} \hspace{0.5mm} g_{\lambda \rho} \hspace{0.25mm}  X^{AB} \hspace{0.25mm} \tilde G^A \cdot G^{B} \,, 
\eeq
that holds for any symmetric  $X^{AB}$ and introduced the following shorthand notation  
\beq \label{eq:twoctypes}
\left \langle Q^{AB} \right \rangle =  \left \langle \bar q  \hspace{0.25mm}  \gamma_5 \hspace{0.25mm}  q  \hspace{0.25mm}  g_s^2  \hspace{0.25mm}   \tilde G^A \cdot G^B  \right \rangle \,, \qquad
\left \langle Q^{ABC} \right \rangle = \left \langle \bar q  \hspace{0.25mm}  t^A \hspace{0.25mm} \gamma_5  \hspace{0.25mm}  q  \hspace{0.25mm}  g_s^2  \hspace{0.25mm}   \tilde G^B \cdot G^C  \right \rangle
\eeq
for the two types of condensates appearing in~(\ref{eq:MAB2}). A calculation similar to the one detailed above leads to 
\beq \label{eq:NAB2}
\begin{split}
N_{\mu \nu}^{AB} & = g^{\mu \nu} \, \gamma_5 \left [  \frac{\delta^{AB}}{72} \, X^{CD}    \left \langle Q^{CD} \right \rangle  + \frac{d^{ABE}}{24} \, X^{CD}    \left \langle Q^{ECD} \right \rangle \right ] +  \, \ldots \;  \,.
\end{split}
\eeq

To  determine the expression~(\ref{eq:quarkweinbergbasic8}), we finally specify the possible colour structures $c^{ABCD}$ that appear in the definition~(\ref{eq:o6o8}) of $O_8$. There are in fact three independent CP-violating four-gluon operators~\cite{Morozov:1985ef,Chang:1991ry,Booth:1992iz} and the corresponding colour structures can be chosen to be
\beq \label{eq:cABCD}
c^{ABCD}_m= \left \{ \hspace{0.25mm} \delta^{AB} \hspace{0.25mm} \delta^{CD},  \, \delta^{AC} \hspace{0.25mm} \delta^{BD}, \,  d^{ABE} \hspace{0.25mm} d^{CDE} \hspace{0.25mm} \right \}\, .
\eeq
Notice that these coefficients are symmetric under the simultaneous exchange of $A \leftrightarrow B$ and $C \leftrightarrow D$ and the pairwise exchange $AB \leftrightarrow CD$.  Using (\ref{eq:MAB2}), (\ref{eq:NAB2}) and the properties of $c^{ABCD}$, we find that~(\ref{eq:quarkweinbergbasic8}) can be written as follows 
\beq \label{eq:SO8ampfinal}
\begin{split}
S^{O_8}_{\rm amp}(z) &   =  -\frac{1}{72 \hspace{0.25mm} \pi^2 \hspace{0.25mm} z^2} \hspace{0.75mm} \gamma_5 \, \bigg [ c^{AB} \left \langle Q^{AB} \right \rangle + c^{ABC} \left \langle Q^{ABC} \right \rangle \bigg ]   \,,
\end{split}
\eeq
with the new colour structures 
\beq \label{eq:newcolour}
c^{AB} =  c^{ACBC}+c^{ACCB} + 3 c^{ABCC} \,, \qquad  c^{ABC} = 3 d_{CDE}  \left ( c^{ADBE}+c^{ADEB} + 3 c^{ABDE} \right ) \,. 
\eeq 
In fact, using the identities (see for instance~\cite{Fadin:2005zj})
\beq \label{eq:dsums}
d_{ACD} \hspace{0.25mm} d_{BCD} = \frac{C_A^2 - 4}{C_A} \, \delta_{AB} \,, \qquad 
d_{ADE} \hspace{0.25mm} d_{BDF} \hspace{0.25mm} d_{CEF} = \frac{C_A^2 - 12}{2 \hspace{0.25mm} C_A} \, d_{ABC} \,, 
\eeq
it is a matter of simple algebra to show that for the three colour structures in~(\ref{eq:cABCD}) the coefficients~(\ref{eq:newcolour}) take  the  explicit form 
\beq \label{eq:newcolorexplicit}
c^{AB} = c_1 \hspace{0.25mm} \delta^{AB} \,, \qquad c_1^m = \left \{26, 12, \frac{10}{3} \right \} \,, \qquad
c^{ABC} = c_2 \hspace{0.25mm} d^{ABC} \,, \qquad c_2^m = \left \{6, 12, 12 \right \} \,.
\eeq

A comparison of~(\ref{eq:SO8ampfinal}) with~(\ref{eq:S06finpos}) and (\ref{eq:SO6pos}) then implies that the Weinberg-induced correction~(\ref{eq:SO8}) to the quark propagator  in position space can be written as 
\beq \label{eq:SO8final}
S^{O_8} (x) =  \frac{1}{288 \hspace{0.25mm} \pi^2} \hspace{0.75mm} i \gamma_5 \, \bigg [ c_1  \left \langle \bar q  \hspace{0.25mm}  i \gamma_5 \hspace{0.25mm}  q  \hspace{0.25mm}  g_s^2  \hspace{0.25mm}   \tilde G \cdot G  \right \rangle + c_2  \left \langle \bar q  \hspace{0.25mm}  t^A \hspace{0.25mm} i \gamma_5  \hspace{0.25mm}  q  \hspace{0.25mm}  g_s^2  \hspace{0.25mm}   d_{ABC} \hspace{0.25mm} \tilde G^B \cdot G^C  \right \rangle \bigg ]  \hspace{0.25mm} \ln \left (- \frac{\mu_{\rm IR}^2 x^2}{4} \right ) \,,
\eeq
with the coefficients $c_1$ and $c_2$  given in (\ref{eq:newcolorexplicit}). Notice that the dimension-seven condensates appearing in~(\ref{eq:SO8final}) are the only non-zero matrix elements that can be constructed out of two quark fields, a QCD field strength tensor and its dual~\cite{Oganesian:1997wz,Polyakov:1998ip}. This finding provides  a sanity check of the calculations leading to $S^{O_8} (x)$.

\subsection{OPE correlator, matching and discussion}
\label{eq:matchingdiscussion8}

To determine the OPE correlator~(\ref{eq:piope6}) corresponding to the dimension-eight Weinberg-type operator $O_8$, we also need values for the two condensates in~(\ref{eq:SO8final}). 
The only estimates that exist at present are based on the  instanton liquid model~\cite{Ioffe:2010zz,Shuryak:1981ff,Diakonov:1983hh,Diakonov:1985eg,Schafer:1996wv}. One obtains~\cite{Polyakov:1998ip}
\beq \label{eq:dim7cond}
 \left \langle \bar q  \hspace{0.25mm}  i \gamma_5 \hspace{0.25mm}  q  \hspace{0.25mm}  g_s^2  \hspace{0.25mm}   \tilde G \cdot G  \right \rangle = \frac{64}{5 \hspace{0.25mm}  \bar \rho^4} \hspace{0.25mm}  \braket{\bar q q} \,, \qquad 
  \left \langle \bar q  \hspace{0.25mm}  t^A \hspace{0.25mm} i \gamma_5  \hspace{0.25mm}  q  \hspace{0.25mm}  g_s^2  \hspace{0.25mm}   d_{ABC} \hspace{0.25mm} \tilde G^B \cdot G^C  \right \rangle  = \frac{32}{15 \hspace{0.25mm}   \bar \rho^4} \hspace{0.25mm} \braket{\bar q q} \,.
\eeq
In the diluted instanton gas model, the quark condensate is given by 
\beq \label{eq:qqinst}
 \braket{\bar q q} = - \frac{3 m_q}{2 \hspace{0.25mm}\pi^2\hspace{0.25mm} \bar \rho^2} \,, 
\eeq 
where $m_q \simeq \Lambda_{\rm QCD} \simeq 0.3 \, {\rm GeV}$ denotes the constituent  quark mass and $\bar \rho \simeq 1/(0.6 \, {\rm GeV})$ is  the  average instanton size. Notice that for the quoted values of $m_q$ and $\bar \rho$ one finds  $\braket{\bar{q}q} \simeq - (0.25 \, {\rm GeV})^3$ in agreement with the standard value for the quark condensate~\cite{Colangelo:2000dp,Ioffe:2010zz,Gubler:2018ctz}. 

Noticing that after employing the relations~(\ref{eq:dim7cond}) the structure of~(\ref{eq:SO8final}) and~(\ref{eq:SO6pos}) are precisely the same, the derivation of $\Pi_{\rm OPE} (q^2)$  and the matching of the phenomenological and the OPE correlators for $O_8$ proceeds as  in Sections~\ref{sec:OPEwein6} and \ref{eq:matchingdiscussion6}, respectively. In particular,  for the coefficient $r(q^2)$ that multiplies the $i \gamma_5$ term in~(\ref{eq:piphenfinal}), we obtain 
\beq \label{eq:r8}
r_8 (\beta) = \frac{4 \left (6 \hspace{0.25mm}  c_1 +  c_2 \right )}{45 \hspace{0.25mm}  \pi^2  \hspace{0.25mm} \bar \rho^4} \hspace{0.25mm} \frac{f_{O} (\beta)}{f_q (\beta)}  \hspace{0.25mm}   \ln \left ( \frac{M^2}{\Lambda_{\rm QCD}^2}  \right )\,,
\eeq
with the functions $f_q (\beta)$ and $f_O (\beta)$ defined in~(\ref{eq:fqf6}). Like in the case of $O_6$, we will  employ $\beta = -1$  in our numerical analysis of the $O_8$ matrix elements, since  this is the appropriate choice for our sum-rule calculations (cf.~the discussion at the end of~Section~\ref{eq:matchingdiscussion6}).

\section{Numerical analysis}
\label{sec:matchingandnumerics}

\subsection{Dimension-six contribution}
\label{sec:numerics6}

Using $\beta = -1$ and inserting~(\ref{eq:r6}) into~(\ref{eq:dnmuntildeg}), we obtain the following expression for the contribution of the dimension-six Weinberg operator to the nEDM
\beq \label{eq:nEDMO6}
\big (d_n \big )_{O_6} = -\mu_n \, \frac{9 g_s  \hspace{0.25mm} m_0^2}{32 \hspace{0.25mm} \pi^2}   \hspace{0.25mm}  \ln \left ( \frac{M^2}{\Lambda_{\rm QCD}^2}  \right ) \,,
\eeq
which differs by the analytic result given in~\cite{Demir:2002gg} by a sign.  

In our numerical analysis we use 
\begin{align}  \label{eq:numericalinput6}
\mu_n & = -1.91 \, \frac{e}{2 m_p}  = -1.02  \, \frac{e}{\rm GeV}  \,, \hspace{-1.5cm}
& g_s  & = \sqrt{4 \pi \alpha_s} = 2.13 \pm 0.03 \,, \nonumber \\[-2.5mm] \\[-2.5mm]
m_0^2 & = \left ( 0.8 \pm 0.2 \right ) \, {\rm GeV}^2 \,, 
& \frac{M}{\Lambda_{\rm QCD}}  & \in \sqrt{2} \; [1, 2] \,, \nonumber
\end{align}
where the input values and errors of $\mu_n$, $m_p$, $g_s$ and $m_0^2$ are taken from~\cite{Colangelo:2000dp,Ioffe:2010zz,Gubler:2018ctz,Tanabashi:2018oca} and the strong coupling constant corresponds to a LO $\alpha_s$ evaluated at a renormalisation scale of $1 \, {\rm GeV}$. We note that our choice $M \in  \sqrt{2} \; [1, 2] \, \Lambda_{\rm QCD} \simeq [0.42, 0.85] \, {\rm GeV}$ covers the full range of Borel masses that has been considered in the related sum-rule calculations of the QCD $\theta$-term and CEDM contributions to~the~nEDM~\cite{Pospelov:2000bw,Hisano:2012sc,Pospelov:1999ha,Pospelov:1999mv}. 

With the input given in~(\ref{eq:numericalinput6}) we find from~(\ref{eq:nEDMO6}) the following numerical result 
\beq \label{eq:nEDMO6num}
\left ( \frac{d_n}{e} \right )_{O_6} = 74  \left ( 1 \pm 0.5 \right ) \, {\rm MeV}  \,, 
\eeq
where the individual uncertainties have been added in quadrature to obtain the final relative error of 50\%. The dominant source of uncertainty in our prediction for $(d_n/e)_{O_6}$ arises from the  variation of the scale ratio $M/\Lambda_{\rm QCD}$ and amounts to almost 90\% of the total error given above. We add that the quoted total uncertainty in~(\ref{eq:nEDMO6}) is larger than the naive expectation of the size of the sum-rule contributions due to excited states (cf.~the discussion at the end of Section~\ref{sec:phenoparam}) and that the sum-rule predictions for  the down-quark and up-quark CEDMs~\cite{Pospelov:2000bw,Hisano:2012sc}  are also accurate to about 50\%. Notice that the central value of our prediction~(\ref{eq:nEDMO6}) differs by a factor of roughly $1/3$ from the numerical result presented in~\cite{Demir:2002gg}. Here a factor of $1/3$ is accounted for by the different normalisation of the effective operator $O_6$, while the flipped overall sign in~(\ref{eq:nEDMO6})  is compensated by the fact that in the latter article the incorrect relation $\mu_n  = 1.91 \, e/(2 m_p)$ has been used to obtain a numerical result. 

\subsection{Dimension-eight contributions}
\label{sec:numerics8}

Inserting~(\ref{eq:r8}) into~(\ref{eq:dnmuntildeg}), we find for the Ioffe interpolating current,~i.e.~$\beta = -1$, the following expression for the nEDM contribution of the dimension-eight Weinberg operators 
\beq \label{eq:nEDMO8}
\big (d_n \big )_{O_8}^m = -\mu_n \,  \frac{72}{5 \hspace{0.25mm}\pi^2  \hspace{0.25mm} \bar \rho^4}  \hspace{0.25mm}   \ln \left ( \frac{M^2}{\Lambda_{\rm QCD}^2}  \right )\;  \left \{ 1, \frac{14}{27}, \frac{16}{81}\right \} \,,
\eeq
where the numbers  in the curly bracket correspond to the three different colour structures in~(\ref{eq:cABCD}). 

The average instanton size that enters~(\ref{eq:nEDMO8}) can be determined in various ways. Including the value of $\bar \rho $  that allows one to reproduce the phenomenological values of the quark and gluon condensates~\cite{Shuryak:1981ff}, that is obtained through variational techniques and the mean field approximation~\cite{Diakonov:1983hh} and that is found in LQCD calculations~\cite{Chu:1994vi,Negele:1998ev,DeGrand:2001tm,Faccioli:2003qz,Bowman:2004xi,Cristoforetti:2006ar}, we arrive at  the combination 
\beq \label{eq:rhobarinput}
\bar \rho = \frac{1}{(0.58 \pm 0.09) \, {\rm GeV}} \,.
\eeq
This prediction has an uncertainty of $15\%$, which we believe to be a rather conservative error in view of the results given in~\cite{Shuryak:1981ff,Diakonov:1983hh,Chu:1994vi,Negele:1998ev,DeGrand:2001tm,Faccioli:2003qz,Bowman:2004xi,Cristoforetti:2006ar}. 

In order to obtain a numerical result for the $O_8$ contribution to  the nEDM, we use the  input given in (\ref{eq:numericalinput6}) and (\ref{eq:rhobarinput}). Adding individual uncertainties in quadrature we find 
\beq \label{eq:dn8numerics}
\left ( \frac{d_n}{e} \right )_{O_8}^m = 2.5 \cdot 10^{-1} \left ( 1 \pm 0.8 \right ) \, {\rm GeV}^3  \; \Big \{ 1, 0.5, 0.2 \Big \}\,. 
\eeq
Here the dominant source of uncertainty stems again from the  variation of $M/\Lambda_{\rm QCD}$ and amounts  close to 60\% of the quoted total error of 80\%.

\section{Conclusions}
\label{sec:conclusions}

In this work, we have calculated the hadronic matrix elements of dimension-six and dimension-eight operators of Weinberg type~(\ref{eq:o6o8})  using QCD sum-rule techniques. Calculations along the same line  of the dimension-four and dimension-five contributions to the nEDM,~i.e.~the QCD $\theta$-term and CEDMs,  have been performed in~\cite{Pospelov:2000bw,Lebedev:2004va,Hisano:2012sc,Pospelov:1999ha,Pospelov:1999mv}. A sum-rule estimate of the dimension-six Weinberg operator  $O_6$ also exists~\cite{Demir:2002gg}, but this article does not provide details on the actual computation, which motivated us to carry out  an independent evaluation. Our determination of the hadronic matrix elements of the dimension-eight term $O_8$ is instead new, and provides the first systematic study of contributions to $d_n$ due to CP-violating four-gluon operators. The main results of our article,~i.e.~the numerical expressions~(\ref{eq:nEDMO6num}) and~(\ref{eq:dn8numerics}), will be used in a companion paper~\cite{Haisch:2019xyi} to derive model-independent bounds on CP-violating Higgs-gluon interactions in BSM scenarios with vanishing  light-quark Yukawa couplings. 

Our sum-rule estimates are based on the observation~\cite{Bigi:1990kz,Demir:2002gg} that  the Weinberg-type contributions to  the nEDM  can be obtained by calculating the $i \gamma_5$ rotation of the nucleon wave function induced by~(\ref{eq:o6o8}) and relating it  to the corresponding rotation of the neutron anomalous magnetic moment~$\mu_n$. In this approximation only diagrams are included that factorise into a propagator with a CP-violating mass insertion and into a part that couples to the external photon field, while non-factorisable vertex corrections are neglected (see Figure~\ref{fig:neutronpropexp}). In addition, we neglect contributions from excited neutron-like states $N^\prime$ in our estimates. These simplifications lead to uncertainties in our predictions that we estimate to be of the order of 35\% using Borel techniques. The OPE computation of the dimension-six and dimension-eight sum rules is described in detail, and includes a discussion of the matching and the appropriate choice of the neutron-interpolating current. The final analytic expressions for the $O_6$ and $O_8$ contributions to $d_n$ are reported in~(\ref{eq:nEDMO6}) and~(\ref{eq:nEDMO8}), respectively,  and our result for $\big |(d_n)_{O_6} \big |$ is found to agree with that given in~\cite{Demir:2002gg}. The hadronic matrix elements of the Weinberg-type operators turn out to be logarithmically sensitive to the ratio $M/\Lambda_{\rm QCD}$ of the Borel mass and the QCD scale. This IR sensitivity provides the dominant theoretical uncertainty of our predictions. By varying $M/\Lambda_{\rm QCD}$ in the range $\sqrt{2} \,[1,2]$ we find uncertainties close to 45\%, which  exceeds the size of the expected effects from vertex diagrams and excited states. Adding individual errors  in quadrature the total uncertainties of our numerical predictions for $(d_n/e)_{O_6}$ and $(d_n/e)_{O_8}$ are 50\% and 80\%. Sum-rule studies of  the  $\theta$-term and CEDM contributions to the nEDM~\cite{Pospelov:2000bw,Lebedev:2004va,Hisano:2012sc,Pospelov:1999ha,Pospelov:1999mv}  have also found uncertainties of a similar magnitude. 

While our sum-rule estimates of the $d_n$ contributions due to the dimension-six and dimension-eight operators of Weinberg type~(\ref{eq:o6o8}) have sizeable uncertainties, we believe that they are more robust than other existing determinations that are based on NDA~\cite{Weinberg:1989dx} or the VIA~\cite{Bigi:1990kz}. In particular, in the  sum-rule approach  there is no sign ambiguity between the prediction for $d_n$ and the hadronic matrix elements of $O_6$ and $O_8$ --- see for instance~\cite{Jung:2013hka,Chien:2015xha,Cirigliano:2016nyn,Dekens:2018bci,Brod:2018pli} for EDM studies that allow for both signs of the $O_6$ contribution. To find out whether our sum-rule estimates  are reliable would require first-principle calculations of the nEDM, which are in principle possible using existing  LQCD methodology. While such calculations  have  gained significant momentum~\cite{Bhattacharya:2015rsa,Bhattacharya:2016rrc,Abramczyk:2017oxr,Dragos:2017wms,Rizik:2018lrz,Kim:2018rce,Bhattacharya:2018qat,Syritsyn:2019vvt}, LQCD simulations involving Weinberg-type operators  are challenging~\cite{Gupta:2019fex,Cirigliano:2019jig}, and it remains to be seen which accuracy such computations can achieve in the near future. Till then phenomenological studies of hadronic EDMs have to rely on the predictions~(\ref{eq:nEDMO6num}) and~(\ref{eq:dn8numerics}) despite all their limitations.  
 
\acknowledgments We thank Jordy~De~Vries for pointing out that having explicit results for the $O_6$ and $O_8$ contributions to the proton EDM~(pEDM) would also be useful.  Following his suggestion, we provide the relevant numerical results in~Appendix~\ref{sec:dp}. We acknowledge helpful discussions with Yoshitaka~Hatta concerning the prediction~(\ref{eq:mup}) of the pEDM. The Feynman diagrams shown in this article have been drawn with the help of the \LaTeX \ package~{\tt feynMF}~\cite{Ohl:1995kr}. Parts of the Dirac and colour algebra calculations have been cross-checked against {\tt FeynCalc}~\cite{Shtabovenko:2016sxi} and {\tt TRACER}~\cite{Jamin:1991dp}. 

\appendix

\section{Weinberg-type contributions to the pEDM}
\label{sec:dp}

Below  we provide  numerical expressions  for the dimension-six and dimension-eight Weinberg-type contributions to the pEDM. Since the arguments given in Section~\ref{sec:pipheno} for the case of the neutron also apply to  the proton, the pEDM $d_p$ and the magnetic moment of the proton $\mu_p$ are related by $d_p = \mu_p \hspace{0.25mm} r(q^2)$. The function~$r(q^2)$ is then found by matching the phenomenological ansatz to the  OPE calculation, where the neutron-interpolating current~(\ref{eq:eta}) has to be replaced by a proton-interpolating current. In fact, the  proton-interpolating current is simply obtained from~(\ref{eq:j1j2}) by interchanging down-quark and up-quark fields, and in the limit of massless quarks the results (\ref{eq:opegen}) and~(\ref{eq:opegen211}) can be shown to hold in the case of the proton as well. As a result the function~$r(q^2)$ that appears in~(\ref{eq:dnmuntildeg}) and in $d_p = \mu_p \hspace{0.25mm} r(q^2)$ is the same. The analytic results for the $O_6$ and $O_8$ contributions to $d_p$ can therefore be derived from~(\ref{eq:nEDMO6}) and (\ref{eq:nEDMO8}), respectively,  by replacing~$\mu_n$ with~$\mu_p$. Experimentally, one has~\cite{Tanabashi:2018oca} 
\beq \label{eq:mup}
\mu_p  = 2.79 \, \frac{e}{2 m_p} = -1.46 \hspace{0.5mm} \mu_n = 1.49  \, \frac{e}{\rm GeV} \,,
\eeq
and it thus follows that in the case of the pEDM the numerical results (\ref{eq:nEDMO6num}) and (\ref{eq:dn8numerics}) read 
\begin{align} 
\left ( \frac{d_p}{e} \right )_{O_6} & = -109  \left ( 1 \pm 0.5 \right ) \, {\rm MeV}  \,, \label{eq:pEDM6} \\[2mm]
\left ( \frac{d_p}{e} \right )_{O_8}^m & = -3.7 \cdot 10^{-1} \left ( 1 \pm 0.8 \right ) \, {\rm GeV}^3  \; \Big \{ 1, 0.5, 0.2 \Big \} \label{eq:pEDM8} \,.
\end{align}
We stress that in (\ref{eq:mup}) the total magnetic moment of the proton and not only the anomalous part  (as claimed in~\cite{Yamanaka:2020kjo}) has to be used. This conclusion follows from generalising (\ref{eq:matj}) to (\ref{eq:upu}) to the case of a non-zero form factor $F_1(0)$ and agrees with the  results obtained in~\cite{Hatta:2020riw}. 

\section{Fixed-point gauge}
\label{sec:fixedpointgauge}

In the case of QCD the fixed-point  or Fock-Schwinger gauge~\cite{Fock:1937dy,Schwinger:1951nm} can  be  defined without loss of generality  for gauge-invariant quantities    by
\beq \label{eq:fpg}
x^\mu G_\mu^A (x) = 0 \,,
\eeq
where it is sufficient to restrict this choice of gauge to classical gluon fields $G_\mu^A$. 

For the gauge choice~(\ref{eq:fpg}) it is easy to show that one can express the gluon field  through the QCD field strength tensor $G^A_{\mu \nu} (x) =\partial_\mu G^A_\nu (x) -\partial_\nu G^A_\mu (x) +g_s f^{ABC} G^B_\mu (x)  \hspace{0.25mm} G^C_\nu (x)$ evaluated at $x = 0$. Here $g_s$ denotes the strong coupling constant. To derive the sought relation, one notices first that 
\beq \label{eq:g1}
\begin{split}
G_\mu^A (x) & = \partial_\mu \left(x^\nu \hspace{0.25mm} G^A_\nu(x) \right) - x^\nu \partial_\mu G_\nu^A(x) \\[2mm] 
& = - x^\nu \left ( G_{\mu \nu}^A (x) + \partial_\nu G_\mu^A (x) - g_s f^{ABC} G^B_\mu (x)  \hspace{0.25mm} G^C_\nu (x)  \right ) \\[2mm] 
& =  x^\nu G_{\nu \mu}^A (x) - x^\nu \partial_\nu G_\mu^A (x)   \,,
\end{split} 
\eeq
where we have employed the gauge condition~(\ref{eq:fpg}) twice and used the anti-symmetry of $G_{\mu \nu}^A$ to obtain the final result. Setting $x^\nu=\alpha y^\nu$ with an arbitrary parameter $\alpha$, one can then write 
\beq \label{eq:diff}
x^\nu G_{\nu \mu}^A (x)  = \alpha y^\nu G^A_{\nu \mu} (\alpha y)= G^A_\mu (\alpha y) + \alpha y^\nu \frac{\partial}{\partial ( \alpha y^\nu )} \, G^A_\mu(\alpha y) = \frac{d}{d\alpha} \left(\alpha G^A_\mu(\alpha y)\right) \,.
\eeq
Now if one integrates both sides of the above relation over $\alpha \in [0, 1]$ and assumes that $G^A_\mu(x)$ is non-singular at $x=0$, one finds 
\beq \label{eq:int}
\int_0^1 \! d\alpha \, \alpha y^\nu G^A_{\nu \mu} (\alpha y) = G^A_\mu(y) \,.
\eeq
Using a similar assumption for the QCD field strength, one can Taylor expand $G^A_{\nu \mu} (\alpha y)$  around $\alpha y =0$ and perform the integration on the left-hand side of~(\ref{eq:int}). It follows that 
\beq \label{eq:fexp1}
G_\mu^A (x) = \frac{1}{2} \hspace{0.25mm} x^\nu G^A_{\nu \mu}(0) + \frac{1}{3} \hspace{0.25mm}  x^\nu x^\rho \partial_\rho G^A_{\nu \mu} (0) + \ldots \,,
\eeq
where we have switched back from the variable $y$ to the variable $x$. 

The latter expression can be further simplified by noting that  as a result of~(\ref{eq:fpg}), the partial derivatives in~(\ref{eq:fexp1}) can be promoted to covariant derivatives $D_\mu = \partial_\mu -ig_s G^A_\mu  t^A$. In consequence, one has 
\beq \label{eq:fexp2}
G_\mu^A (x) = \frac{1}{2} \hspace{0.25mm} x^\nu G^A_{\nu \mu}(0) + \frac{1}{3} \hspace{0.25mm}  x^\nu x^\rho D_\rho G^A_{\nu \mu} (0) + \ldots \,, 
\eeq
in the fixed-point gauge of QCD. The same result also holds in the case of EM with $G_\mu^A$ replaced by~$A_\mu$ and $G_{\mu \nu}^A$ replaced by $F_{\mu \nu}$.

Finally, notice that due to~(\ref{eq:fpg}) the Taylor expansion of the classical quark field $q$ can also be formulated in terms of covariant  rather than partial derivatives. One has 
\beq \label{eq:qexp}
q(x) =  q(0) + x^\mu D_\mu \hspace{0.25mm} q(0) + \frac{1}{2}  \hspace{0.25mm}  x^\mu  x^\nu  D_\mu  D_\nu \hspace{0.25mm} q(0) +  \ldots \,. 
\eeq 

\section{Borel transforms}
\label{sec:borel}

We define the Borel transformation of a function $F(Q^2)$ with $Q^2 = -q^2$ in the following way 
\beq \label{eq:defborel}
\mathcal{B}\left[F(Q^2)\right] = \lim _{\text{\scriptsize$ \begin{matrix} Q^2,n \rightarrow \infty \\ Q^2/n = M^2 = {\rm const.} \end{matrix}$}} \frac{(Q^2)^n}{(n-1)!} \left( - \frac{d}{d Q^2}\right)^n F(Q^2) \,.
\eeq
The auxiliary parameter $M$ is called the Borel mass and occurs in the final result of most sum-rule calculations. A collection of useful Borel transforms can be found for instance in~\cite{Novikov:1983gd,Reinders:1984sr,gubler} and reads
\beq \label{eq:borelrules}
\begin{split}
\mathcal{B}\left[\left(Q^2\right)^k\right] & = 0 \,, \\[2mm]
\mathcal{B}\left[\left(\frac{1}{Q^2}\right)^k\right] & = \frac{1}{(k-1)!} \left(\frac{1}{M^2}\right)^k \,,  \\[2mm]
\mathcal{B}\left[\left(\frac{1}{s+Q^2}\right)^k \right] &= \frac{1}{(k-1)!} \left(\frac{1}{M^2}\right)^k e^{-\frac{s}{M^2}} \,, \\[2mm]
\mathcal{B}\left[\left(Q^2\right)^k \ln \left(Q^2\right)\right] & = (-1)^{k+1} \Gamma(k+1) (M^2)^k \,, \\[2mm]
\mathcal{B}\left[\left(\frac{1}{Q^2}\right)^k \left(\ln\left(\frac{Q^2}{\mu^2}\right)\right)^{-\epsilon}\right] & = \frac{1}{\Gamma(k)} \left(\frac{1}{M^2}\right)^k \left(\ln\left(\frac{Q^2}{\mu^2}\right)\right)^{-\epsilon} \hspace{0.5mm} \left[1+ \mathcal{O}\left( \left(\ln\left(\frac{Q^2}{\mu^2}\right)\right)^{-1} \right) \right] \,.
\end{split}
\eeq
Here $ k \in \mathbb{N}_+$ and $\Gamma (z)$ denotes the Euler gamma function. From~(\ref{eq:borelrules}) it is clear that all polynomial contributions to sum-rule correlation functions vanish after Borel transformation. 

\section{OPE for the quark propagator}
\label{sec:quarkOPE}

The free position-space propagator $S^{(0)} (x)$ of a massless quark is easily derived by taking the Fourier transform of the well-known momentum-space representation
\beq \label{eq:S0def}
S^{(0)} (x) = \int \! {\dbar}^{\; 4} p \, e^{-ipx} \, \frac{i \slashed{p}}{p^2} = \frac{1}{(2\pi)^4} \left ( -i\slashed{\partial}_E \right )  \int \! d^4p_E \, \frac{e^{ip_E x_E}}{p_E^2}\, ,
\eeq
where colour indices are implicit and we have applied a Wick rotation to Euclidean space using $x^2\rightarrow -x^2_E$ and $p^2\rightarrow -p^2_E$. The four-dimensional integration measure can be written as $d^4 p_E =  \bar{p}^{3} \hspace{0.25mm} d\bar{p}\hspace{0.5mm} d\Omega_4$, where  the magnitude of the four-dimensional Euclidean momentum vector has been denoted by $\bar{p} = |p_E|$. The differential solid angle is given by
\beq \label{eq:omega4}
\int \! d\Omega_4 = \int_0^{2\pi}  d\phi \int_0^\pi  d\theta_1  d\theta_{2} \, \sin \theta_{1} \sin ^{2} \theta_{2}   \,.
\eeq
It follows that $d\Omega_4= d\Omega_3 \sin^2 \theta_2$ and~(\ref{eq:S0def}) hence takes the form 
\beq \label{eq:S0fin}
\begin{split}
S^{(0)}(x) & =\frac{1}{4\pi^3} \, (-i\slashed{\partial}_E) \int_0^\infty \! d\bar{p} \bar{p} \int_0^\pi d\theta_2 \, \sin^2 \theta_2 \, e^{i\bar{p}\bar{x}\cos \theta_2 } \\[2mm]
& = \frac{1}{4\pi^3} \, (-i\slashed{\partial}_E) \int_0^\infty \! d\bar{p} \bar{p} \; \frac{\pi J_1 (\bar{p}\bar{x})}{\bar{p}\bar{x}}
 = \frac{1}{4\pi^2} \, (-i\slashed{\partial}_E) \, \frac{1}{\bar{x}^2}  = \frac{i \slashed{x}}{2\pi^2 x^4} \,.
\end{split}
\eeq
Here $\bar x = |x_E|$, $J_1(z)$ denotes the Bessel function of first kind with index $1$ and in the final step we have rotated back from Euclidean to Minkowski space noting that $\slashed{x}_E \rightarrow \slashed{x}$. 

In order to determine the non-perturbative contributions $S^q (x)$ to the quark propagator, one needs to evaluate the correlator $\big \langle \Omega  | T \big  [ q_a^i (x) \hspace{0.25mm} \bar q_b^j (0)  \big ] | \Omega \big \rangle$. Using the expansion of the classic quark field~(\ref{eq:qexp}), one obtains 
\beq \label{eq:Sqcalc}
\big \langle \Omega | T \big  [ q_a^i (x) \hspace{0.25mm} \bar q_b^j (0)  \big ]  | \Omega \big \rangle =  \big \langle  q_a^i (0) \hspace{0.25mm} \bar q_b^j (0) \big \rangle  + \ldots = -\frac{1}{12} \, \delta_{ab}  \hspace{0.25mm} \delta^{ij} \, \langle \bar q q \rangle + \ldots \,, 
\eeq
where the fields of the condensate $\langle \bar q q \rangle$ are evaluated at $x = 0$. Notice that the minus sign in the final result comes from the exchange of the fermion fields and the numerical prefactor can be determined by contracting the expression in the middle and on the right  with $ \delta^{ab}  \hspace{0.25mm} \delta_{ij}$. Ignoring colour and spinor indices, the expansion~(\ref{eq:Sqcalc}) thus leads to the expression for $S^q(x)$ as given in~(\ref{eq:firstsecond}). 

\section{Fourier transforms}
\label{sec:ft}

We define the Fourier transform of a  function $F(p)$ by
\beq\label{eq:fourierdef}
{\cal F} \hspace{0.25mm} \Big [ F(p) \Big ]  = \left ( \frac{\mu^2_{\rm IR}}{4\hspace{0.25mm}   \pi \hspace{0.25mm} e^{\gamma_E}} \right ) ^{-\epsilon_{\rm IR}}  \int \! \dbar^{\, \,  4+2 \epsilon_{\rm IR}} p  \, e^{-ipx} \, F(p) \,, 
\eeq
where we have performed the integration in $d=4+2\epsilon_{\rm IR}$ space-time dimensions with $\epsilon_{\rm IR} > 0$ to regulate the IR divergences that appear in some of the Fourier integrals that we have encountered  in Sections~\ref{sec:OPEwein6} and \ref{sec:o8quarkprop}. The symbol  $\gamma_E$ denotes the Euler's constant and $\mu_{\rm IR}$ is a mass scale needed to restore the correct dimensionality of~(\ref{eq:fourierdef}). 

In the case that $F(p)$ is  polynomial in $1/p^2$, a simple calculation along the lines of  the computation performed in Section~\ref{sec:quarkOPE} leads to 
\beq\label{eq:FTp2}
{\cal F} \hspace{0.25mm} \left  [ \frac{1}{\left (p^2 \right )^k}  \right  ]  = \frac{i}{4^k \hspace{0.25mm} \pi^2}  \hspace{0.25mm} \left ( - \frac{\mu_{\rm IR}^2 x^2}{4 \hspace{0.25mm} e^{\gamma_E}} \right )^{-\epsilon_{\rm IR}}  \frac{\Gamma \left ( 2 - k + \epsilon_{\rm IR} \right ) }{ \Gamma \left ( k \right ) }   \hspace{0.25mm} \left ( x^2 \right )^{k-2} \,, 
\eeq
where $ k \in \mathbb{N}_+$. The Fourier  transforms of type~(\ref{eq:FTp2}) relevant for our article are 
\begin{align} 
{\cal F} \hspace{0.25mm} \left  [ \frac{1}{p^2}  \right  ]  & = \frac{i}{4  \hspace{0.25mm} \pi^2 x^2} \,, \label{eq:FT1} \\[2mm]
{\cal F} \hspace{0.25mm} \left  [ \frac{1}{p^4}  \right  ]  & = \frac{i}{16 \hspace{0.25mm} \pi^2} \left [ \frac{1}{\epsilon_{\rm IR}} - \ln \left ( -\frac{\mu_{\rm IR}^2 x^2}{4} \right )  \right ] \,. \label{eq:FT2}  
\end{align}

Tensor Fourier integrals with a polynomial denominator of the form $(p^2)^k$ can be obtained from (\ref{eq:FTp2}) by taking derivatives 
\beq \label{eq:tensorFTgeneral}
{\cal F} \hspace{0.25mm} \left [  \frac{p_\mu p_\nu \cdots}{\left (p^2 \right )^k}   \right  ]  =\big ( i \partial_\mu \big )  \hspace{0.5mm} \big ( i \partial_\nu \big) \, \cdots \,   {\cal F} \hspace{0.25mm} \left [  \frac{1}{\left (p^2 \right )^k}   \right  ] \,.
\eeq
This procedure allows  one to derive  for example 
\beq \label{eq:FT4}
{\cal F} \hspace{0.25mm} \left [  \frac{p_\mu p_\nu}{p^4}   \right  ]  =  -\frac{ i}{4 \hspace{0.25mm} \pi^2 x^4} \left ( x_{\mu} x_{\nu} - \frac{g_{\mu \nu}}{2} x^2 \right )  \,,
\eeq
a relation that has been used in Sections~\ref{sec:OPEwein6} and \ref{sec:o8quarkprop}. 

In our sum-rule calculations we have also encountered ultraviolet~(UV) divergent Fourier transformations of a function $F(x)$. To regulate UV divergences we work  in $d=4-2\epsilon_{\rm UV}$ space-time dimensions with $\epsilon_{\rm UV} > 0$, introduce the mass scale $\mu_{\rm UV}$    and define 
\beq\label{eq:invfourierdef}
 {\cal F} \hspace{0.25mm} \Big [ F(x) \Big ]  = \left ( \frac{\mu^2_{\rm UV}  \hspace{0.25mm} e^{\gamma_E} }{4\hspace{0.25mm}   \pi}  \right )^{-\epsilon_{\rm UV}}  \int \! d^{\, 4 - 2 \epsilon_{\rm UV}} x  \, e^{ipx} \, F(x) \,.
\eeq

If $F(x)$ is polynomial in $1/x^2$, it is straightforward to evaluate (\ref{eq:invfourierdef}). For $ k \in \mathbb{N}_+$ we obtain 
\beq\label{eq:FTx2}
{\cal F} \hspace{0.25mm} \left  [ \frac{1}{\left (x^2 \right )^k}  \right  ]  = -\frac{i \hspace{0.25mm}  \pi^2}{4^{k-2}}  \hspace{0.25mm} \left ( - \frac{\mu_{\rm UV}^2 \hspace{0.25mm} e^{\gamma_E}}{ p^2  } \right )^{-\epsilon_{\rm UV}}  \frac{\Gamma \left ( 2 - k - \epsilon_{\rm UV} \right ) }{ \Gamma \left ( k \right ) }   \hspace{0.25mm} \left ( p^2 \right )^{k-2} \,.
\eeq
The Fourier integrals of the form~(\ref{eq:FTx2}) that occur in our article are   
\begin{align} 
{\cal F} \hspace{0.25mm} \left  [ \frac{1}{x^2}  \right  ]  & = -\frac{4 \hspace{0.25mm} i \hspace{0.25mm} \pi^2}{p^2} \,, \label{eq:FT1add} \\[2mm]
{\cal F} \hspace{0.25mm} \left  [ \frac{1}{x^6}  \right  ]  & = -\frac{i \hspace{0.25mm} \pi^2   p^2}{8}  \left [ \frac{1}{\epsilon_{\rm UV}} - \ln \left ( -\frac{\mu_{\rm UV}^2}{p^2} \right ) - 1 \right ] \label{eq:FT3add}  \,.
\end{align}

We also encounter in our work Fourier transforms that are both IR and UV divergent. They are of the type ${\cal F} \hspace{0.25mm} \Big [ 1/\big ((x^2)^l  \big ) \hspace{1mm} {\cal F} \hspace{0.25mm} \Big [ 1/\big ((p^2)^k  \big )   \Big ]   \Big ] $ with $k,l  \in \mathbb{N}_+$. Using the result given in~(\ref{eq:FTp2}) and~(\ref{eq:FTx2}) these double integrals are readily computed. We find  
\beq \label{eq:doubleFT}
\begin{split}
 {\cal F} \hspace{0.25mm} \left [ \frac{ 1}{\left ( x^2 \right )^l}  \, {\cal F} \left [ \frac{ 1}{\left ( p^2 \right )^k}  \right ] \right ] & = \frac{1}{4^l}  \hspace{0.25mm} \left ( - \frac{\mu_{\rm IR}^2}{ p^2  \hspace{0.25mm} e^{\gamma_E}} \right )^{-\epsilon_{\rm IR}} \left ( - \frac{\mu_{\rm UV}^2  \hspace{0.25mm} e^{\gamma_E}}{ p^2 } \right )^{-\epsilon_{\rm UV}}  \\[2mm] 
 &\phantom{xx} \times \frac{\Gamma \left ( 2 - k +\epsilon_{\rm IR} \right ) \Gamma \left ( k - l  - \epsilon_{\rm IR}  - \epsilon_{\rm UV}  \right )  }{ \Gamma \left ( k  \right )  \Gamma \left ( 2 - k + l + \epsilon_{\rm IR} \right ) }   \hspace{0.25mm} \left ( p^2 \right )^{l - k} \,.
\end{split}
\eeq
The only Fourier integral of the form~(\ref{eq:doubleFT}) that is necessary to compute the two-loop contributions to the OPE correlation functions considered in this paper is 
\beq \label{eq:doubleFTneeded}
 {\cal F} \hspace{0.25mm} \left [ \frac{ 1}{x^6}  \, {\cal F} \left [ \frac{ 1}{p^4 }  \right ] \right ]= \frac{p^2}{128}  \hspace{0.25mm}  \ln \left ( - \frac{\mu_{\rm IR}^2}{ p^2}  \right ) \hspace{0.5mm}  \ln \left ( - \frac{\mu_{\rm UV}^2}{ p^2}  \right ) + \ldots \,,
\eeq
where the ellipses represent  terms that vanish after Borel transformation (cf.~Appendix~\ref{sec:borel}), meaning that these contributions do not enter the analytic expressions~(\ref{eq:nEDMO6}) and~(\ref{eq:nEDMO8}).  


\begin{thebibliography}{99}%
\makeatletter
\providecommand \@ifxundefined [1]{%
 \@ifx{#1\undefined}
}%
\providecommand \@ifnum [1]{%
 \ifnum #1\expandafter \@firstoftwo
 \else \expandafter \@secondoftwo
 \fi
}%
\providecommand \@ifx [1]{%
 \ifx #1\expandafter \@firstoftwo
 \else \expandafter \@secondoftwo
 \fi
}%
\providecommand \natexlab [1]{#1}%
\providecommand \enquote  [1]{``#1''}%
\providecommand \bibnamefont  [1]{#1}%
\providecommand \bibfnamefont [1]{#1}%
\providecommand \citenamefont [1]{#1}%
\providecommand \href@noop [0]{\@secondoftwo}%
\providecommand \href [0]{\begingroup \@sanitize@url \@href}%
\providecommand \@href[1]{\@@startlink{#1}\@@href}%
\providecommand \@@href[1]{\endgroup#1\@@endlink}%
\providecommand \@sanitize@url [0]{\catcode `\\12\catcode `\$12\catcode
  `\&12\catcode `\#12\catcode `\^12\catcode `\_12\catcode `\%12\relax}%
\providecommand \@@startlink[1]{}%
\providecommand \@@endlink[0]{}%
\providecommand \url  [0]{\begingroup\@sanitize@url \@url }%
\providecommand \@url [1]{\endgroup\@href {#1}{\urlprefix }}%
\providecommand \urlprefix  [0]{URL }%
\providecommand \Eprint [0]{\href }%
\providecommand \doibase [0]{http://dx.doi.org/}%
\providecommand \selectlanguage [0]{\@gobble}%
\providecommand \bibinfo  [0]{\@secondoftwo}%
\providecommand \bibfield  [0]{\@secondoftwo}%
\providecommand \translation [1]{[#1]}%
\providecommand \BibitemOpen [0]{}%
\providecommand \bibitemStop [0]{}%
\providecommand \bibitemNoStop [0]{.\EOS\space}%
\providecommand \EOS [0]{\spacefactor3000\relax}%
\providecommand \BibitemShut  [1]{\csname bibitem#1\endcsname}%
\let\auto@bib@innerbib\@empty
\bibitem [{\citenamefont {Pospelov}\ and\ \citenamefont
  {Ritz}(2005)}]{Pospelov:2005pr}%
  \BibitemOpen
  \bibfield  {author} {\bibinfo {author} {\bibfnamefont {M.}~\bibnamefont
  {Pospelov}}\ and\ \bibinfo {author} {\bibfnamefont {A.}~\bibnamefont
  {Ritz}},\ }\href {\doibase 10.1016/j.aop.2005.04.002} {\bibfield  {journal}
  {\bibinfo  {journal} {Annals Phys.}\ }\textbf {\bibinfo {volume} {318}},\
  \bibinfo {pages} {119} (\bibinfo {year} {2005})},\ \Eprint
  {http://arxiv.org/abs/hep-ph/0504231} {arXiv:hep-ph/0504231 [hep-ph]}
  \BibitemShut {NoStop}%
\bibitem [{\citenamefont {Li}\ \emph {et~al.}(2010)\citenamefont {Li},
  \citenamefont {Profumo},\ and\ \citenamefont {Ramsey-Musolf}}]{Li:2010ax}%
  \BibitemOpen
  \bibfield  {author} {\bibinfo {author} {\bibfnamefont {Y.}~\bibnamefont
  {Li}}, \bibinfo {author} {\bibfnamefont {S.}~\bibnamefont {Profumo}}, \ and\
  \bibinfo {author} {\bibfnamefont {M.}~\bibnamefont {Ramsey-Musolf}},\ }\href
  {\doibase 10.1007/JHEP08(2010)062} {\bibfield  {journal} {\bibinfo  {journal}
  {JHEP}\ }\textbf {\bibinfo {volume} {08}},\ \bibinfo {pages} {062} (\bibinfo
  {year} {2010})},\ \Eprint {http://arxiv.org/abs/1006.1440} {arXiv:1006.1440
  [hep-ph]} \BibitemShut {NoStop}%
\bibitem [{\citenamefont {McKeen}\ \emph {et~al.}(2012)\citenamefont {McKeen},
  \citenamefont {Pospelov},\ and\ \citenamefont {Ritz}}]{McKeen:2012av}%
  \BibitemOpen
  \bibfield  {author} {\bibinfo {author} {\bibfnamefont {D.}~\bibnamefont
  {McKeen}}, \bibinfo {author} {\bibfnamefont {M.}~\bibnamefont {Pospelov}}, \
  and\ \bibinfo {author} {\bibfnamefont {A.}~\bibnamefont {Ritz}},\ }\href
  {\doibase 10.1103/PhysRevD.86.113004} {\bibfield  {journal} {\bibinfo
  {journal} {Phys. Rev.}\ }\textbf {\bibinfo {volume} {D86}},\ \bibinfo {pages}
  {113004} (\bibinfo {year} {2012})},\ \Eprint {http://arxiv.org/abs/1208.4597}
  {arXiv:1208.4597 [hep-ph]} \BibitemShut {NoStop}%
\bibitem [{\citenamefont {Engel}\ \emph {et~al.}(2013)\citenamefont {Engel},
  \citenamefont {Ramsey-Musolf},\ and\ \citenamefont {van
  Kolck}}]{Engel:2013lsa}%
  \BibitemOpen
  \bibfield  {author} {\bibinfo {author} {\bibfnamefont {J.}~\bibnamefont
  {Engel}}, \bibinfo {author} {\bibfnamefont {M.~J.}\ \bibnamefont
  {Ramsey-Musolf}}, \ and\ \bibinfo {author} {\bibfnamefont {U.}~\bibnamefont
  {van Kolck}},\ }\href {\doibase 10.1016/j.ppnp.2013.03.003} {\bibfield
  {journal} {\bibinfo  {journal} {Prog. Part. Nucl. Phys.}\ }\textbf {\bibinfo
  {volume} {71}},\ \bibinfo {pages} {21} (\bibinfo {year} {2013})},\ \Eprint
  {http://arxiv.org/abs/1303.2371} {arXiv:1303.2371 [nucl-th]} \BibitemShut
  {NoStop}%
\bibitem [{\citenamefont {Chang}\ \emph {et~al.}(2013)\citenamefont {Chang},
  \citenamefont {Pan},\ and\ \citenamefont {Xu}}]{Chang:2013cia}%
  \BibitemOpen
  \bibfield  {author} {\bibinfo {author} {\bibfnamefont {W.-F.}\ \bibnamefont
  {Chang}}, \bibinfo {author} {\bibfnamefont {W.-P.}\ \bibnamefont {Pan}}, \
  and\ \bibinfo {author} {\bibfnamefont {F.}~\bibnamefont {Xu}},\ }\href
  {\doibase 10.1103/PhysRevD.88.033004} {\bibfield  {journal} {\bibinfo
  {journal} {Phys. Rev.}\ }\textbf {\bibinfo {volume} {D88}},\ \bibinfo {pages}
  {033004} (\bibinfo {year} {2013})},\ \Eprint {http://arxiv.org/abs/1303.7035}
  {arXiv:1303.7035 [hep-ph]} \BibitemShut {NoStop}%
\bibitem [{\citenamefont {Jung}\ and\ \citenamefont
  {Pich}(2014)}]{Jung:2013hka}%
  \BibitemOpen
  \bibfield  {author} {\bibinfo {author} {\bibfnamefont {M.}~\bibnamefont
  {Jung}}\ and\ \bibinfo {author} {\bibfnamefont {A.}~\bibnamefont {Pich}},\
  }\href {\doibase 10.1007/JHEP04(2014)076} {\bibfield  {journal} {\bibinfo
  {journal} {JHEP}\ }\textbf {\bibinfo {volume} {04}},\ \bibinfo {pages} {076}
  (\bibinfo {year} {2014})},\ \Eprint {http://arxiv.org/abs/1308.6283}
  {arXiv:1308.6283 [hep-ph]} \BibitemShut {NoStop}%
\bibitem [{\citenamefont {Gripaios}\ and\ \citenamefont
  {Sutherland}(2014)}]{Gripaios:2013lea}%
  \BibitemOpen
  \bibfield  {author} {\bibinfo {author} {\bibfnamefont {B.}~\bibnamefont
  {Gripaios}}\ and\ \bibinfo {author} {\bibfnamefont {D.}~\bibnamefont
  {Sutherland}},\ }\href {\doibase 10.1103/PhysRevD.89.076004} {\bibfield
  {journal} {\bibinfo  {journal} {Phys. Rev.}\ }\textbf {\bibinfo {volume}
  {D89}},\ \bibinfo {pages} {076004} (\bibinfo {year} {2014})},\ \Eprint
  {http://arxiv.org/abs/1309.7822} {arXiv:1309.7822 [hep-ph]} \BibitemShut
  {NoStop}%
\bibitem [{\citenamefont {Brod}\ \emph {et~al.}(2013)\citenamefont {Brod},
  \citenamefont {Haisch},\ and\ \citenamefont {Zupan}}]{Brod:2013cka}%
  \BibitemOpen
  \bibfield  {author} {\bibinfo {author} {\bibfnamefont {J.}~\bibnamefont
  {Brod}}, \bibinfo {author} {\bibfnamefont {U.}~\bibnamefont {Haisch}}, \ and\
  \bibinfo {author} {\bibfnamefont {J.}~\bibnamefont {Zupan}},\ }\href
  {\doibase 10.1007/JHEP11(2013)180} {\bibfield  {journal} {\bibinfo  {journal}
  {JHEP}\ }\textbf {\bibinfo {volume} {11}},\ \bibinfo {pages} {180} (\bibinfo
  {year} {2013})},\ \Eprint {http://arxiv.org/abs/1310.1385} {arXiv:1310.1385
  [hep-ph]} \BibitemShut {NoStop}%
\bibitem [{\citenamefont {Inoue}\ \emph {et~al.}(2014)\citenamefont {Inoue},
  \citenamefont {Ramsey-Musolf},\ and\ \citenamefont {Zhang}}]{Inoue:2014nva}%
  \BibitemOpen
  \bibfield  {author} {\bibinfo {author} {\bibfnamefont {S.}~\bibnamefont
  {Inoue}}, \bibinfo {author} {\bibfnamefont {M.~J.}\ \bibnamefont
  {Ramsey-Musolf}}, \ and\ \bibinfo {author} {\bibfnamefont {Y.}~\bibnamefont
  {Zhang}},\ }\href {\doibase 10.1103/PhysRevD.89.115023} {\bibfield  {journal}
  {\bibinfo  {journal} {Phys. Rev.}\ }\textbf {\bibinfo {volume} {D89}},\
  \bibinfo {pages} {115023} (\bibinfo {year} {2014})},\ \Eprint
  {http://arxiv.org/abs/1403.4257} {arXiv:1403.4257 [hep-ph]} \BibitemShut
  {NoStop}%
\bibitem [{\citenamefont {Gorbahn}\ and\ \citenamefont
  {Haisch}(2014)}]{Gorbahn:2014sha}%
  \BibitemOpen
  \bibfield  {author} {\bibinfo {author} {\bibfnamefont {M.}~\bibnamefont
  {Gorbahn}}\ and\ \bibinfo {author} {\bibfnamefont {U.}~\bibnamefont
  {Haisch}},\ }\href {\doibase 10.1007/JHEP06(2014)033} {\bibfield  {journal}
  {\bibinfo  {journal} {JHEP}\ }\textbf {\bibinfo {volume} {06}},\ \bibinfo
  {pages} {033} (\bibinfo {year} {2014})},\ \Eprint
  {http://arxiv.org/abs/1404.4873} {arXiv:1404.4873 [hep-ph]} \BibitemShut
  {NoStop}%
\bibitem [{\citenamefont {Altmannshofer}\ \emph {et~al.}(2015)\citenamefont
  {Altmannshofer}, \citenamefont {Brod},\ and\ \citenamefont
  {Schmaltz}}]{Altmannshofer:2015qra}%
  \BibitemOpen
  \bibfield  {author} {\bibinfo {author} {\bibfnamefont {W.}~\bibnamefont
  {Altmannshofer}}, \bibinfo {author} {\bibfnamefont {J.}~\bibnamefont {Brod}},
  \ and\ \bibinfo {author} {\bibfnamefont {M.}~\bibnamefont {Schmaltz}},\
  }\href {\doibase 10.1007/JHEP05(2015)125} {\bibfield  {journal} {\bibinfo
  {journal} {JHEP}\ }\textbf {\bibinfo {volume} {05}},\ \bibinfo {pages} {125}
  (\bibinfo {year} {2015})},\ \Eprint {http://arxiv.org/abs/1503.04830}
  {arXiv:1503.04830 [hep-ph]} \BibitemShut {NoStop}%
\bibitem [{\citenamefont {Dwivedi}\ \emph {et~al.}(2015)\citenamefont
  {Dwivedi}, \citenamefont {Ghosh}, \citenamefont {Mukhopadhyaya},\ and\
  \citenamefont {Shivaji}}]{Dwivedi:2015nta}%
  \BibitemOpen
  \bibfield  {author} {\bibinfo {author} {\bibfnamefont {S.}~\bibnamefont
  {Dwivedi}}, \bibinfo {author} {\bibfnamefont {D.~K.}\ \bibnamefont {Ghosh}},
  \bibinfo {author} {\bibfnamefont {B.}~\bibnamefont {Mukhopadhyaya}}, \ and\
  \bibinfo {author} {\bibfnamefont {A.}~\bibnamefont {Shivaji}},\ }\href
  {\doibase 10.1103/PhysRevD.92.095015} {\bibfield  {journal} {\bibinfo
  {journal} {Phys. Rev.}\ }\textbf {\bibinfo {volume} {D92}},\ \bibinfo {pages}
  {095015} (\bibinfo {year} {2015})},\ \Eprint
  {http://arxiv.org/abs/1505.05844} {arXiv:1505.05844 [hep-ph]} \BibitemShut
  {NoStop}%
\bibitem [{\citenamefont {Chien}\ \emph {et~al.}(2016)\citenamefont {Chien},
  \citenamefont {Cirigliano}, \citenamefont {Dekens}, \citenamefont
  {de~Vries},\ and\ \citenamefont {Mereghetti}}]{Chien:2015xha}%
  \BibitemOpen
  \bibfield  {author} {\bibinfo {author} {\bibfnamefont {Y.~T.}\ \bibnamefont
  {Chien}}, \bibinfo {author} {\bibfnamefont {V.}~\bibnamefont {Cirigliano}},
  \bibinfo {author} {\bibfnamefont {W.}~\bibnamefont {Dekens}}, \bibinfo
  {author} {\bibfnamefont {J.}~\bibnamefont {de~Vries}}, \ and\ \bibinfo
  {author} {\bibfnamefont {E.}~\bibnamefont {Mereghetti}},\ }\href {\doibase
  10.1007/JHEP02(2016)011} {\bibfield  {journal} {\bibinfo  {journal} {JHEP}\
  }\textbf {\bibinfo {volume} {02}},\ \bibinfo {pages} {011} (\bibinfo {year}
  {2016})},\ \Eprint {http://arxiv.org/abs/1510.00725} {arXiv:1510.00725
  [hep-ph]} \BibitemShut {NoStop}%
\bibitem [{\citenamefont {Cirigliano}\ \emph
  {et~al.}(2016{\natexlab{a}})\citenamefont {Cirigliano}, \citenamefont
  {Dekens}, \citenamefont {de~Vries},\ and\ \citenamefont
  {Mereghetti}}]{Cirigliano:2016njn}%
  \BibitemOpen
  \bibfield  {author} {\bibinfo {author} {\bibfnamefont {V.}~\bibnamefont
  {Cirigliano}}, \bibinfo {author} {\bibfnamefont {W.}~\bibnamefont {Dekens}},
  \bibinfo {author} {\bibfnamefont {J.}~\bibnamefont {de~Vries}}, \ and\
  \bibinfo {author} {\bibfnamefont {E.}~\bibnamefont {Mereghetti}},\ }\href
  {\doibase 10.1103/PhysRevD.94.016002} {\bibfield  {journal} {\bibinfo
  {journal} {Phys. Rev.}\ }\textbf {\bibinfo {volume} {D94}},\ \bibinfo {pages}
  {016002} (\bibinfo {year} {2016}{\natexlab{a}})},\ \Eprint
  {http://arxiv.org/abs/1603.03049} {arXiv:1603.03049 [hep-ph]} \BibitemShut
  {NoStop}%
\bibitem [{\citenamefont {Cirigliano}\ \emph
  {et~al.}(2016{\natexlab{b}})\citenamefont {Cirigliano}, \citenamefont
  {Dekens}, \citenamefont {de~Vries},\ and\ \citenamefont
  {Mereghetti}}]{Cirigliano:2016nyn}%
  \BibitemOpen
  \bibfield  {author} {\bibinfo {author} {\bibfnamefont {V.}~\bibnamefont
  {Cirigliano}}, \bibinfo {author} {\bibfnamefont {W.}~\bibnamefont {Dekens}},
  \bibinfo {author} {\bibfnamefont {J.}~\bibnamefont {de~Vries}}, \ and\
  \bibinfo {author} {\bibfnamefont {E.}~\bibnamefont {Mereghetti}},\ }\href
  {\doibase 10.1103/PhysRevD.94.034031} {\bibfield  {journal} {\bibinfo
  {journal} {Phys. Rev.}\ }\textbf {\bibinfo {volume} {D94}},\ \bibinfo {pages}
  {034031} (\bibinfo {year} {2016}{\natexlab{b}})},\ \Eprint
  {http://arxiv.org/abs/1605.04311} {arXiv:1605.04311 [hep-ph]} \BibitemShut
  {NoStop}%
\bibitem [{\citenamefont {Yamanaka}\ \emph {et~al.}(2017)\citenamefont
  {Yamanaka}, \citenamefont {Sahoo}, \citenamefont {Yoshinaga}, \citenamefont
  {Sato}, \citenamefont {Asahi},\ and\ \citenamefont {Das}}]{Yamanaka:2017mef}%
  \BibitemOpen
  \bibfield  {author} {\bibinfo {author} {\bibfnamefont {N.}~\bibnamefont
  {Yamanaka}}, \bibinfo {author} {\bibfnamefont {B.~K.}\ \bibnamefont {Sahoo}},
  \bibinfo {author} {\bibfnamefont {N.}~\bibnamefont {Yoshinaga}}, \bibinfo
  {author} {\bibfnamefont {T.}~\bibnamefont {Sato}}, \bibinfo {author}
  {\bibfnamefont {K.}~\bibnamefont {Asahi}}, \ and\ \bibinfo {author}
  {\bibfnamefont {B.~P.}\ \bibnamefont {Das}},\ }\href {\doibase
  10.1140/epja/i2017-12237-2} {\bibfield  {journal} {\bibinfo  {journal} {Eur.
  Phys. J.}\ }\textbf {\bibinfo {volume} {A53}},\ \bibinfo {pages} {54}
  (\bibinfo {year} {2017})},\ \Eprint {http://arxiv.org/abs/1703.01570}
  {arXiv:1703.01570 [hep-ph]} \BibitemShut {NoStop}%
\bibitem [{\citenamefont {Yanase}\ \emph {et~al.}(2019)\citenamefont {Yanase},
  \citenamefont {Yoshinaga}, \citenamefont {Higashiyama},\ and\ \citenamefont
  {Yamanaka}}]{Yanase:2018qqq}%
  \BibitemOpen
  \bibfield  {author} {\bibinfo {author} {\bibfnamefont {K.}~\bibnamefont
  {Yanase}}, \bibinfo {author} {\bibfnamefont {N.}~\bibnamefont {Yoshinaga}},
  \bibinfo {author} {\bibfnamefont {K.}~\bibnamefont {Higashiyama}}, \ and\
  \bibinfo {author} {\bibfnamefont {N.}~\bibnamefont {Yamanaka}},\ }\href
  {\doibase 10.1103/PhysRevD.99.075021} {\bibfield  {journal} {\bibinfo
  {journal} {Phys. Rev.}\ }\textbf {\bibinfo {volume} {D99}},\ \bibinfo {pages}
  {075021} (\bibinfo {year} {2019})},\ \Eprint
  {http://arxiv.org/abs/1805.00419} {arXiv:1805.00419 [nucl-th]} \BibitemShut
  {NoStop}%
\bibitem [{\citenamefont {Dekens}\ \emph {et~al.}(2019)\citenamefont {Dekens},
  \citenamefont {de~Vries}, \citenamefont {Jung},\ and\ \citenamefont
  {Vos}}]{Dekens:2018bci}%
  \BibitemOpen
  \bibfield  {author} {\bibinfo {author} {\bibfnamefont {W.}~\bibnamefont
  {Dekens}}, \bibinfo {author} {\bibfnamefont {J.}~\bibnamefont {de~Vries}},
  \bibinfo {author} {\bibfnamefont {M.}~\bibnamefont {Jung}}, \ and\ \bibinfo
  {author} {\bibfnamefont {K.~K.}\ \bibnamefont {Vos}},\ }\href {\doibase
  10.1007/JHEP01(2019)069} {\bibfield  {journal} {\bibinfo  {journal} {JHEP}\
  }\textbf {\bibinfo {volume} {01}},\ \bibinfo {pages} {069} (\bibinfo {year}
  {2019})},\ \Eprint {http://arxiv.org/abs/1809.09114} {arXiv:1809.09114
  [hep-ph]} \BibitemShut {NoStop}%
\bibitem [{\citenamefont {Cesarotti}\ \emph {et~al.}(2019)\citenamefont
  {Cesarotti}, \citenamefont {Lu}, \citenamefont {Nakai}, \citenamefont
  {Parikh},\ and\ \citenamefont {Reece}}]{Cesarotti:2018huy}%
  \BibitemOpen
  \bibfield  {author} {\bibinfo {author} {\bibfnamefont {C.}~\bibnamefont
  {Cesarotti}}, \bibinfo {author} {\bibfnamefont {Q.}~\bibnamefont {Lu}},
  \bibinfo {author} {\bibfnamefont {Y.}~\bibnamefont {Nakai}}, \bibinfo
  {author} {\bibfnamefont {A.}~\bibnamefont {Parikh}}, \ and\ \bibinfo {author}
  {\bibfnamefont {M.}~\bibnamefont {Reece}},\ }\href {\doibase
  10.1007/JHEP05(2019)059} {\bibfield  {journal} {\bibinfo  {journal} {JHEP}\
  }\textbf {\bibinfo {volume} {05}},\ \bibinfo {pages} {059} (\bibinfo {year}
  {2019})},\ \Eprint {http://arxiv.org/abs/1810.07736} {arXiv:1810.07736
  [hep-ph]} \BibitemShut {NoStop}%
\bibitem [{\citenamefont {Panico}\ \emph {et~al.}(2019)\citenamefont {Panico},
  \citenamefont {Pomarol},\ and\ \citenamefont {Riembau}}]{Panico:2018hal}%
  \BibitemOpen
  \bibfield  {author} {\bibinfo {author} {\bibfnamefont {G.}~\bibnamefont
  {Panico}}, \bibinfo {author} {\bibfnamefont {A.}~\bibnamefont {Pomarol}}, \
  and\ \bibinfo {author} {\bibfnamefont {M.}~\bibnamefont {Riembau}},\ }\href
  {\doibase 10.1007/JHEP04(2019)090} {\bibfield  {journal} {\bibinfo  {journal}
  {JHEP}\ }\textbf {\bibinfo {volume} {04}},\ \bibinfo {pages} {090} (\bibinfo
  {year} {2019})},\ \Eprint {http://arxiv.org/abs/1810.09413} {arXiv:1810.09413
  [hep-ph]} \BibitemShut {NoStop}%
\bibitem [{\citenamefont {Brod}\ and\ \citenamefont
  {Stamou}(2018)}]{Brod:2018pli}%
  \BibitemOpen
  \bibfield  {author} {\bibinfo {author} {\bibfnamefont {J.}~\bibnamefont
  {Brod}}\ and\ \bibinfo {author} {\bibfnamefont {E.}~\bibnamefont {Stamou}},\
  }\href@noop {} {\  (\bibinfo {year} {2018})},\ \Eprint
  {http://arxiv.org/abs/1810.12303} {arXiv:1810.12303 [hep-ph]} \BibitemShut
  {NoStop}%
\bibitem [{\citenamefont {Brod}\ and\ \citenamefont
  {Skodras}(2019)}]{Brod:2018lbf}%
  \BibitemOpen
  \bibfield  {author} {\bibinfo {author} {\bibfnamefont {J.}~\bibnamefont
  {Brod}}\ and\ \bibinfo {author} {\bibfnamefont {D.}~\bibnamefont {Skodras}},\
  }\href {\doibase 10.1007/JHEP01(2019)233} {\bibfield  {journal} {\bibinfo
  {journal} {JHEP}\ }\textbf {\bibinfo {volume} {01}},\ \bibinfo {pages} {233}
  (\bibinfo {year} {2019})},\ \Eprint {http://arxiv.org/abs/1811.05480}
  {arXiv:1811.05480 [hep-ph]} \BibitemShut {NoStop}%
\bibitem [{\citenamefont {Cirigliano}\ \emph
  {et~al.}(2019{\natexlab{a}})\citenamefont {Cirigliano}, \citenamefont
  {Crivellin}, \citenamefont {Dekens}, \citenamefont {de~Vries}, \citenamefont
  {Hoferichter},\ and\ \citenamefont {Mereghetti}}]{Cirigliano:2019vfc}%
  \BibitemOpen
  \bibfield  {author} {\bibinfo {author} {\bibfnamefont {V.}~\bibnamefont
  {Cirigliano}}, \bibinfo {author} {\bibfnamefont {A.}~\bibnamefont
  {Crivellin}}, \bibinfo {author} {\bibfnamefont {W.}~\bibnamefont {Dekens}},
  \bibinfo {author} {\bibfnamefont {J.}~\bibnamefont {de~Vries}}, \bibinfo
  {author} {\bibfnamefont {M.}~\bibnamefont {Hoferichter}}, \ and\ \bibinfo
  {author} {\bibfnamefont {E.}~\bibnamefont {Mereghetti}},\ }\href {\doibase
  10.1103/PhysRevLett.123.051801} {\bibfield  {journal} {\bibinfo  {journal}
  {Phys. Rev. Lett.}\ }\textbf {\bibinfo {volume} {123}},\ \bibinfo {pages}
  {051801} (\bibinfo {year} {2019}{\natexlab{a}})},\ \Eprint
  {http://arxiv.org/abs/1903.03625} {arXiv:1903.03625 [hep-ph]} \BibitemShut
  {NoStop}%
\bibitem [{\citenamefont {Baron}\ \emph {et~al.}(2014)\citenamefont {Baron}
  \emph {et~al.}}]{Baron:2013eja}%
  \BibitemOpen
  \bibfield  {author} {\bibinfo {author} {\bibfnamefont {J.}~\bibnamefont
  {Baron}} \emph {et~al.} (\bibinfo {collaboration} {ACME}),\ }\href {\doibase
  10.1126/science.1248213} {\bibfield  {journal} {\bibinfo  {journal}
  {Science}\ }\textbf {\bibinfo {volume} {343}},\ \bibinfo {pages} {269}
  (\bibinfo {year} {2014})},\ \Eprint {http://arxiv.org/abs/1310.7534}
  {arXiv:1310.7534 [physics.atom-ph]} \BibitemShut {NoStop}%
\bibitem [{\citenamefont {Andreev}\ \emph {et~al.}(2018)\citenamefont {Andreev}
  \emph {et~al.}}]{Andreev:2018ayy}%
  \BibitemOpen
  \bibfield  {author} {\bibinfo {author} {\bibfnamefont {V.}~\bibnamefont
  {Andreev}} \emph {et~al.} (\bibinfo {collaboration} {ACME}),\ }\href
  {\doibase 10.1038/s41586-018-0599-8} {\bibfield  {journal} {\bibinfo
  {journal} {Nature}\ }\textbf {\bibinfo {volume} {562}},\ \bibinfo {pages}
  {355} (\bibinfo {year} {2018})}\BibitemShut {NoStop}%
\bibitem [{\citenamefont {Baker}\ \emph {et~al.}(2006)\citenamefont {Baker}
  \emph {et~al.}}]{Baker:2006ts}%
  \BibitemOpen
  \bibfield  {author} {\bibinfo {author} {\bibfnamefont {C.~A.}\ \bibnamefont
  {Baker}} \emph {et~al.},\ }\href {\doibase 10.1103/PhysRevLett.97.131801}
  {\bibfield  {journal} {\bibinfo  {journal} {Phys. Rev. Lett.}\ }\textbf
  {\bibinfo {volume} {97}},\ \bibinfo {pages} {131801} (\bibinfo {year}
  {2006})},\ \Eprint {http://arxiv.org/abs/hep-ex/0602020}
  {arXiv:hep-ex/0602020 [hep-ex]} \BibitemShut {NoStop}%
\bibitem [{\citenamefont {Pendlebury}\ \emph {et~al.}(2015)\citenamefont
  {Pendlebury} \emph {et~al.}}]{Afach:2015sja}%
  \BibitemOpen
  \bibfield  {author} {\bibinfo {author} {\bibfnamefont {J.~M.}\ \bibnamefont
  {Pendlebury}} \emph {et~al.},\ }\href {\doibase 10.1103/PhysRevD.92.092003}
  {\bibfield  {journal} {\bibinfo  {journal} {Phys. Rev.}\ }\textbf {\bibinfo
  {volume} {D92}},\ \bibinfo {pages} {092003} (\bibinfo {year} {2015})},\
  \Eprint {http://arxiv.org/abs/1509.04411} {arXiv:1509.04411 [hep-ex]}
  \BibitemShut {NoStop}%
\bibitem [{\citenamefont {Griffith}\ \emph {et~al.}(2009)\citenamefont
  {Griffith}, \citenamefont {Swallows}, \citenamefont {Loftus}, \citenamefont
  {Romalis}, \citenamefont {Heckel},\ and\ \citenamefont
  {Fortson}}]{Griffith:2009zz}%
  \BibitemOpen
  \bibfield  {author} {\bibinfo {author} {\bibfnamefont {W.~C.}\ \bibnamefont
  {Griffith}}, \bibinfo {author} {\bibfnamefont {M.~D.}\ \bibnamefont
  {Swallows}}, \bibinfo {author} {\bibfnamefont {T.~H.}\ \bibnamefont
  {Loftus}}, \bibinfo {author} {\bibfnamefont {M.~V.}\ \bibnamefont {Romalis}},
  \bibinfo {author} {\bibfnamefont {B.~R.}\ \bibnamefont {Heckel}}, \ and\
  \bibinfo {author} {\bibfnamefont {E.~N.}\ \bibnamefont {Fortson}},\ }\href
  {\doibase 10.1103/PhysRevLett.102.101601} {\bibfield  {journal} {\bibinfo
  {journal} {Phys. Rev. Lett.}\ }\textbf {\bibinfo {volume} {102}},\ \bibinfo
  {pages} {101601} (\bibinfo {year} {2009})},\ \Eprint
  {http://arxiv.org/abs/0901.2328} {arXiv:0901.2328 [physics.atom-ph]}
  \BibitemShut {NoStop}%
\bibitem [{\citenamefont {Graner}\ \emph {et~al.}(2016)\citenamefont {Graner},
  \citenamefont {Chen}, \citenamefont {Lindahl},\ and\ \citenamefont
  {Heckel}}]{Graner:2016ses}%
  \BibitemOpen
  \bibfield  {author} {\bibinfo {author} {\bibfnamefont {B.}~\bibnamefont
  {Graner}}, \bibinfo {author} {\bibfnamefont {Y.}~\bibnamefont {Chen}},
  \bibinfo {author} {\bibfnamefont {E.~G.}\ \bibnamefont {Lindahl}}, \ and\
  \bibinfo {author} {\bibfnamefont {B.~R.}\ \bibnamefont {Heckel}},\ }\href
  {\doibase 10.1103/PhysRevLett.119.119901, 10.1103/PhysRevLett.116.161601}
  {\bibfield  {journal} {\bibinfo  {journal} {Phys. Rev. Lett.}\ }\textbf
  {\bibinfo {volume} {116}},\ \bibinfo {pages} {161601} (\bibinfo {year}
  {2016})},\ \bibinfo {note} {[Erratum: Phys. Rev. Lett. {\bf 119}, no. 11, 119901
  (2017)]},\ \Eprint {http://arxiv.org/abs/1601.04339} {arXiv:1601.04339
  [physics.atom-ph]} \BibitemShut {NoStop}%
\bibitem [{\citenamefont {Skripnikov}\ \emph {et~al.}(2013)\citenamefont
  {Skripnikov}, \citenamefont {Petrov},\ and\ \citenamefont
  {Titov}}]{etde_22253800}%
  \BibitemOpen
  \bibfield  {author} {\bibinfo {author} {\bibfnamefont {L.~V.}\ \bibnamefont
  {Skripnikov}}, \bibinfo {author} {\bibfnamefont {A.~N.}\ \bibnamefont
  {Petrov}}, \ and\ \bibinfo {author} {\bibfnamefont {A.~V.}\ \bibnamefont
  {Titov}},\ }\href {\doibase 10.1063/1.4843955} {\bibfield  {journal}
  {\bibinfo  {journal} {Journal of Chemical Physics}\ }\textbf {\bibinfo
  {volume} {139}} (\bibinfo {year} {2013}),\ 10.1063/1.4843955}\BibitemShut
  {NoStop}%
\bibitem [{\citenamefont {Fleig}\ and\ \citenamefont
  {Nayak}(2014)}]{Fleig:2014uaa}%
  \BibitemOpen
  \bibfield  {author} {\bibinfo {author} {\bibfnamefont {T.}~\bibnamefont
  {Fleig}}\ and\ \bibinfo {author} {\bibfnamefont {M.~K.}\ \bibnamefont
  {Nayak}},\ }\href {\doibase 10.1016/j.jms.2014.03.017} {\bibfield  {journal}
  {\bibinfo  {journal} {J. Molec. Spectrosc.}\ }\textbf {\bibinfo {volume}
  {300}},\ \bibinfo {pages} {16} (\bibinfo {year} {2014})},\ \Eprint
  {http://arxiv.org/abs/1401.2284} {arXiv:1401.2284 [physics.atom-ph]}
  \BibitemShut {NoStop}%
\bibitem [{\citenamefont {Bhattacharya}\ \emph
  {et~al.}(2015{\natexlab{a}})\citenamefont {Bhattacharya}, \citenamefont
  {Cirigliano}, \citenamefont {Gupta}, \citenamefont {Lin},\ and\ \citenamefont
  {Yoon}}]{Bhattacharya:2015esa}%
  \BibitemOpen
  \bibfield  {author} {\bibinfo {author} {\bibfnamefont {T.}~\bibnamefont
  {Bhattacharya}}, \bibinfo {author} {\bibfnamefont {V.}~\bibnamefont
  {Cirigliano}}, \bibinfo {author} {\bibfnamefont {R.}~\bibnamefont {Gupta}},
  \bibinfo {author} {\bibfnamefont {H.-W.}\ \bibnamefont {Lin}}, \ and\
  \bibinfo {author} {\bibfnamefont {B.}~\bibnamefont {Yoon}},\ }\href {\doibase
  10.1103/PhysRevLett.115.212002} {\bibfield  {journal} {\bibinfo  {journal}
  {Phys. Rev. Lett.}\ }\textbf {\bibinfo {volume} {115}},\ \bibinfo {pages}
  {212002} (\bibinfo {year} {2015}{\natexlab{a}})},\ \Eprint
  {http://arxiv.org/abs/1506.04196} {arXiv:1506.04196 [hep-lat]} \BibitemShut
  {NoStop}%
\bibitem [{\citenamefont {Bhattacharya}\ \emph
  {et~al.}(2015{\natexlab{b}})\citenamefont {Bhattacharya}, \citenamefont
  {Cirigliano}, \citenamefont {Cohen}, \citenamefont {Gupta}, \citenamefont
  {Joseph}, \citenamefont {Lin},\ and\ \citenamefont
  {Yoon}}]{Bhattacharya:2015wna}%
  \BibitemOpen
  \bibfield  {author} {\bibinfo {author} {\bibfnamefont {T.}~\bibnamefont
  {Bhattacharya}}, \bibinfo {author} {\bibfnamefont {V.}~\bibnamefont
  {Cirigliano}}, \bibinfo {author} {\bibfnamefont {S.}~\bibnamefont {Cohen}},
  \bibinfo {author} {\bibfnamefont {R.}~\bibnamefont {Gupta}}, \bibinfo
  {author} {\bibfnamefont {A.}~\bibnamefont {Joseph}}, \bibinfo {author}
  {\bibfnamefont {H.-W.}\ \bibnamefont {Lin}}, \ and\ \bibinfo {author}
  {\bibfnamefont {B.}~\bibnamefont {Yoon}} (\bibinfo {collaboration} {PNDME}),\
  }\href {\doibase 10.1103/PhysRevD.92.094511} {\bibfield  {journal} {\bibinfo
  {journal} {Phys. Rev.}\ }\textbf {\bibinfo {volume} {D92}},\ \bibinfo {pages}
  {094511} (\bibinfo {year} {2015}{\natexlab{b}})},\ \Eprint
  {http://arxiv.org/abs/1506.06411} {arXiv:1506.06411 [hep-lat]} \BibitemShut
  {NoStop}%
\bibitem [{\citenamefont {Gupta}\ \emph {et~al.}(2018)\citenamefont {Gupta},
  \citenamefont {Yoon}, \citenamefont {Bhattacharya}, \citenamefont
  {Cirigliano}, \citenamefont {Jang},\ and\ \citenamefont
  {Lin}}]{Gupta:2018lv}%
  \BibitemOpen
  \bibfield  {author} {\bibinfo {author} {\bibfnamefont {R.}~\bibnamefont
  {Gupta}}, \bibinfo {author} {\bibfnamefont {B.}~\bibnamefont {Yoon}},
  \bibinfo {author} {\bibfnamefont {T.}~\bibnamefont {Bhattacharya}}, \bibinfo
  {author} {\bibfnamefont {V.}~\bibnamefont {Cirigliano}}, \bibinfo {author}
  {\bibfnamefont {Y.-C.}\ \bibnamefont {Jang}}, \ and\ \bibinfo {author}
  {\bibfnamefont {H.-W.}\ \bibnamefont {Lin}},\ }\href {\doibase
  10.1103/PhysRevD.98.091501} {\bibfield  {journal} {\bibinfo  {journal} {Phys.
  Rev.}\ }\textbf {\bibinfo {volume} {D98}},\ \bibinfo {pages} {091501}
  (\bibinfo {year} {2018})},\ \Eprint {http://arxiv.org/abs/1808.07597}
  {arXiv:1808.07597 [hep-lat]} \BibitemShut {NoStop}%
\bibitem [{\citenamefont {Pospelov}\ and\ \citenamefont
  {Ritz}(2001)}]{Pospelov:2000bw}%
  \BibitemOpen
  \bibfield  {author} {\bibinfo {author} {\bibfnamefont {M.}~\bibnamefont
  {Pospelov}}\ and\ \bibinfo {author} {\bibfnamefont {A.}~\bibnamefont
  {Ritz}},\ }\href {\doibase 10.1103/PhysRevD.63.073015} {\bibfield  {journal}
  {\bibinfo  {journal} {Phys. Rev.}\ }\textbf {\bibinfo {volume} {D63}},\
  \bibinfo {pages} {073015} (\bibinfo {year} {2001})},\ \Eprint
  {http://arxiv.org/abs/hep-ph/0010037} {arXiv:hep-ph/0010037 [hep-ph]}
  \BibitemShut {NoStop}%
\bibitem [{\citenamefont {Lebedev}\ \emph {et~al.}(2004)\citenamefont
  {Lebedev}, \citenamefont {Olive}, \citenamefont {Pospelov},\ and\
  \citenamefont {Ritz}}]{Lebedev:2004va}%
  \BibitemOpen
  \bibfield  {author} {\bibinfo {author} {\bibfnamefont {O.}~\bibnamefont
  {Lebedev}}, \bibinfo {author} {\bibfnamefont {K.~A.}\ \bibnamefont {Olive}},
  \bibinfo {author} {\bibfnamefont {M.}~\bibnamefont {Pospelov}}, \ and\
  \bibinfo {author} {\bibfnamefont {A.}~\bibnamefont {Ritz}},\ }\href {\doibase
  10.1103/PhysRevD.70.016003} {\bibfield  {journal} {\bibinfo  {journal} {Phys.
  Rev.}\ }\textbf {\bibinfo {volume} {D70}},\ \bibinfo {pages} {016003}
  (\bibinfo {year} {2004})},\ \Eprint {http://arxiv.org/abs/hep-ph/0402023}
  {arXiv:hep-ph/0402023 [hep-ph]} \BibitemShut {NoStop}%
\bibitem [{\citenamefont {Hisano}\ \emph {et~al.}(2012)\citenamefont {Hisano},
  \citenamefont {Lee}, \citenamefont {Nagata},\ and\ \citenamefont
  {Shimizu}}]{Hisano:2012sc}%
  \BibitemOpen
  \bibfield  {author} {\bibinfo {author} {\bibfnamefont {J.}~\bibnamefont
  {Hisano}}, \bibinfo {author} {\bibfnamefont {J.~Y.}\ \bibnamefont {Lee}},
  \bibinfo {author} {\bibfnamefont {N.}~\bibnamefont {Nagata}}, \ and\ \bibinfo
  {author} {\bibfnamefont {Y.}~\bibnamefont {Shimizu}},\ }\href {\doibase
  10.1103/PhysRevD.85.114044} {\bibfield  {journal} {\bibinfo  {journal} {Phys.
  Rev.}\ }\textbf {\bibinfo {volume} {D85}},\ \bibinfo {pages} {114044}
  (\bibinfo {year} {2012})},\ \Eprint {http://arxiv.org/abs/1204.2653}
  {arXiv:1204.2653 [hep-ph]} \BibitemShut {NoStop}%
\bibitem [{\citenamefont {Weinberg}(1989)}]{Weinberg:1989dx}%
  \BibitemOpen
  \bibfield  {author} {\bibinfo {author} {\bibfnamefont {S.}~\bibnamefont
  {Weinberg}},\ }\href {\doibase 10.1103/PhysRevLett.63.2333} {\bibfield
  {journal} {\bibinfo  {journal} {Phys. Rev. Lett.}\ }\textbf {\bibinfo
  {volume} {63}},\ \bibinfo {pages} {2333} (\bibinfo {year}
  {1989})}\BibitemShut {NoStop}%
\bibitem [{\citenamefont {Bigi}\ and\ \citenamefont
  {Uraltsev}(1991)}]{Bigi:1990kz}%
  \BibitemOpen
  \bibfield  {author} {\bibinfo {author} {\bibfnamefont {I.~I.~Y.}\
  \bibnamefont {Bigi}}\ and\ \bibinfo {author} {\bibfnamefont {N.~G.}\
  \bibnamefont {Uraltsev}},\ }\href {\doibase 10.1016/0550-3213(91)90339-Y}
  {\bibfield  {journal} {\bibinfo  {journal} {Nucl. Phys.}\ }\textbf {\bibinfo
  {volume} {B353}},\ \bibinfo {pages} {321} (\bibinfo {year}
  {1991})}\BibitemShut {NoStop}%
\bibitem [{\citenamefont {Demir}\ \emph {et~al.}(2003)\citenamefont {Demir},
  \citenamefont {Pospelov},\ and\ \citenamefont {Ritz}}]{Demir:2002gg}%
  \BibitemOpen
  \bibfield  {author} {\bibinfo {author} {\bibfnamefont {D.~A.}\ \bibnamefont
  {Demir}}, \bibinfo {author} {\bibfnamefont {M.}~\bibnamefont {Pospelov}}, \
  and\ \bibinfo {author} {\bibfnamefont {A.}~\bibnamefont {Ritz}},\ }\href
  {\doibase 10.1103/PhysRevD.67.015007} {\bibfield  {journal} {\bibinfo
  {journal} {Phys. Rev.}\ }\textbf {\bibinfo {volume} {D67}},\ \bibinfo {pages}
  {015007} (\bibinfo {year} {2003})},\ \Eprint
  {http://arxiv.org/abs/hep-ph/0208257} {arXiv:hep-ph/0208257 [hep-ph]}
  \BibitemShut {NoStop}%
\bibitem [{\citenamefont {Bhattacharya}\ \emph
  {et~al.}(2015{\natexlab{c}})\citenamefont {Bhattacharya}, \citenamefont
  {Cirigliano}, \citenamefont {Gupta}, \citenamefont {Mereghetti},\ and\
  \citenamefont {Yoon}}]{Bhattacharya:2015rsa}%
  \BibitemOpen
  \bibfield  {author} {\bibinfo {author} {\bibfnamefont {T.}~\bibnamefont
  {Bhattacharya}}, \bibinfo {author} {\bibfnamefont {V.}~\bibnamefont
  {Cirigliano}}, \bibinfo {author} {\bibfnamefont {R.}~\bibnamefont {Gupta}},
  \bibinfo {author} {\bibfnamefont {E.}~\bibnamefont {Mereghetti}}, \ and\
  \bibinfo {author} {\bibfnamefont {B.}~\bibnamefont {Yoon}},\ }\href {\doibase
  10.1103/PhysRevD.92.114026} {\bibfield  {journal} {\bibinfo  {journal} {Phys.
  Rev.}\ }\textbf {\bibinfo {volume} {D92}},\ \bibinfo {pages} {114026}
  (\bibinfo {year} {2015}{\natexlab{c}})},\ \Eprint
  {http://arxiv.org/abs/1502.07325} {arXiv:1502.07325 [hep-ph]} \BibitemShut
  {NoStop}%
\bibitem [{\citenamefont {Bhattacharya}\ \emph {et~al.}(2016)\citenamefont
  {Bhattacharya}, \citenamefont {Cirigliano}, \citenamefont {Gupta},\ and\
  \citenamefont {Yoon}}]{Bhattacharya:2016rrc}%
  \BibitemOpen
  \bibfield  {author} {\bibinfo {author} {\bibfnamefont {T.}~\bibnamefont
  {Bhattacharya}}, \bibinfo {author} {\bibfnamefont {V.}~\bibnamefont
  {Cirigliano}}, \bibinfo {author} {\bibfnamefont {R.}~\bibnamefont {Gupta}}, \
  and\ \bibinfo {author} {\bibfnamefont {B.}~\bibnamefont {Yoon}},\ }\bibfield
  {booktitle} {\emph {\bibinfo {booktitle} {{Proceedings, 34th International
  Symposium on Lattice Field Theory (Lattice 2016): Southampton, UK, July
  24-30, 2016}}},\ }\href {\doibase 10.22323/1.256.0225} {\bibfield  {journal}
  {\bibinfo  {journal} {PoS}\ }\textbf {\bibinfo {volume} {LATTICE2016}},\
  \bibinfo {pages} {225} (\bibinfo {year} {2016})},\ \Eprint
  {http://arxiv.org/abs/1612.08438} {arXiv:1612.08438 [hep-lat]} \BibitemShut
  {NoStop}%
\bibitem [{\citenamefont {Abramczyk}\ \emph {et~al.}(2017)\citenamefont
  {Abramczyk}, \citenamefont {Aoki}, \citenamefont {Blum}, \citenamefont
  {Izubuchi}, \citenamefont {Ohki},\ and\ \citenamefont
  {Syritsyn}}]{Abramczyk:2017oxr}%
  \BibitemOpen
  \bibfield  {author} {\bibinfo {author} {\bibfnamefont {M.}~\bibnamefont
  {Abramczyk}}, \bibinfo {author} {\bibfnamefont {S.}~\bibnamefont {Aoki}},
  \bibinfo {author} {\bibfnamefont {T.}~\bibnamefont {Blum}}, \bibinfo {author}
  {\bibfnamefont {T.}~\bibnamefont {Izubuchi}}, \bibinfo {author}
  {\bibfnamefont {H.}~\bibnamefont {Ohki}}, \ and\ \bibinfo {author}
  {\bibfnamefont {S.}~\bibnamefont {Syritsyn}},\ }\href {\doibase
  10.1103/PhysRevD.96.014501} {\bibfield  {journal} {\bibinfo  {journal} {Phys.
  Rev.}\ }\textbf {\bibinfo {volume} {D96}},\ \bibinfo {pages} {014501}
  (\bibinfo {year} {2017})},\ \Eprint {http://arxiv.org/abs/1701.07792}
  {arXiv:1701.07792 [hep-lat]} \BibitemShut {NoStop}%
\bibitem [{\citenamefont {Dragos}\ \emph {et~al.}(2018)\citenamefont {Dragos},
  \citenamefont {Luu}, \citenamefont {Shindler},\ and\ \citenamefont
  {de~Vries}}]{Dragos:2017wms}%
  \BibitemOpen
  \bibfield  {author} {\bibinfo {author} {\bibfnamefont {J.}~\bibnamefont
  {Dragos}}, \bibinfo {author} {\bibfnamefont {T.}~\bibnamefont {Luu}},
  \bibinfo {author} {\bibfnamefont {A.}~\bibnamefont {Shindler}}, \ and\
  \bibinfo {author} {\bibfnamefont {J.}~\bibnamefont {de~Vries}},\ }\bibfield
  {booktitle} {\emph {\bibinfo {booktitle} {{Proceedings, 35th International
  Symposium on Lattice Field Theory (Lattice 2017): Granada, Spain, June 18-24,
  2017}}},\ }\href {\doibase 10.1051/epjconf/201817506018} {\bibfield
  {journal} {\bibinfo  {journal} {EPJ Web Conf.}\ }\textbf {\bibinfo {volume}
  {175}},\ \bibinfo {pages} {06018} (\bibinfo {year} {2018})},\ \Eprint
  {http://arxiv.org/abs/1711.04730} {arXiv:1711.04730 [hep-lat]} \BibitemShut
  {NoStop}%
\bibitem [{\citenamefont {Rizik}\ \emph {et~al.}(2018)\citenamefont {Rizik},
  \citenamefont {Monahan},\ and\ \citenamefont {Shindler}}]{Rizik:2018lrz}%
  \BibitemOpen
  \bibfield  {author} {\bibinfo {author} {\bibfnamefont {M.}~\bibnamefont
  {Rizik}}, \bibinfo {author} {\bibfnamefont {C.}~\bibnamefont {Monahan}}, \
  and\ \bibinfo {author} {\bibfnamefont {A.}~\bibnamefont {Shindler}},\
  }\bibfield  {booktitle} {\emph {\bibinfo {booktitle} {{Proceedings, 36th
  International Symposium on Lattice Field Theory (Lattice 2018): East Lansing,
  MI, United States, July 22-28, 2018}}},\ }\href {\doibase
  10.22323/1.334.0215} {\bibfield  {journal} {\bibinfo  {journal} {PoS}\
  }\textbf {\bibinfo {volume} {LATTICE2018}},\ \bibinfo {pages} {215} (\bibinfo
  {year} {2018})},\ \Eprint {http://arxiv.org/abs/1810.05637} {arXiv:1810.05637
  [hep-lat]} \BibitemShut {NoStop}%
\bibitem [{\citenamefont {Kim}\ \emph {et~al.}(2019)\citenamefont {Kim},
  \citenamefont {Dragos}, \citenamefont {Shindler}, \citenamefont {Luu},\ and\
  \citenamefont {de~Vries}}]{Kim:2018rce}%
  \BibitemOpen
  \bibfield  {author} {\bibinfo {author} {\bibfnamefont {J.}~\bibnamefont
  {Kim}}, \bibinfo {author} {\bibfnamefont {J.}~\bibnamefont {Dragos}},
  \bibinfo {author} {\bibfnamefont {A.}~\bibnamefont {Shindler}}, \bibinfo
  {author} {\bibfnamefont {T.}~\bibnamefont {Luu}}, \ and\ \bibinfo {author}
  {\bibfnamefont {J.}~\bibnamefont {de~Vries}},\ }\bibfield  {booktitle} {\emph
  {\bibinfo {booktitle} {{Proceedings, 36th International Symposium on Lattice
  Field Theory (Lattice 2018): East Lansing, MI, United States, July 22-28,
  2018}}},\ }\href {\doibase 10.22323/1.334.0260} {\bibfield  {journal}
  {\bibinfo  {journal} {PoS}\ }\textbf {\bibinfo {volume} {LATTICE2018}},\
  \bibinfo {pages} {260} (\bibinfo {year} {2019})},\ \Eprint
  {http://arxiv.org/abs/1810.10301} {arXiv:1810.10301 [hep-lat]} \BibitemShut
  {NoStop}%
\bibitem [{\citenamefont {Bhattacharya}\ \emph {et~al.}(2018)\citenamefont
  {Bhattacharya}, \citenamefont {Yoon}, \citenamefont {Gupta},\ and\
  \citenamefont {Cirigliano}}]{Bhattacharya:2018qat}%
  \BibitemOpen
  \bibfield  {author} {\bibinfo {author} {\bibfnamefont {T.}~\bibnamefont
  {Bhattacharya}}, \bibinfo {author} {\bibfnamefont {B.}~\bibnamefont {Yoon}},
  \bibinfo {author} {\bibfnamefont {R.}~\bibnamefont {Gupta}}, \ and\ \bibinfo
  {author} {\bibfnamefont {V.}~\bibnamefont {Cirigliano}}\ }(\bibinfo {year}
  {2018})\ \Eprint {http://arxiv.org/abs/1812.06233} {arXiv:1812.06233
  [hep-lat]} \BibitemShut {NoStop}%
\bibitem [{\citenamefont {Syritsyn}\ \emph {et~al.}(2019)\citenamefont
  {Syritsyn}, \citenamefont {Izubuchi},\ and\ \citenamefont
  {Ohki}}]{Syritsyn:2019vvt}%
  \BibitemOpen
  \bibfield  {author} {\bibinfo {author} {\bibfnamefont {S.}~\bibnamefont
  {Syritsyn}}, \bibinfo {author} {\bibfnamefont {T.}~\bibnamefont {Izubuchi}},
  \ and\ \bibinfo {author} {\bibfnamefont {H.}~\bibnamefont {Ohki}},\ }in\
  \href@noop {} {\emph {\bibinfo {booktitle} {{13th Conference on Quark
  Confinement and the Hadron Spectrum (Confinement XIII) Maynooth, Ireland,
  July 31-August 6, 2018}}}}\ (\bibinfo {year} {2019})\ \Eprint
  {http://arxiv.org/abs/1901.05455} {arXiv:1901.05455 [hep-lat]} \BibitemShut
  {NoStop}%
\bibitem [{\citenamefont {Gupta}(2019)}]{Gupta:2019fex}%
  \BibitemOpen
  \bibfield  {author} {\bibinfo {author} {\bibfnamefont {R.}~\bibnamefont
  {Gupta}},\ }\bibfield  {booktitle} {\emph {\bibinfo {booktitle} {{23rd
  International Symposium on Spin Physics (SPIN 2018) Ferrara, Italy, September
  10-14, 2018}}},\ }\href {\doibase 10.22323/1.346.0095} {\bibfield  {journal}
  {\bibinfo  {journal} {PoS}\ }\textbf {\bibinfo {volume} {SPIN2018}},\
  \bibinfo {pages} {095} (\bibinfo {year} {2019})},\ \Eprint
  {http://arxiv.org/abs/1904.00323} {arXiv:1904.00323 [hep-lat]} \BibitemShut
  {NoStop}%
\bibitem [{\citenamefont {Cirigliano}\ \emph
  {et~al.}(2019{\natexlab{b}})\citenamefont {Cirigliano}, \citenamefont
  {Davoudi}, \citenamefont {Bhattacharya}, \citenamefont {Izubuchi},
  \citenamefont {Shanahan}, \citenamefont {Syritsyn},\ and\ \citenamefont
  {Wagman}}]{Cirigliano:2019jig}%
  \BibitemOpen
  \bibfield  {author} {\bibinfo {author} {\bibfnamefont {V.}~\bibnamefont
  {Cirigliano}}, \bibinfo {author} {\bibfnamefont {Z.}~\bibnamefont {Davoudi}},
  \bibinfo {author} {\bibfnamefont {T.}~\bibnamefont {Bhattacharya}}, \bibinfo
  {author} {\bibfnamefont {T.}~\bibnamefont {Izubuchi}}, \bibinfo {author}
  {\bibfnamefont {P.~E.}\ \bibnamefont {Shanahan}}, \bibinfo {author}
  {\bibfnamefont {S.}~\bibnamefont {Syritsyn}}, \ and\ \bibinfo {author}
  {\bibfnamefont {M.~L.}\ \bibnamefont {Wagman}} (\bibinfo {collaboration}
  {USQCD}),\ }\href@noop {} {\  (\bibinfo {year} {2019}{\natexlab{b}})},\
  \Eprint {http://arxiv.org/abs/1904.09704} {arXiv:1904.09704 [hep-lat]}
  \BibitemShut {NoStop}%
\bibitem [{Hew(2012)}]{Hewett:2012ns}%
  \BibitemOpen
  \bibfield  {author} {\bibinfo {author} {\bibfnamefont {J.~L.}~\bibnamefont
  {Hewett}},\ }
  \href {\doibase 10.2172/1042577} (\bibinfo {year} {2012}),\ \Eprint
  {http://arxiv.org/abs/1205.2671} {arXiv:1205.2671 [hep-ex]} \BibitemShut
  {NoStop}%
\bibitem [{\citenamefont
  {Schmidt-Wellenburg}(2016)}]{Schmidt-Wellenburg:2016nfv}%
  \BibitemOpen
  \bibfield  {author} {\bibinfo {author} {\bibfnamefont {P.}~\bibnamefont
  {Schmidt-Wellenburg}},\ }\href@noop {} {\  (\bibinfo {year} {2016})},\
  \Eprint {http://arxiv.org/abs/1607.06609} {arXiv:1607.06609 [hep-ex]}
  \BibitemShut {NoStop}%
\bibitem [{\citenamefont {Ito}\ \emph {et~al.}(2018)\citenamefont {Ito} \emph
  {et~al.}}]{Ito:2017ywc}%
  \BibitemOpen
  \bibfield  {author} {\bibinfo {author} {\bibfnamefont {T.~M.}\ \bibnamefont
  {Ito}} \emph {et~al.},\ }\href {\doibase 10.1103/PhysRevC.97.012501}
  {\bibfield  {journal} {\bibinfo  {journal} {Phys. Rev.}\ }\textbf {\bibinfo
  {volume} {C97}},\ \bibinfo {pages} {012501} (\bibinfo {year} {2018})},\
  \Eprint {http://arxiv.org/abs/1710.05182} {arXiv:1710.05182
  [physics.ins-det]} \BibitemShut {NoStop}%
\bibitem [{\citenamefont {Morozov}(1984)}]{Morozov:1985ef}%
  \BibitemOpen
  \bibfield  {author} {\bibinfo {author} {\bibfnamefont {A.~{\relax Yu}.}\
  \bibnamefont {Morozov}},\ }\href@noop {} {\bibfield  {journal} {\bibinfo
  {journal} {Sov. J. Nucl. Phys.}\ }\textbf {\bibinfo {volume} {40}},\ \bibinfo
  {pages} {505} (\bibinfo {year} {1984})},\ \bibinfo {note} {[Yad. Fiz. 40, 788
  (1984)]}\BibitemShut {NoStop}%
\bibitem [{\citenamefont {Braaten}\ \emph
  {et~al.}(1990{\natexlab{a}})\citenamefont {Braaten}, \citenamefont {Li},\
  and\ \citenamefont {Yuan}}]{Braaten:1990gq}%
  \BibitemOpen
  \bibfield  {author} {\bibinfo {author} {\bibfnamefont {E.}~\bibnamefont
  {Braaten}}, \bibinfo {author} {\bibfnamefont {C.-S.}\ \bibnamefont {Li}}, \
  and\ \bibinfo {author} {\bibfnamefont {T.-C.}\ \bibnamefont {Yuan}},\ }\href
  {\doibase 10.1103/PhysRevLett.64.1709} {\bibfield  {journal} {\bibinfo
  {journal} {Phys. Rev. Lett.}\ }\textbf {\bibinfo {volume} {64}},\ \bibinfo
  {pages} {1709} (\bibinfo {year} {1990}{\natexlab{a}})}\BibitemShut {NoStop}%
\bibitem [{\citenamefont {Braaten}\ \emph
  {et~al.}(1990{\natexlab{b}})\citenamefont {Braaten}, \citenamefont {Li},\
  and\ \citenamefont {Yuan}}]{Braaten:1990zt}%
  \BibitemOpen
  \bibfield  {author} {\bibinfo {author} {\bibfnamefont {E.}~\bibnamefont
  {Braaten}}, \bibinfo {author} {\bibfnamefont {C.~S.}\ \bibnamefont {Li}}, \
  and\ \bibinfo {author} {\bibfnamefont {T.~C.}\ \bibnamefont {Yuan}},\ }\href
  {\doibase 10.1103/PhysRevD.42.276} {\bibfield  {journal} {\bibinfo  {journal}
  {Phys. Rev.}\ }\textbf {\bibinfo {volume} {D42}},\ \bibinfo {pages} {276}
  (\bibinfo {year} {1990}{\natexlab{b}})}\BibitemShut {NoStop}%
\bibitem [{\citenamefont {Chang}\ \emph {et~al.}(1992)\citenamefont {Chang},
  \citenamefont {Kephart}, \citenamefont {Keung},\ and\ \citenamefont
  {Yuan}}]{Chang:1991ry}%
  \BibitemOpen
  \bibfield  {author} {\bibinfo {author} {\bibfnamefont {D.}~\bibnamefont
  {Chang}}, \bibinfo {author} {\bibfnamefont {T.~W.}\ \bibnamefont {Kephart}},
  \bibinfo {author} {\bibfnamefont {W.-Y.}\ \bibnamefont {Keung}}, \ and\
  \bibinfo {author} {\bibfnamefont {T.~C.}\ \bibnamefont {Yuan}},\ }\href
  {\doibase 10.1103/PhysRevLett.68.439} {\bibfield  {journal} {\bibinfo
  {journal} {Phys. Rev. Lett.}\ }\textbf {\bibinfo {volume} {68}},\ \bibinfo
  {pages} {439} (\bibinfo {year} {1992})}\BibitemShut {NoStop}%
\bibitem [{\citenamefont {Booth}(1992)}]{Booth:1992iz}%
  \BibitemOpen
  \bibfield  {author} {\bibinfo {author} {\bibfnamefont {M.~J.}\ \bibnamefont
  {Booth}},\ }\href {\doibase 10.1103/PhysRevD.45.2518} {\bibfield  {journal}
  {\bibinfo  {journal} {Phys. Rev.}\ }\textbf {\bibinfo {volume} {D45}},\
  \bibinfo {pages} {2518} (\bibinfo {year} {1992})}\BibitemShut {NoStop}%
\bibitem [{\citenamefont {Haisch}\ and\ \citenamefont
  {Hala}(2019)}]{Haisch:2019xyi}%
  \BibitemOpen
  \bibfield  {author} {\bibinfo {author} {\bibfnamefont {U.}~\bibnamefont
  {Haisch}}\ and\ \bibinfo {author} {\bibfnamefont {A.}~\bibnamefont {Hala}},\
  }\href {\doibase 10.1007/JHEP11(2019)117} {\bibfield  {journal} {\bibinfo
  {journal} {JHEP}\ }\textbf {\bibinfo {volume} {11}},\ \bibinfo {pages} {117}
  (\bibinfo {year} {2019})},\ \Eprint {http://arxiv.org/abs/1909.09373}
  {arXiv:1909.09373 [hep-ph]} \BibitemShut {NoStop}%
\bibitem [{\citenamefont {Pospelov}\ and\ \citenamefont
  {Ritz}(1999)}]{Pospelov:1999ha}%
  \BibitemOpen
  \bibfield  {author} {\bibinfo {author} {\bibfnamefont {M.}~\bibnamefont
  {Pospelov}}\ and\ \bibinfo {author} {\bibfnamefont {A.}~\bibnamefont
  {Ritz}},\ }\href {\doibase 10.1103/PhysRevLett.83.2526} {\bibfield  {journal}
  {\bibinfo  {journal} {Phys. Rev. Lett.}\ }\textbf {\bibinfo {volume} {83}},\
  \bibinfo {pages} {2526} (\bibinfo {year} {1999})},\ \Eprint
  {http://arxiv.org/abs/hep-ph/9904483} {arXiv:hep-ph/9904483 [hep-ph]}
  \BibitemShut {NoStop}%
\bibitem [{\citenamefont {Pospelov}\ and\ \citenamefont
  {Ritz}(2000)}]{Pospelov:1999mv}%
  \BibitemOpen
  \bibfield  {author} {\bibinfo {author} {\bibfnamefont {M.}~\bibnamefont
  {Pospelov}}\ and\ \bibinfo {author} {\bibfnamefont {A.}~\bibnamefont
  {Ritz}},\ }\href {\doibase 10.1016/S0550-3213(99)00817-2} {\bibfield
  {journal} {\bibinfo  {journal} {Nucl. Phys.}\ }\textbf {\bibinfo {volume}
  {B573}},\ \bibinfo {pages} {177} (\bibinfo {year} {2000})},\ \Eprint
  {http://arxiv.org/abs/hep-ph/9908508} {arXiv:hep-ph/9908508 [hep-ph]}
  \BibitemShut {NoStop}%
\bibitem [{\citenamefont {Shifman}\ \emph {et~al.}(1979)\citenamefont
  {Shifman}, \citenamefont {Vainshtein},\ and\ \citenamefont
  {Zakharov}}]{Shifman:1978bx}%
  \BibitemOpen
  \bibfield  {author} {\bibinfo {author} {\bibfnamefont {M.~A.}\ \bibnamefont
  {Shifman}}, \bibinfo {author} {\bibfnamefont {A.~I.}\ \bibnamefont
  {Vainshtein}}, \ and\ \bibinfo {author} {\bibfnamefont {V.~I.}\ \bibnamefont
  {Zakharov}},\ }\href {\doibase 10.1016/0550-3213(79)90022-1} {\bibfield
  {journal} {\bibinfo  {journal} {Nucl. Phys.}\ }\textbf {\bibinfo {volume}
  {B147}},\ \bibinfo {pages} {385} (\bibinfo {year} {1979})}\BibitemShut
  {NoStop}%
\bibitem [{\citenamefont {Ioffe}\ and\ \citenamefont
  {Smilga}(1984)}]{Ioffe:1983ju}%
  \BibitemOpen
  \bibfield  {author} {\bibinfo {author} {\bibfnamefont {B.~L.}\ \bibnamefont
  {Ioffe}}\ and\ \bibinfo {author} {\bibfnamefont {A.~V.}\ \bibnamefont
  {Smilga}},\ }\href {\doibase 10.1016/0550-3213(84)90364-X} {\bibfield
  {journal} {\bibinfo  {journal} {Nucl. Phys.}\ }\textbf {\bibinfo {volume}
  {B232}},\ \bibinfo {pages} {109} (\bibinfo {year} {1984})}\BibitemShut
  {NoStop}%
\bibitem [{\citenamefont {Balitsky}\ and\ \citenamefont
  {Yung}(1983)}]{Balitsky:1983xk}%
  \BibitemOpen
  \bibfield  {author} {\bibinfo {author} {\bibfnamefont {I.~I.}\ \bibnamefont
  {Balitsky}}\ and\ \bibinfo {author} {\bibfnamefont {A.~V.}\ \bibnamefont
  {Yung}},\ }\href {\doibase 10.1016/0370-2693(83)90676-7} {\bibfield
  {journal} {\bibinfo  {journal} {Phys. Lett.}\ }\textbf {\bibinfo {volume}
  {129B}},\ \bibinfo {pages} {328} (\bibinfo {year} {1983})}\BibitemShut
  {NoStop}%
\bibitem [{\citenamefont {Novikov}\ \emph {et~al.}(1984)\citenamefont
  {Novikov}, \citenamefont {Shifman}, \citenamefont {Vainshtein},\ and\
  \citenamefont {Zakharov}}]{Novikov:1983gd}%
  \BibitemOpen
  \bibfield  {author} {\bibinfo {author} {\bibfnamefont {V.~A.}\ \bibnamefont
  {Novikov}}, \bibinfo {author} {\bibfnamefont {M.~A.}\ \bibnamefont
  {Shifman}}, \bibinfo {author} {\bibfnamefont {A.~I.}\ \bibnamefont
  {Vainshtein}}, \ and\ \bibinfo {author} {\bibfnamefont {V.~I.}\ \bibnamefont
  {Zakharov}},\ }\href@noop {} {\bibfield  {journal} {\bibinfo  {journal}
  {Fortsch. Phys.}\ }\textbf {\bibinfo {volume} {32}},\ \bibinfo {pages} {585}
  (\bibinfo {year} {1984})}\BibitemShut {NoStop}%
\bibitem [{\citenamefont {Reinders}\ \emph {et~al.}(1985)\citenamefont
  {Reinders}, \citenamefont {Rubinstein},\ and\ \citenamefont
  {Yazaki}}]{Reinders:1984sr}%
  \BibitemOpen
  \bibfield  {author} {\bibinfo {author} {\bibfnamefont {L.~J.}\ \bibnamefont
  {Reinders}}, \bibinfo {author} {\bibfnamefont {H.}~\bibnamefont
  {Rubinstein}}, \ and\ \bibinfo {author} {\bibfnamefont {S.}~\bibnamefont
  {Yazaki}},\ }\href {\doibase 10.1016/0370-1573(85)90065-1} {\bibfield
  {journal} {\bibinfo  {journal} {Phys. Rept.}\ }\textbf {\bibinfo {volume}
  {127}},\ \bibinfo {pages} {1} (\bibinfo {year} {1985})}\BibitemShut {NoStop}%
\bibitem [{\citenamefont {Colangelo}\ and\ \citenamefont
  {Khodjamirian}(2000)}]{Colangelo:2000dp}%
  \BibitemOpen
  \bibfield  {author} {\bibinfo {author} {\bibfnamefont {P.}~\bibnamefont
  {Colangelo}}\ and\ \bibinfo {author} {\bibfnamefont {A.}~\bibnamefont
  {Khodjamirian}},\ }\href {\doibase 10.1142/9789812810458_0033} {\bibfield
  {journal} {\bibinfo  {journal} {At The Frontier of Particle Physics},
  \bibinfo {pages} {1495}} (\bibinfo {year} {2000})},\ \Eprint
  {http://arxiv.org/abs/hep-ph/0010175} {arXiv:hep-ph/0010175 [hep-ph]}
  \BibitemShut {NoStop}%
\bibitem [{\citenamefont {Ioffe}\ \emph {et~al.}(2010)\citenamefont {Ioffe},
  \citenamefont {Fadin},\ and\ \citenamefont {Lipatov}}]{Ioffe:2010zz}%
  \BibitemOpen
  \bibfield  {author} {\bibinfo {author} {\bibfnamefont {B.~L.}\ \bibnamefont
  {Ioffe}}, \bibinfo {author} {\bibfnamefont {V.~S.}\ \bibnamefont {Fadin}}, \
  and\ \bibinfo {author} {\bibfnamefont {L.~N.}\ \bibnamefont {Lipatov}},
  }\href {\doibase 10.1017/CBO9780511711817} {\emph {\bibinfo {title} {{Quantum
  chromodynamics: Perturbative and nonperturbative aspects}}}},\ Vol.~\bibinfo
  {volume} {30}\ (\bibinfo  {publisher} {Cambridge Univ. Press},\ \bibinfo
  {year} {2010})\BibitemShut {NoStop}%
\bibitem [{\citenamefont {Chan}\ \emph {et~al.}(1997)\citenamefont {Chan},
  \citenamefont {Henley},\ and\ \citenamefont {Meissner}}]{Chan:1997fw}%
  \BibitemOpen
  \bibfield  {author} {\bibinfo {author} {\bibfnamefont {C.-T.}\ \bibnamefont
  {Chan}}, \bibinfo {author} {\bibfnamefont {E.~M.}\ \bibnamefont {Henley}}, \
  and\ \bibinfo {author} {\bibfnamefont {T.}~\bibnamefont {Meissner}},\
  }\href@noop {} {\  (\bibinfo {year} {1997})},\ \Eprint
  {http://arxiv.org/abs/hep-ph/9905317} {arXiv:hep-ph/9905317 [hep-ph]}
  \BibitemShut {NoStop}%
\bibitem [{\citenamefont {Tanabashi}\ \emph {et~al.}(2018)\citenamefont
  {Tanabashi} \emph {et~al.}}]{Tanabashi:2018oca}%
  \BibitemOpen
  \bibfield  {author} {\bibinfo {author} {\bibfnamefont {M.}~\bibnamefont
  {Tanabashi}} \emph {et~al.} (\bibinfo {collaboration} {Particle Data
  Group}),\ }\href {\doibase 10.1103/PhysRevD.98.030001} {\bibfield  {journal}
  {\bibinfo  {journal} {Phys. Rev.}\ }\textbf {\bibinfo {volume} {D98}},\
  \bibinfo {pages} {030001} (\bibinfo {year} {2018})}\BibitemShut {NoStop}%
\bibitem [{\citenamefont {Chung}\ \emph {et~al.}(1984)\citenamefont {Chung},
  \citenamefont {Dosch}, \citenamefont {Kremer},\ and\ \citenamefont
  {Schall}}]{Chung:1984gr}%
  \BibitemOpen
  \bibfield  {author} {\bibinfo {author} {\bibfnamefont {Y.}~\bibnamefont
  {Chung}}, \bibinfo {author} {\bibfnamefont {H.~G.}\ \bibnamefont {Dosch}},
  \bibinfo {author} {\bibfnamefont {M.}~\bibnamefont {Kremer}}, \ and\ \bibinfo
  {author} {\bibfnamefont {D.}~\bibnamefont {Schall}},\ }\href {\doibase
  10.1007/BF01557473} {\bibfield  {journal} {\bibinfo  {journal} {Z. Phys.}\
  }\textbf {\bibinfo {volume} {C25}},\ \bibinfo {pages} {151} (\bibinfo {year}
  {1984})}\BibitemShut {NoStop}%
\bibitem [{\citenamefont {Leinweber}(1995)}]{Leinweber:1994nm}%
  \BibitemOpen
  \bibfield  {author} {\bibinfo {author} {\bibfnamefont {D.~B.}\ \bibnamefont
  {Leinweber}},\ }\href {\doibase 10.1103/PhysRevD.51.6383} {\bibfield
  {journal} {\bibinfo  {journal} {Phys. Rev.}\ }\textbf {\bibinfo {volume}
  {D51}},\ \bibinfo {pages} {6383} (\bibinfo {year} {1995})},\ \Eprint
  {http://arxiv.org/abs/nucl-th/9406001} {arXiv:nucl-th/9406001 [nucl-th]}
  \BibitemShut {NoStop}%
\bibitem [{\citenamefont {Leinweber}(1997)}]{Leinweber:1995fn}%
  \BibitemOpen
  \bibfield  {author} {\bibinfo {author} {\bibfnamefont {D.~B.}\ \bibnamefont
  {Leinweber}},\ }\href {\doibase 10.1006/aphy.1996.5641} {\bibfield  {journal}
  {\bibinfo  {journal} {Annals Phys.}\ }\textbf {\bibinfo {volume} {254}},\
  \bibinfo {pages} {328} (\bibinfo {year} {1997})},\ \Eprint
  {http://arxiv.org/abs/nucl-th/9510051} {arXiv:nucl-th/9510051 [nucl-th]}
  \BibitemShut {NoStop}%
\bibitem [{\citenamefont {Peccei}\ and\ \citenamefont
  {Quinn}(1977)}]{Peccei:1977hh}%
  \BibitemOpen
  \bibfield  {author} {\bibinfo {author} {\bibfnamefont {R.~D.}\ \bibnamefont
  {Peccei}}\ and\ \bibinfo {author} {\bibfnamefont {H.~R.}\ \bibnamefont
  {Quinn}},\ }\href {\doibase 10.1103/PhysRevLett.38.1440} {\bibfield
  {journal} {\bibinfo  {journal} {Phys. Rev. Lett.}\ }\textbf {\bibinfo
  {volume} {38}},\ \bibinfo {pages} {1440} (\bibinfo {year} {1977})} \BibitemShut {NoStop}%
\bibitem [{\citenamefont {Gubler}\ and\ \citenamefont
  {Satow}(2019)}]{Gubler:2018ctz}%
  \BibitemOpen
  \bibfield  {author} {\bibinfo {author} {\bibfnamefont {P.}~\bibnamefont
  {Gubler}}\ and\ \bibinfo {author} {\bibfnamefont {D.}~\bibnamefont {Satow}},\
  }\href {\doibase 10.1016/j.ppnp.2019.02.005} {\bibfield  {journal} {\bibinfo
  {journal} {Prog. Part. Nucl. Phys.}\ }\textbf {\bibinfo {volume} {106}},\
  \bibinfo {pages} {1} (\bibinfo {year} {2019})},\ \Eprint
  {http://arxiv.org/abs/1812.00385} {arXiv:1812.00385 [hep-ph]} \BibitemShut
  {NoStop}%
\bibitem [{\citenamefont {Gubler}(2013)}]{gubler}%
  \BibitemOpen
  \bibfield  {author} {\bibinfo {author} {\bibfnamefont {P.}~\bibnamefont
  {Gubler}},\ }\href {\doibase 10.1007/978-4-431-54318-3} {\emph {\bibinfo
  {title} {A Bayesian Analysis of QCD Sum Rules}}},\ \bibinfo {edition} {1st}\
  ed.\ (\bibinfo  {publisher} {Springer Japan},\ \bibinfo {year}
  {2013})\BibitemShut {NoStop}%
\bibitem [{\citenamefont {Belyaev}\ and\ \citenamefont
  {Ioffe}(1982)}]{Belyaev:1982sa}%
  \BibitemOpen
  \bibfield  {author} {\bibinfo {author} {\bibfnamefont {V.~M.}\ \bibnamefont
  {Belyaev}}\ and\ \bibinfo {author} {\bibfnamefont {B.~L.}\ \bibnamefont
  {Ioffe}},\ }\href@noop {} {\bibfield  {journal} {\bibinfo  {journal} {Sov.
  Phys. JETP}\ }\textbf {\bibinfo {volume} {56}},\ \bibinfo {pages} {493}
  (\bibinfo {year} {1982})},\ \bibinfo {note} {[Zh. Eksp. Teor. Fiz.83, 876
  (1982)]}\BibitemShut {NoStop}%
\bibitem [{\citenamefont {Ioffe}(1981)}]{Ioffe:1981kw}%
  \BibitemOpen
  \bibfield  {author} {\bibinfo {author} {\bibfnamefont {B.~L.}\ \bibnamefont
  {Ioffe}},\ }\href {\doibase 10.1016/0550-3213(81)90259-5}
  {\bibfield  {journal} {\bibinfo  {journal}
  {Nucl. Phys.}\ }\textbf {\bibinfo {volume} {B188}},\ \bibinfo {pages} {317}
  (\bibinfo {year} {1981})},\ \bibinfo {note} {[Erratum: Nucl. Phys. {\bf B191}, 591
  (1981)]}\BibitemShut {NoStop}%
\bibitem [{\citenamefont {Ioffe}(1983)}]{Ioffe:1982ce}%
  \BibitemOpen
  \bibfield  {author} {\bibinfo {author} {\bibfnamefont {B.~L.}\ \bibnamefont
  {Ioffe}},\ }\href {\doibase 10.1007/BF01571709} {\bibfield  {journal}
  {\bibinfo  {journal} {Z. Phys.}\ }\textbf {\bibinfo {volume} {C18}},\
  \bibinfo {pages} {67} (\bibinfo {year} {1983})}\BibitemShut {NoStop}%
\bibitem [{\citenamefont {Ioffe}\ and\ \citenamefont
  {Smilga}(1983)}]{Ioffe:1983cj}%
  \BibitemOpen
  \bibfield  {author} {\bibinfo {author} {\bibfnamefont {B.~L.}\ \bibnamefont
  {Ioffe}}\ and\ \bibinfo {author} {\bibfnamefont {A.~V.}\ \bibnamefont
  {Smilga}},\ }\href@noop {} {\bibfield  {journal} {\bibinfo  {journal} {JETP
  Lett.}\ }\textbf {\bibinfo {volume} {37}},\ \bibinfo {pages} {298} (\bibinfo
  {year} {1983})},\ \bibinfo {note} {[Erratum: JETP Lett. {\bf 38}, 48
  (1983)]}\BibitemShut {NoStop}%
\bibitem [{\citenamefont {Aw}\ \emph {et~al.}(1999)\citenamefont {Aw},
  \citenamefont {Banerjee},\ and\ \citenamefont {Forkel}}]{Aw:1999gz}%
  \BibitemOpen
  \bibfield  {author} {\bibinfo {author} {\bibfnamefont {M.}~\bibnamefont
  {Aw}}, \bibinfo {author} {\bibfnamefont {M.~K.}\ \bibnamefont {Banerjee}}, \
  and\ \bibinfo {author} {\bibfnamefont {H.}~\bibnamefont {Forkel}},\ }\href
  {\doibase 10.1016/S0370-2693(99)00358-5} {\bibfield  {journal} {\bibinfo
  {journal} {Phys. Lett.}\ }\textbf {\bibinfo {volume} {B454}},\ \bibinfo
  {pages} {147} (\bibinfo {year} {1999})},\ \Eprint
  {http://arxiv.org/abs/hep-ph/9902458} {arXiv:hep-ph/9902458 [hep-ph]}
  \BibitemShut {NoStop}%
\bibitem [{\citenamefont {Nieves}\ and\ \citenamefont
  {Pal}(2004)}]{Nieves:2003in}%
  \BibitemOpen
  \bibfield  {author} {\bibinfo {author} {\bibfnamefont {J.~F.}\ \bibnamefont
  {Nieves}}\ and\ \bibinfo {author} {\bibfnamefont {P.~B.}\ \bibnamefont
  {Pal}},\ }\href {\doibase 10.1119/1.1757445} {\bibfield  {journal} {\bibinfo
  {journal} {Am. J. Phys.}\ }\textbf {\bibinfo {volume} {72}},\ \bibinfo
  {pages} {1100} (\bibinfo {year} {2004})},\ \Eprint
  {http://arxiv.org/abs/hep-ph/0306087} {arXiv:hep-ph/0306087 [hep-ph]}
  \BibitemShut {NoStop}%
\bibitem [{\citenamefont {Nishi}(2005)}]{Nishi:2004st}%
  \BibitemOpen
  \bibfield  {author} {\bibinfo {author} {\bibfnamefont {C.~C.}\ \bibnamefont
  {Nishi}},\ }\href {\doibase 10.1119/1.2074087} {\bibfield  {journal}
  {\bibinfo  {journal} {Am. J. Phys.}\ }\textbf {\bibinfo {volume} {73}},\
  \bibinfo {pages} {1160} (\bibinfo {year} {2005})},\ \Eprint
  {http://arxiv.org/abs/hep-ph/0412245} {arXiv:hep-ph/0412245 [hep-ph]}
  \BibitemShut {NoStop}%
\bibitem [{\citenamefont {Fadin}\ and\ \citenamefont
  {Fiore}(2005)}]{Fadin:2005zj}%
  \BibitemOpen
  \bibfield  {author} {\bibinfo {author} {\bibfnamefont {V.~S.}\ \bibnamefont
  {Fadin}}\ and\ \bibinfo {author} {\bibfnamefont {R.}~\bibnamefont {Fiore}},\
  }\href {\doibase 10.1103/PhysRevD.72.014018} {\bibfield  {journal} {\bibinfo
  {journal} {Phys. Rev.}\ }\textbf {\bibinfo {volume} {D72}},\ \bibinfo {pages}
  {014018} (\bibinfo {year} {2005})},\ \Eprint
  {http://arxiv.org/abs/hep-ph/0502045} {arXiv:hep-ph/0502045 [hep-ph]}
  \BibitemShut {NoStop}%
\bibitem [{\citenamefont {Oganesian}(1998)}]{Oganesian:1997wz}%
  \BibitemOpen
  \bibfield  {author} {\bibinfo {author} {\bibfnamefont {A.~G.}\ \bibnamefont
  {Oganesian}},\ }\bibfield  {booktitle} {\emph {\bibinfo {booktitle}
  {{Nonaccelerator new physics. Proceedings, 1st International Workshop,
  NANP'97, Dubna, Russia, July 7-11, 1997}}},\ }\href@noop {} {\bibfield
  {journal} {\bibinfo  {journal} {Phys. Atom. Nucl.}\ }\textbf {\bibinfo
  {volume} {61}},\ \bibinfo {pages} {1359} (\bibinfo {year} {1998})},\ \bibinfo
  {note} {[Yad. Fiz. 61, 1464 (1998)]},\ \Eprint
  {http://arxiv.org/abs/hep-ph/9704435} {arXiv:hep-ph/9704435 [hep-ph]}
  \BibitemShut {NoStop}%
\bibitem [{\citenamefont {Polyakov}\ and\ \citenamefont
  {Weiss}(1998)}]{Polyakov:1998ip}%
  \BibitemOpen
  \bibfield  {author} {\bibinfo {author} {\bibfnamefont {M.~V.}\ \bibnamefont
  {Polyakov}}\ and\ \bibinfo {author} {\bibfnamefont {C.}~\bibnamefont
  {Weiss}},\ }\href {\doibase 10.1103/PhysRevD.57.4471} {\bibfield  {journal}
  {\bibinfo  {journal} {Phys. Rev.}\ }\textbf {\bibinfo {volume} {D57}},\
  \bibinfo {pages} {4471} (\bibinfo {year} {1998})},\ \Eprint
  {http://arxiv.org/abs/hep-ph/9710534} {arXiv:hep-ph/9710534 [hep-ph]}
  \BibitemShut {NoStop}%
\bibitem [{\citenamefont {Shuryak}(1982)}]{Shuryak:1981ff}%
  \BibitemOpen
  \bibfield  {author} {\bibinfo {author} {\bibfnamefont {E.~V.}\ \bibnamefont
  {Shuryak}},\ }\href {\doibase 10.1016/0550-3213(82)90478-3} {\bibfield
  {journal} {\bibinfo  {journal} {Nucl. Phys.}\ }\textbf {\bibinfo {volume}
  {B203}},\ \bibinfo {pages} {93} (\bibinfo {year} {1982})}\BibitemShut
  {NoStop}%
\bibitem [{\citenamefont {Diakonov}\ and\ \citenamefont
  {Petrov}(1984)}]{Diakonov:1983hh}%
  \BibitemOpen
  \bibfield  {author} {\bibinfo {author} {\bibfnamefont {D.}~\bibnamefont
  {Diakonov}}\ and\ \bibinfo {author} {\bibfnamefont {V.~{\relax Yu}.}\
  \bibnamefont {Petrov}},\ }\href {\doibase 10.1016/0550-3213(84)90432-2}
  {\bibfield  {journal} {\bibinfo  {journal} {Nucl. Phys.}\ }\textbf {\bibinfo
  {volume} {B245}},\ \bibinfo {pages} {259} (\bibinfo {year}
  {1984})}\BibitemShut {NoStop}%
\bibitem [{\citenamefont {Diakonov}\ and\ \citenamefont
  {Petrov}(1986)}]{Diakonov:1985eg}%
  \BibitemOpen
  \bibfield  {author} {\bibinfo {author} {\bibfnamefont {D.}~\bibnamefont
  {Diakonov}}\ and\ \bibinfo {author} {\bibfnamefont {V.~{\relax Yu}.}\
  \bibnamefont {Petrov}},\ }\href {\doibase 10.1016/0550-3213(86)90011-8}
  {\bibfield  {journal} {\bibinfo  {journal} {Nucl. Phys.}\ }\textbf {\bibinfo
  {volume} {B272}},\ \bibinfo {pages} {457} (\bibinfo {year}
  {1986})}\BibitemShut {NoStop}%
\bibitem [{\citenamefont {Sch{\"a}fer}\ and\ \citenamefont
  {Shuryak}(1998)}]{Schafer:1996wv}%
  \BibitemOpen
  \bibfield  {author} {\bibinfo {author} {\bibfnamefont {T.}~\bibnamefont
  {Sch{\"a}fer}}\ and\ \bibinfo {author} {\bibfnamefont {E.~V.}\ \bibnamefont
  {Shuryak}},\ }\href {\doibase 10.1103/RevModPhys.70.323} {\bibfield
  {journal} {\bibinfo  {journal} {Rev. Mod. Phys.}\ }\textbf {\bibinfo {volume}
  {70}},\ \bibinfo {pages} {323} (\bibinfo {year} {1998})},\ \Eprint
  {http://arxiv.org/abs/hep-ph/9610451} {arXiv:hep-ph/9610451 [hep-ph]}
  \BibitemShut {NoStop}%
\bibitem [{\citenamefont {Chu}\ \emph {et~al.}(1994)\citenamefont {Chu},
  \citenamefont {Grandy}, \citenamefont {Huang},\ and\ \citenamefont
  {Negele}}]{Chu:1994vi}%
  \BibitemOpen
  \bibfield  {author} {\bibinfo {author} {\bibfnamefont {M.~C.}\ \bibnamefont
  {Chu}}, \bibinfo {author} {\bibfnamefont {J.~M.}\ \bibnamefont {Grandy}},
  \bibinfo {author} {\bibfnamefont {S.}~\bibnamefont {Huang}}, \ and\ \bibinfo
  {author} {\bibfnamefont {J.~W.}\ \bibnamefont {Negele}},\ }\href {\doibase
  10.1103/PhysRevD.49.6039} {\bibfield  {journal} {\bibinfo  {journal} {Phys.
  Rev.}\ }\textbf {\bibinfo {volume} {D49}},\ \bibinfo {pages} {6039} (\bibinfo
  {year} {1994})},\ \Eprint {http://arxiv.org/abs/hep-lat/9312071}
  {arXiv:hep-lat/9312071 [hep-lat]} \BibitemShut {NoStop}%
\bibitem [{\citenamefont {Negele}(1999)}]{Negele:1998ev}%
  \BibitemOpen
  \bibfield  {author} {\bibinfo {author} {\bibfnamefont {J.~W.}\ \bibnamefont
  {Negele}},\ }\bibfield  {booktitle} {\emph {\bibinfo {booktitle} {{Lattice
  Field Theory. Proceedings: 16th International Symposium, Lattice '98,
  Boulder, USA, Jul 13-18, 1998}}},\ }\href {\doibase
  10.1016/S0920-5632(99)85010-5} {\bibfield  {journal} {\bibinfo  {journal}
  {Nucl. Phys. Proc. Suppl.}\ }\textbf {\bibinfo {volume} {73}},\ \bibinfo
  {pages} {92} (\bibinfo {year} {1999})},\ \Eprint
  {http://arxiv.org/abs/hep-lat/9810053} {arXiv:hep-lat/9810053 [hep-lat]}
  \BibitemShut {NoStop}%
\bibitem [{\citenamefont {DeGrand}(2001)}]{DeGrand:2001tm}%
  \BibitemOpen
  \bibfield  {author} {\bibinfo {author} {\bibfnamefont {T.~A.}\ \bibnamefont
  {DeGrand}},\ }\href {\doibase 10.1103/PhysRevD.64.094508} {\bibfield
  {journal} {\bibinfo  {journal} {Phys. Rev.}\ }\textbf {\bibinfo {volume}
  {D64}},\ \bibinfo {pages} {094508} (\bibinfo {year} {2001})},\ \Eprint
  {http://arxiv.org/abs/hep-lat/0106001} {arXiv:hep-lat/0106001 [hep-lat]}
  \BibitemShut {NoStop}%
\bibitem [{\citenamefont {Faccioli}\ and\ \citenamefont
  {DeGrand}(2003)}]{Faccioli:2003qz}%
  \BibitemOpen
  \bibfield  {author} {\bibinfo {author} {\bibfnamefont {P.}~\bibnamefont
  {Faccioli}}\ and\ \bibinfo {author} {\bibfnamefont {T.~A.}\ \bibnamefont
  {DeGrand}},\ }\href {\doibase 10.1103/PhysRevLett.91.182001} {\bibfield
  {journal} {\bibinfo  {journal} {Phys. Rev. Lett.}\ }\textbf {\bibinfo
  {volume} {91}},\ \bibinfo {pages} {182001} (\bibinfo {year} {2003})},\
  \Eprint {http://arxiv.org/abs/hep-ph/0304219} {arXiv:hep-ph/0304219 [hep-ph]}
  \BibitemShut {NoStop}%
\bibitem [{\citenamefont {Bowman}\ \emph {et~al.}(2004)\citenamefont {Bowman},
  \citenamefont {Heller}, \citenamefont {Leinweber}, \citenamefont {Williams},\
  and\ \citenamefont {Zhang}}]{Bowman:2004xi}%
  \BibitemOpen
  \bibfield  {author} {\bibinfo {author} {\bibfnamefont {P.~O.}\ \bibnamefont
  {Bowman}}, \bibinfo {author} {\bibfnamefont {U.~M.}\ \bibnamefont {Heller}},
  \bibinfo {author} {\bibfnamefont {D.~B.}\ \bibnamefont {Leinweber}}, \bibinfo
  {author} {\bibfnamefont {A.~G.}\ \bibnamefont {Williams}}, \ and\ \bibinfo
  {author} {\bibfnamefont {J.-b.}\ \bibnamefont {Zhang}},\ }\bibfield
  {booktitle} {\emph {\bibinfo {booktitle} {{Lattice hadron physics.
  Proceedings, 2nd Topical Workshop, LHP 2003, Cairns, Australia, July 22-30,
  2003}}},\ }\href {\doibase 10.1016/S0920-5632(03)02454-X} {\bibfield
  {journal} {\bibinfo  {journal} {Nucl. Phys. Proc. Suppl.}\ }\textbf {\bibinfo
  {volume} {128}},\ \bibinfo {pages} {23} (\bibinfo {year} {2004})},\ \Eprint
  {http://arxiv.org/abs/hep-lat/0403002} {arXiv:hep-lat/0403002 [hep-lat]}
  \BibitemShut {NoStop}%
\bibitem [{\citenamefont {Cristoforetti}\ \emph {et~al.}(2007)\citenamefont
  {Cristoforetti}, \citenamefont {Faccioli}, \citenamefont {Traini},\ and\
  \citenamefont {Negele}}]{Cristoforetti:2006ar}%
  \BibitemOpen
  \bibfield  {author} {\bibinfo {author} {\bibfnamefont {M.}~\bibnamefont
  {Cristoforetti}}, \bibinfo {author} {\bibfnamefont {P.}~\bibnamefont
  {Faccioli}}, \bibinfo {author} {\bibfnamefont {M.~C.}\ \bibnamefont
  {Traini}}, \ and\ \bibinfo {author} {\bibfnamefont {J.~W.}\ \bibnamefont
  {Negele}},\ }\href {\doibase 10.1103/PhysRevD.75.034008} {\bibfield
  {journal} {\bibinfo  {journal} {Phys. Rev.}\ }\textbf {\bibinfo {volume}
  {D75}},\ \bibinfo {pages} {034008} (\bibinfo {year} {2007})},\ \Eprint
  {http://arxiv.org/abs/hep-ph/0605256} {arXiv:hep-ph/0605256 [hep-ph]}
  \BibitemShut {NoStop}%
\bibitem [{\citenamefont {Ohl}(1995)}]{Ohl:1995kr}%
  \BibitemOpen
  \bibfield  {author} {\bibinfo {author} {\bibfnamefont {T.}~\bibnamefont
  {Ohl}},\ }\href {\doibase 10.1016/0010-4655(95)90137-S} {\bibfield  {journal}
  {\bibinfo  {journal} {Comput. Phys. Commun.}\ }\textbf {\bibinfo {volume}
  {90}},\ \bibinfo {pages} {340} (\bibinfo {year} {1995})},\ \Eprint
  {http://arxiv.org/abs/hep-ph/9505351} {arXiv:hep-ph/9505351 [hep-ph]}
  \BibitemShut {NoStop}%
\bibitem [{\citenamefont {Shtabovenko}\ \emph {et~al.}(2016)\citenamefont
  {Shtabovenko}, \citenamefont {Mertig},\ and\ \citenamefont
  {Orellana}}]{Shtabovenko:2016sxi}%
  \BibitemOpen
  \bibfield  {author} {\bibinfo {author} {\bibfnamefont {V.}~\bibnamefont
  {Shtabovenko}}, \bibinfo {author} {\bibfnamefont {R.}~\bibnamefont {Mertig}},
  \ and\ \bibinfo {author} {\bibfnamefont {F.}~\bibnamefont {Orellana}},\
  }\href {\doibase 10.1016/j.cpc.2016.06.008} {\bibfield  {journal} {\bibinfo
  {journal} {Comput. Phys. Commun.}\ }\textbf {\bibinfo {volume} {207}},\
  \bibinfo {pages} {432} (\bibinfo {year} {2016})},\ \Eprint
  {http://arxiv.org/abs/1601.01167} {arXiv:1601.01167 [hep-ph]} \BibitemShut
  {NoStop}%
\bibitem [{\citenamefont {Jamin}\ and\ \citenamefont
  {Lautenbacher}(1993)}]{Jamin:1991dp}%
  \BibitemOpen
  \bibfield  {author} {\bibinfo {author} {\bibfnamefont {M.}~\bibnamefont
  {Jamin}}\ and\ \bibinfo {author} {\bibfnamefont {M.~E.}\ \bibnamefont
  {Lautenbacher}},\ }\href {\doibase 10.1016/0010-4655(93)90097-V} {\bibfield
  {journal} {\bibinfo  {journal} {Comput. Phys. Commun.}\ }\textbf {\bibinfo
  {volume} {74}},\ \bibinfo {pages} {265} (\bibinfo {year} {1993})}\BibitemShut
  {NoStop}%
\bibitem [{\citenamefont {Yamanaka}\ and\ \citenamefont
  {Hiyama}(2020)}]{Yamanaka:2020kjo}%
  \BibitemOpen
  \bibfield  {author} {\bibinfo {author} {\bibfnamefont {N.}~\bibnamefont
  {Yamanaka}}\ and\ \bibinfo {author} {\bibfnamefont {E.}~\bibnamefont
  {Hiyama}},\ }\href@noop {} {\  (\bibinfo {year} {2020})},\ \Eprint
  {http://arxiv.org/abs/2011.02531} {arXiv:2011.02531 [hep-ph]} \BibitemShut
  {NoStop}%
\bibitem [{\citenamefont {Hatta}(2020)}]{Hatta:2020riw}%
  \BibitemOpen
  \bibfield  {author} {\bibinfo {author} {\bibfnamefont {Y.}~\bibnamefont
  {Hatta}},\ }\href@noop {} {\  (\bibinfo {year} {2020})},\ \Eprint
  {http://arxiv.org/abs/2012.01865} {arXiv:2012.01865 [hep-ph]} \BibitemShut
  {NoStop}%
\bibitem [{\citenamefont {Fock}(1937)}]{Fock:1937dy}%
  \BibitemOpen
  \bibfield  {author} {\bibinfo {author} {\bibfnamefont {V.}~\bibnamefont
  {Fock}},\ }\href@noop {} {\bibfield  {journal} {\bibinfo  {journal} {Phys. Z.
  Sowjetunion}\ }\textbf {\bibinfo {volume} {12}},\ \bibinfo {pages} {404}
  (\bibinfo {year} {1937})}\BibitemShut {NoStop}%
\bibitem [{\citenamefont {Schwinger}(1951)}]{Schwinger:1951nm}%
  \BibitemOpen
  \bibfield  {author} {\bibinfo {author} {\bibfnamefont {J.~S.}\ \bibnamefont
  {Schwinger}},\ }\href {\doibase 10.1103/PhysRev.82.664} {\bibfield  {journal}
  {\bibinfo  {journal} {Phys. Rev.}\ }\textbf {\bibinfo {volume} {82}},\
  \bibinfo {pages} {664} (\bibinfo {year} {1951})}\BibitemShut {NoStop}%
\end{thebibliography}

%

\end{document}